\documentclass[11pt]{article}  
\usepackage{amsfonts}

\usepackage{geometry}

\usepackage{animate} 
\usepackage{amsfonts} 
\usepackage{graphicx}
\usepackage{color}
\usepackage{rotating}
\usepackage{amsmath}
\usepackage{amssymb}
\usepackage{amsthm}
\usepackage{float}
\usepackage{url}
\usepackage{subfigure}

\usepackage{natbib} 

\usepackage{enumerate}
\usepackage{enumitem}

\usepackage{dsfont}

\usepackage{algorithm}
\usepackage{algorithmic}

\usepackage{array,arydshln}

\usepackage{soul} 

\usepackage{hyperref} 

\usepackage{xcolor}
\hypersetup{
	colorlinks,
	linkcolor= {blue!90!black}, 
	citecolor={blue!60!black},
	urlcolor={blue!80!black}
}

\newcommand\bb {\mathbf b}

\newcommand\be {\mathbf e}

\newcommand\bi {\mathbf i}

\newcommand\bene {\mathbf n}

\newcommand\bu {\mathbf u}

\newcommand\bx {\mathbf x}
\newcommand\by {\mathbf y}
\newcommand\bz {\mathbf z}

\newcommand\bB {\mathbf B}

\newcommand\bH {\mathbf H}
\newcommand\bI {\mathbf I}

\newcommand\bU {\mathbf U}

\newcommand\bX {\mathbf X}

\newcommand\indica {\mathbb{I}}

\newcommand\wb {\widehat{{b}}}
\newcommand\wbb {\widehat{\bb}}

\newcommand\wm {\widehat{m}}

\newcommand\werre {\widehat{r}}

\newcommand\wbB {\widehat{\bB}}

\newcommand\wF {\widehat{F}}

\newcommand\wG {\widehat{G}}

\newcommand\wtw {\widetilde{w}}

\newcommand\itB {{\mathcal{B}}}

\newcommand\itC {{\mathcal{C}}}
\newcommand\itD {{\mathcal{D}}}

\newcommand\itH {{\mathcal{H}}}

\newcommand\itK {{\mathcal{K}}}

\newcommand\itS {{\mathcal{S}}}

\newcommand\itV {{\mathcal{V}}}

\newcommand\balfa {\mbox{\boldmath $\alpha$}}

\newcommand\bmu {\mbox{\boldmath $\mu$}}

\newcommand\bSi {\mbox{\boldmath $\Sigma$}}

\newcommand\weps {\widehat{\epsilon}}

\newcommand\wgamma {\widehat{\gamma}}

\newcommand\wlam {\widehat{\lambda}}

\newcommand\wsigma {\widehat{\sigma}}

\def\real{\mathbb{R}}

\def\newX{\mathbb{X}}
\def\newK{\mathbb{K}}
\def\newY{\mathbb{Y}}
\def\wnewK{\widehat{\newK}}

\newcommand{\esp}{\mathbb{E}}
\newcommand{\prob}{\mathbb{P}}

\newcommand{\diag}{\mbox{\sc diag}}

\newcommand{\trasp}{^\top}
\newcommand{\tras}{\top}

\newcommand\bcero {{\bf{0}}}
\newcommand\buno {{\bf{1}}}

\def\dst{\displaystyle}

\def\argmin{\mathop{\mbox{argmin}}}

\newcommand{\identidad}{\mbox{\bf I}}

\newcommand\noi{\noindent}
\def\dst{\displaystyle}

\def\square{\ifmmode\sqr\else{$\sqr$}\fi}
\def\sqr{\vcenter{
         \hrule height.1mm
         \hbox{\vrule width.1mm height2.2mm\kern2.18mm
\vrule width.1mm}
         \hrule height.1mm}}

\newcommand{\ini}{\mbox{\footnotesize \sc ini}}

\newcommand{\rob}{\mbox{\footnotesize \sc r}}

\newcommand\outlier {\mbox{{\footnotesize\sc o}}}

\usepackage[utf8]{inputenc}

\begin{document}
	
	\title{Robust Nonparametric Regression \\ for Compositional Data: the Simplicial--Real case} 
	\author{Ana M. Bianco$^1$, Graciela Boente$^1$,\\ 
		Wenceslao Gonz\'alez--Manteiga$^2$ \\
		Francisco Gude Sampedro$^2$ and Ana P\'erez--Gonz\'alez$^3$\\
		$^1$ Universidad de Buenos Aires and CONICET\\
		$^2$ Universidad de Santiago de Compostela\\
		$^3$ Universidad de Vigo
	}
	\date{}
	\maketitle

\begin{abstract}
	Statistical analysis on compositional data has gained a lot of attention due to their great potential of applications. A feature of these data is that they are multivariate  vectors that lie in the simplex, that is, the components of each vector are positive and sum up a constant value. This fact poses a challenge to the analyst due to the internal dependency of the components which exhibit a spurious negative correlation. Since classical multivariate techniques are not appropriate in this scenario, it is necessary  to endow the simplex of a suitable algebraic-geometrical structure, which is a starting point to develop adequate methodology and strategies to handle compositions. We centered our attention  on regression problems with real responses and compositional covariates and we adopt a nonparametric approach due to the flexibility it provides. Aware of the potential damage that outliers may produce,  we introduce a robust estimator in the framework of nonparametric regression for compositional data. The performance of the estimators is investigated by means of a numerical study where different contamination schemes are simulated. Through a real data analysis the advantages of using a robust procedure is illustrated.
	
\end{abstract}

\section{Introduction}\label{sec:intro}

Statistical analysis in contexts other than   the Euclidean space $\real^D$ has become an important area of research in modern statistics.  Traditional methods, which are designed to handle Euclidean spaces equipped with the usual inner product, often fail to leverage the underlying geometric structures inherent in non--standard domains. In fact, these procedures typically assume that data points are drawn from a space  where the classical notion of distance in  $\real^D$ and operations, such as averaging or summation, are applicable as in $\real^D$. However, when dealing with data that lie in more complex spaces, such as manifolds, these assumptions may not  still  be   valid. In these situations, standard statistical procedures may perform poorly, affecting our ability to draw accurate results or   to make reliable predictions. Therefore, it is necessary to embed the sample space in a suitable geometrical structure to get all the advantages of an Euclidean vector space.

A relevant class among non--standard settings is that of compositional data,  where only the proportions between variables are informative. Unlike traditional observations, compositional data are constrained by a fixed sum, and the interpretation of the data depends on the relationships between the components rather than their absolute values. Typical examples of compositional data include the composition of a rock   expressed in terms of the  proportion or percentage of different compounds, household expenses  expressed as percentages of the monthly budget, nutrient information on food labels  and  proportions of pollutants in air, among others. 
This enumeration emphasizes the interdisciplinary nature of this topic, making it clear why this research area  has been of increasing interest in recent years.  Formally speaking, a compositional datum is a  $D-$vector $\bx=(x_1,\dots,x_D)\trasp$ belonging to the simplex 
\begin{equation}
\label{eq:simplex}
\itS^D= \left\{\bx=(x_1,\dots,x_D): x_i \ge 0, 1 \le i\le D, \sum_{i=1}^D x_i= \kappa \right\} \, ,
\end{equation}
where $\kappa$ is a positive constant. This is the simplicial sample space.

Our motivation stems from the need to understand the relationship between dietary macronutrient profiles and glycaemic control, specifically focusing on glucose levels. 
Understanding how macronutrient composition, i.e., the relative proportions of carbohydrates, lipids and proteins, affects the average blood glucose levels  is crucial for managing conditions such as different status of diabetes.
Given the increasing prevalence of metabolic disorders, this research area  has received more and more attention in both  clinical practice and  public health.

 The data for this analysis come from a subset of  non--diabetic participants enrolled in the A Estrada Glycation and Inflammation
Study (AEGIS). This is a cross-sectional study performed   between   2012 and 2015 in the municipality of A Estrada, in Northwestern  Spain.  Over the course of six consecutive days, these subjects recorded their diet while glucose levels were continuously registered. From the recorded  information, a glycaemic index defined as the area under the curve of glycaemia during the week was computed for each individual. 
Our aim is to analyse the relationship between this index and the percentages of the three macronutrients ingested during the same period.  To this end, we require regression techniques that account for compositional covariates constrained to lie within a simplex, as well as real-valued responses.

Historically, the first approach to simplicial-real regression was   parametric  employing linear models defined through a proper inner product. \citet{Aitchison:BaconShone:1984} in a seminal paper proposed  removing the constant-sum constraint by transforming the compositions  using  log--contrast models based on transformations of the compositional data. 

 It is also worth emphasizing that linear regression models with compositional covariates differ fundamentally from standard linear regression models defined in unconstrained Euclidean spaces. In the latter setting, multicollinearity among real-valued covariates is typically addressed through variable selection or regularization techniques. Although multicollinearity is generally viewed as a consequence of redundant information among the covariates and poor conditioning of the design matrix, the dependence among compositional covariates is an intrinsic property of the simplex. 

 Compositional data possess a geometry that must be explicitly incorporated into the statistical model rather than ignored. This is precisely the role of Aitchison geometry, which preserves the relative information conveyed by compositional data. Consequently, linear regression with compositional covariates differs from the standard Euclidean framework in two fundamental aspects. First, the components of a composition are intrinsically dependent because of the simplex structure. Second, the regression model is formulated using the Aitchison inner product, rather than the usual Euclidean inner product in $\real^D$. 
However, it is worth noticing that in this setting the   interpretation  of linear models with compositional covariates is not as direct as with regular real covariates. Moreover, traditional linear models may be too restrictive to accurately capture the relationship between covariates and responses, so alternative approaches are needed. Specifically, when dealing with complex or non--linear data, a data-driven perspective offers a compelling option.

    In order to explore the relation between the glycaemic index and the macronutrients in a  robust and   flexible way,  we consider    nonparametric regression models   with positive  compositional covariates  that  provide a malleable framework, as they do not confine the regression function to a finite-dimensional space, allowing for a more adaptable fit to the data.
   In this flexible  setting, 
\citet{Marzio:etal:2015}  pioneered nonparametric estimators of the regression function by  redesigning  concepts originally developed for standard Euclidean spaces. These estimators rely on weighted averages, which, while offering flexibility, are sensitive to the influence of  outlying values in the response variable, also known as \lq \lq vertical outliers\rq \rq. This vulnerability to anomalous points motivates the development of robust methods, as we will discuss further in this paper.  It is worth mentioning that in nonparametric regression with Euclidean covariates, robust procedures do not deal with isolated  covariates or outliers in the covariate space, since in that case the set of neighbours of the outlying point, let us say $\bx_{i_0}$, may reduce to $\bx_{i_0}$ itself, see the discussion in Section 5 of \citet{Salibian:2023}. The same behaviour arises when compositional covariates are considered. For that reason, only vertical outliers may be considered in nonparametric regression settings.  

 The use of robust techniques results crucial in the   AEGIS dataset, since the  outlier detection rule  based on the robust fit residuals allows    to identify    patients who, from a clinical perspective, exhibit a different underlying condition.   


In our contribution we first overview the basic geometrical notions related to the simplex, the natural sample space of compositional data. Unlike the regular case, the basic operations of sum and scalar multiplication are not close in this domain, so it is a necessary step to endow the simplex an algebraic-geometrical structure to get an Euclidean vector space. Secondly, we review the main probabilistic models on the simplex. 

We then focus on our purpose which is to provide robust estimators for the nonparametric regression model with real response and compositional regressors.

The organization of the manuscript is as follows. In Section \ref{sec:basic} we introduce Aitchison geometry and three isomorphisms that transform the simplex in the real space. We also briefly outline probabilistic models on the simplex as well. In Section \ref{sec:regresion} we recall the definition of simplicial kernels, we overview the nonparametric approach followed by \citet{Marzio:etal:2015}  and just after, we introduce our two robust proposals.  Furthermore, in this section we also describe a bandwidth selection method with the purpose of being resistant to outliers. Finally, in Section \ref{sec:monte}  we report the results of a numerical experiment that was carried out in order to compare,  under different models and contamination schemes, the performance of our robust  estimators  with their classical  counterparts based on least squares. Section \ref{sec:realdata} illustrates the advantages of using a robust procedure on a real data set.

\section{Basic concepts}\label{sec:basic}
Henceforth, we will assume that $\kappa=1$ in \eqref{eq:simplex}, i.e., $\itS^D$ refers to 
$$\itS^D= \left\{\bx=(x_1,\dots,x_D): x_i \ge 0, 1 \le i\le D, \sum_{i=1}^Dx_i= 1 \right\}\,.$$

\subsection{Key geometrical notions in the simplex}

It is well known that when a random element, such as a vector or a variable, is constrained to take values in a space $\chi \subset \real^D$, ordinary operations such as addition or multiplication by a scalar may lead to non compatible results, in the sense that the new element may lie outside $\chi$. For, instance, the real line with the classical sum and multiplication by scalars is a vector space and the ordinary inner product and Euclidean distance preserve compatibility with them. However, this geometrical structure is not adequate for the positive real line, see \citet{MateuFigueras:Pawlowsky:2008} for an illustrative description of these problems. This elementary example shows that  when dealing with a constrained domain it is necessary to embed the sample space in a suitable geometrical structure.

Aitchison geometry gives the simplex the structure of a real vector space. It adapts the fundamental notions of translation, scaling, inner product and orthogonal projection that we know in the classical geometry to this particular framework.   Perturbation and powering are the basic operations that play the role of addition  (or translation)
 and  scaling, respectively, \citep[see][]{Aitchison:1982, Aitchison:2003}. Given $\bx, \by \in \itS^D$ and $\alpha\in \real$, the perturbation between them and the powering are defined, respectively,  as
$$\bx \oplus \by= \itC(x_1\,y_1,\dots, x_D\,y_D) \quad \mbox{ and }  \quad  \alpha\odot \bx= \itC(x_1^{\alpha},\dots,x_D^{\alpha} ) \,,$$
where for $\bu\in \real^D$ such that $u_j>0$, for all $1\le j\le D$,  the closure operator is 
$$\itC(\bu)=\itC(u_1 ,\dots, u_D )=\left(\frac{  u_1}{\sum_{j=1}^D u_j}, \dots, \frac{  u_D}{\sum_{j=1}^D u_j}\right)\trasp\,.$$
Note that for the perturbation the neutral element is $\bene=\itC( 1,  \dots, 1)=(1/D,\dots, 1/D)\trasp$ and the opposite of $\bx$ is $\itC( 1/x_1, 1/x_2,\dots, 1/x_D)$, while that repeated perturbation is a particular case of powering, for example, $\bx \oplus \bx = 2 \odot \bx$. The simplex with these two operations becomes a vector space and $(\itS^D, \oplus)$ is a commutative group.
Besides, the difference perturbation, which is the opposite operation of perturbation,  is then defined as 
$$\bx \ominus \by= \bx \oplus \left((-1)\odot\by\right)=\itC\left(\frac{x_1}{y_1}, \dots,\frac{x_D}{y_D} \right)\,,$$
while the Aitchison inner product is given by
$$\langle \bx, \by\rangle_a = \frac{1}{D}\sum_{i<j}   \log\left(\frac{x_i}{x_j}\right)\,\log\left(\frac{y_i}{y_j}\right) = \sum_{j=1}^D \log\left(\frac{x_j}{g_D(\bx)}\right)\,\log\left(\frac{y_j}{g_D(\by)}\right)\,,$$
where $g_D(\bx)= \left(\prod_{\ell=1}^D x_\ell\right)^{1/D}$ is the geometric mean of components of $\bx$. 
Hence, with this set of operations the simplex gains the structure of a $D-1$ dimensional Euclidean vector space. The Aitchison norm and distance, related to this product, are defined as
$$\|\bx\|_a= \langle \bx, \bx\rangle_a= \sum_{j=1}^D \left\{\log\left(\frac{x_j}{g_D(\bx)}\right)\right\}^2\,,$$ 
$$ d_a^2 (\bx, \by) = \frac{1}{D} \sum_{i<j}  \left\{\log\left(\frac{x_i}{x_j}\right)-\log\left(\frac{y_i}{y_j}\right)\right\}^2 = \sum_{j=1}^D\left\{\log\left(\frac{x_j}{g_D(\bx)}\right)-\log\left(\frac{y_j}{g_D(\by)}\right)\right\}^2\,.$$
Finally, to it is easy to see that
$$d_a  (\bx, \by)=\|\bx\ominus \by\|_a\qquad \mbox{and}\qquad \|\bx\|_a=d_a  (\bx, \bene)\,.$$
It is worth mentioning that with this geometrical structure $\itS^D$ verifies the usual properties that relates the distance with the sum and scaling operations, thus in the simplex we have  for any $\bx, \by, \bz \in \itS^D$ and $\alpha \in \real$
$$d_a(\bz \oplus \bx, \bz \oplus \by) = d_a(\bx, \by)  \quad \mbox{ and } \quad d_a(\alpha \odot \bx, \alpha \odot \by) = |\alpha| d_a(\bx, \by)\,.$$
A simple way to analyze compositional data relies on  isomorphisms that transform from the simplex to the real space. The first transformation introduced from  $\itS^D$ to $\real^D$, due to \citet{Aitchison:1982}, is  the additive logratio transformation (\textit{alr})  given by
\begin{equation}\label{eq:alr}
	alr\left(\bx\right)=\left\{\log\left(\frac{x_1}{x_D}\right),\dots,\log\left(\frac{x_{D-1}}{x_D}\right)\right\}
\end{equation}
The $alr$ transformation is one-to one and its inverse denoted $inv.alr: \real^{D-1} \to \itS^D$ is given by
\begin{equation*}
	inv.alr(\by)=inv.alr\left(y_1,\ldots,y_{D-1}\right)= \itC\left\{\exp y_1,\dots,\exp y_{D-1},1\right)
\end{equation*}
Since $alr$  transforms  the simplicial space in the whole $\real^{D-1}$, transformed data may be analyze with standard multivariate statistical techniques and then inference conclusions can be transferred to the original data. 
However, a drawback of $alr$ is that it does not preserve distances. An isometrical  version is obtained  using the geometric mean as divisor, resulting the centred log-ratio transformation (\textit{clr}) which is given by $clr:\itS^D\to \real^D$ and 
\begin{equation*}
clr(\bx)= \left(\log\left(\frac{x_1}{g_D(\bx)}\right),\dots,\log\left(\frac{x_{D}}{g_D(\bx)}\right)\right)\,.
	\label{clr}
\end{equation*}
Note that, for any $\bx\in \itS^D$,  $clr(\bx)$ is a constrained vector of $\real^D$  taking values in $\itV$, where $\itV$ stands for the hyperplane  $\itV = \{\bz\in \real^D: \sum_{j=1}^D z_j=0\}$, meaning that $clr(\itS^D)=\itV$. 

Therefore, inverse of the ${clr}$ transformation, denoted $inv.clr$, may be defined with domain in  $\itV $ as
$$ inv.clr(\by)= \itC\left( \exp(\by)\right)\,.$$
As noted by \citet{Pawlowsky:etal:2015} $clr$ transformation is symmetric in the components, but at the price of being constrained in the image sample: the trasnformed sample lies in a plane, in fact  the components sum up zero, implying that the covariance matrix of $clr(\bx)$ is singular. Note that for any $\bx\in \itS^D$, we have that $clr(\bx)\in \real^D$ is orthogonal to the vector $\buno_D$.

\citet{Egozcue:etal:2003} introduced the isometric log-ratio ($ilr$) transformation of D-parts compositions. Such transformation gives
coordinates in $\real^{D-1}$ ($ilr-$coordinates), and has some advantages with respect to the previously proposed transformations. 
Let the set $\left(\be_1, \ldots, \be_{D-1}\right)$ be an orthonormal basis of $\itS^D$  and denote as $\bU$ the
associated basis-contrast matrix, i.e., the $D \times (D - 1)$ matrix with $i-$th column given by centered log-ratio (clr), $clr(\be_i)$, $i \in  {1, \dots, D-1}$. Since  $\buno_D\trasp \; clr(\be_j)=0$ and the orthonormal basis satisfies $\langle \be_i,\be_j\rangle_a= \delta_{ij} $, where $\delta_{ij}$ is the Kroenecker delta, we get that
$$ \bU\trasp \bU= \bI_{D-1} \quad \mbox{ and } \quad  \bU \bU\trasp= \bI_{D}- \frac 1D \buno_D   \buno_D\trasp \,,$$
where $\bI_k$ is the $k-$dimensional identity matrix and $\buno_k$ is the $k-$dimensional vector with components equal to 1.
The $ilr$ transformation related to $\bU$, lets say  $ilr=ilr_{\bU}$, where $ilr: \itS^D \to \real^{D-1}$,  is the one-to-one linear transformation that relates the vector $\bx^* \in \real^{D-1}$ to $\bx \in \itS^D$, as follows
\begin{equation} \label{eq:irl}
	\bx^*= ilr(\bx)=\bU\trasp {clr}(\bx)=\bU\trasp \log (\bx)\;.
\end{equation}
Then, the inverse of the $ilr$ transformation  related to $\bU$, denoted $inv.ilr:\real^{D-1}\to \itS^D$,  is defined as
$$inv.ilr(\bz)=\itC\left(\exp\{\bU  \bz \}\right)\,,$$
see for instance  \citet{Tsagris:etal:2023} for further details.

From now on, $\|\cdot\|$, $\langle \cdot, \cdot\rangle$ stand for the usual Euclidean norm and inner product in $\real^{D-1}$.

As mentioned, the $ilr$ transformation defines an isomorphism between $\itS^D$ and $\real^{D-1}$,
that is, for $\bx, \by \in \itS^D$, $\alpha\in \real$, we have that
$$ilr(\bx\oplus \by)= ilr(\bx)+ilr(\by) \quad \mbox{and}\quad ilr( \alpha\odot \bx)=\alpha\, ilr(\bx)$$ 
and exhibits the property of being isometric meaning that
$$ d_a (\bx, \by) =\| ilr(\bx) - ilr(\by) \| \,,$$
It is also worth noticing that $\|\bx\|_a=\|ilr(\bx)\|$. Furthermore, if $\bx^*= ilr(\bx)$ and $\by^*= ilr(\by)$, then
$$\langle \bx, \by\rangle_a = \langle \bx^*, \by^*\rangle  = \sum_{j=1}^{D-1} x^*_j y^*_j $$
meaning that Aitchison norms and distances are translated into ordinary ones in ilr-space.

\subsection{Probability models on the simplex}

Probability models are fundamental in order to endow a behavioral pattern to a random phenomenon and they are a very useful tool in Statistics since they give a support to the statistical techniques and analysis. 
We will introduce the building blocks related to this topic, more details can be found, for instance, in Chapter 6 in \citet{Pawlowsky:etal:2015}.

We  recall that a random variable $\bX$ is a random composition if all its possible values lie in $\itS^D$, which we assumed is endowed with the Aitchison geometry (perturbation and powering operations plus Aitchison  inner product, norm and distance).

So, let $\bX=(X_1,\dots,X_D)\trasp$ be a random composition with possible values in $\itS^D$  and for a chosen orthonormal basis of $\itS^D$  consider the contrast matrix $\bU$. Therefore, as seen in \eqref{eq:irl} the $irl-$coordinates related to the selected basis are
$$ \bX^*=(X_1^*,\dots,X_{D-1}^*)\trasp =irl(\bX)= \bU\trasp \log(\bX)\,,$$
where  $ \bX^*$  is a random variable with sample space in the real $D-1$ dimensional space.
We will deal with continuous random variables in $\itS^*$, that is given $\bX$ we assume that there exists a nonnegative real function $f^*: \real^{D-1} \to \real$, defined almost everywhere, such that for any Borelian set $B \subset \real^{D-1}$ 
$$\prob( ilr(\bX) \in B )= \int_B f^*(\bx^*) d\bx^* \,.$$
In this case, $f^*$ is known as the probability density function of the $irl-$coordinates, while $f(\bx)= f^*(irl(\bx))$, defined for any $\bx$ in the simplex, is the probability density function of $\bX$ on $\itS^D$. It is worth noticing that even when the expression of $f^*$ depends on the chosen orthonormal basis, the pdf on the simplex doen not.

Now,  note that probability models are in many cases presented in terms of a probability density function (pdf) that is referred to a measure, usually the reference is the Lebesgue measure which is  natural in a real space. When the simplex endowed with the Aitchison geometry is the sample space, the natural reference measure is the Aitchison one, that introduces an alternative expression of the pdf's even when we are dealing with the same probability law.  The Aitchison measure on the simplex is built from the Lebesgue measure on the $ilr-$coordinates space. 
In fact, let $\lambda$ denote the Lebesgue measure and consider an orthonormal basis on $\itS^D$, its the corresponding $ilr-$transformation into coordinates and $\itB$, an open interval on $\real^{D-1}$, given by
$$ \itB=\left\{ [x_1,\dots,x_{D-1}]  \in \real^{D-1} \mbox{ such that }  a_i < x_i < b_i,  1 \le i\le D-1\right\}\,.$$
If the corresponding $ilr-$interval on $\itS^D$ is $\itB^*$, i.e., $\itB^*=ilr.inv(\itB)$, the Aitchison measure of $\itB^*$ is $\lambda_a(\itB^*)=\lambda(\itB)$.

In the next paragraphs, we recall the definition of two important distributions on the simplex: the Dirichlet distribution and the normal distribution on the simplex.

The Dirichlet distribution, labelled $\itD(\balfa)$ with $\balfa=(\alpha_1,\dots,\alpha_D)\trasp$, is obtained as the closure independent  gamma-distributed random variables that are equally scaled. More precisely, we say that the  variable $\bX \in \itS^D$ has a multivariate density function of the Dirichlet  distribution if
	\begin{equation}
		f(\bx;\balfa)=\frac{\Gamma\left(\sum_{j=1}^D \alpha_j \right)}{\prod_{j=1}^D \Gamma \left(\alpha_j\right)}\prod_{j=1}^D x_j^{\alpha_j-1} \indica_{\{ \bx \in \itS^D \}}\,,
		\label{eq:dirichletdistribution}
	\end{equation}
where $\alpha_j>0$ for $j=1,\dots,D$ and $\Gamma$ is the Gamma function. This expression of the pdf is with respect to the Lebesgue measure.
It is strictly supported on the simplex and it reduces to the beta distribution when $D=2$. The Dirichlet distribution for compositional data was proposed in \citet{Connor:Mosimann:1969}, later on numerous publications analyze and apply  this distribution, see  \citet{hijazi:jernigan:2009} and \citet{Tsagris:Stewart:2018} and the book by \citet{Aitchison:2003}, among others. The Dirichlet distribution can also be formulated in terms of the Aitchison measure, the corresponding representation can be found, for instance, in \citet{Pawlowsky:etal:2015}.

The Logistic Normal distribution was introduced by \citet{Aitchison:1980}. The idea beneath this distribution is to assume that the $arl-$representation of the random composition follows a multivariate normal distribution. The original definition of this distribution is from the $alr-$coordinates of the compositional variates and a posterior application or change of variable theorem.  Given $\bmu \in \real^{D-1}$ and   a symmetric positive definite matrix $\bSi\in \real^{(D-1)\times(D-1)}$, the vector $\bX$ follows an additive logistic distribution  $N_{\itS^D}\left(\mu,\Sigma\right)$  in the simplex space if $alr(\bx)=\log\left(\bx^{(-D)}/x_D\right)$ where $\bx^{(-D)}=(x_1,\dots, x_{D-1})\trasp$,  follows a normal $N_{D-1}\left(\bmu,\bSi\right)$.  However, the same distribution is obtained if it is assumed that the $irl-$coordinates have a normal joint density, which is easier to handle due to the orthonormal basis that mediates in the $irl-$transformation. 
Thus, we say that the random composition $\bX$ follows a Logistic Normal distribution on the simplex  if $\bX^*$ is normally distributed on $\real^{D-1}$, i.e. $\bX^* \sim N_{D-1}(\bmu,\bSi)$, with density function with respect to the Lebesgue measure in $\real^{D-1}$ given by
\begin{equation}
	f\left(\bx^*\right)= {\left(2\pi\right)^{-\left(D-1\right)/2}|\bSi|^{-1/2}} \exp\left[-\frac{1}{2}\left(\bx^*- \bmu\right)\trasp\bSi^{-1}\left(\bx^*-\bmu\right)\right]\,.
\end{equation}
In this case, we denote $\bX \sim N_{\itS^D}(\bmu,\bSi)$. Equivalently,  the density function with respect to the Aitchison measure in $\itS^D$ is of the form
\begin{equation*}
	f\left(\bx\right)= {\left(2\pi\right)^{-\left(D-1\right)/2}|\bSi|^{-1/2}} \exp\left[-\frac{1}{2}\left(ilr\left(\bx\right)- \bmu\right)\trasp\bSi^{-1}\left(ilr\left(\bx\right)-\bmu\right)\right]\,,
\end{equation*}
for $\bx \in \itS^D$. In this representation,  $\bmu$ and $\bSi$ depend on the chosen $irl-$transformation, that is on the selected contrast matrix $\bU$.

See also \citet{MateuFigueras:Pawlowsky:2008} and \citet{Egozcue:etal:2012} for more details. 
 
 
\section{Regression  model}{\label{sec:regresion}}

Modeling is an important task in statistical  analysis, where regression techniques play a central role. The interest focuses on the relationship between variables with the aim of prediction or either understanding the way in which a set of regressor variables ($\bX$), namely the covariates, affect the behaviour of a dependent variable, i.e., the response ($Y$).   Regression problems  involving compositions deserve a particular approach due to the structure of these data. We concentrate on the case where we deal with compositional covariates and real scale response. The first to appear was the parametric treatment of the problem. However, in this setting the interpretability of models is more complex than with regular variables.  As mentioned  in \citet{Greenacre:2021}, even if a linear model is postulated, the explanation of the effect of the coefficients is not as direct as with real-valued variables. In fact, the traditional interpretation that the magnitude of the impact of a change in one unit in a certain variable, remaining the others fixed, is the corresponding coefficient is no longer valid in this context, where the modification of one element of the composition necessarily alters the others due to constraint sum.  

In the setting of linear models,  \citet{Scheffe:1958} early considered the problem of prediction when dealing with mixtures of components and in \citet{Scheffe:1963} design of experiments is treated in the framework of compositions. 
\citet{Aitchison:BaconShone:1984}  considered log--contrast models based on the $alr-$transformation given in \eqref{eq:alr} to circumvent the constant sum. They  proposed to study  linear models of the form 
$$ Y= \theta_0 + \sum_{j=1}^{D-1} \theta_j \log{\frac{x_j}{x_D}} +\epsilon \,,$$
where $\epsilon$ represents the error term and they also included quadratic log-contrast versions of these models. A drawback of this approach is that variables are not treated symmetrically since the practitioner has to chose a component for the common divisor.
Further advances in the Euclidean geometry of the simplex such as the introduction of the $irl-$transformation given in \eqref{eq:irl} and also on the logistic-normal distribution described above, settled down the basis of expressions of the
regression model in coordinates and the use of ordinary least squares, see \citet{Egozcue:etal:2012} for details. 

There exists a vast literature  on the parametric-driven approach, \citet{Coenders:Pawlowsly:2020} reviews many of the alternative parametrizations that were proposed for the simplicial--real case, without neglecting aspects that concern their interpretation.

\subsection{Nonparametric Simplicial-Real model}{\label{sec:propuesta}}

As when dealing with real covariates, linear models may be too restrictive to provide an accurate fit to the responses. In such situations, data--driven perspective emerges as an alternative to parametric-driven modeling and in this new context, nonparametric regression models provide a flexible tool since now the regression function is not restricted to lie in a finite-dimensional space. 

\citet{Marzio:etal:2015} introduced nonparametric estimators of the regression function by adapting the ideas developed in the Euclidean real space. As we will see, the estimators depend on weighted means and hence, they are not protected from outliers, motivating our proposals.

In the sequel, we briefly overview the proposals given by \citet{Marzio:etal:2015}. Throughout this section, we assume that we have a random sample $(y_i, \bx_i)$, $1\le i\le n$,  such that $y_i\in \real$, $\bx_i\sim \itS^D$ and $(y_i, \bx_i)\sim (Y, \bX)$ where  $(Y, \bX)$ satisfies the regression model
\begin{equation}
	\label{eq:modelosimpl-real}
	Y=m(\bX)+\sigma(\bX)\epsilon\;,
\end{equation}
where the error $\epsilon$ is independent of the simplicial covariates $\bX$ and have  a {symmetric} distribution $F_0(\cdot)$, that is, we assume that the error's scale equals 1 to identify the scale function. Hence, when second moments exist, we have that  $\esp(Y|\bX)=m(\bX)$ and $\sigma^2(\bX)=\esp((Y-m(\bX))^2|\bX)$ is   the conditional variance function.

\subsection{The classical approaches}{\label{sec:clasico}}
To define nonparametric regression estimators,  \citet{Marzio:etal:2015} introduced simplicial kernels as follows. Kernels were introduced in \citet{Aitchison:Lauder:1985} and \citet{Chacon:etal:2011} in the framework of density estimation on the simplex.

Let $\itK$ be a continuous even function with maximum at 0 such that
$\int \itK(x) dx< \infty$.
The $D-$simplicial kernel can be defined as
$$ K(\bu) =\frac{\itK\left(\left\|\bu\right\|_a \right)}{\int_{\real^{D}} \itK\left(\left\|\bu\right\|_a\right) d \lambda_a(\bu)}\,,$$
where $\left\|\bu\right\|_a$ stands for  $\left\|\bu\right\|_a= \left\|\bu^*\right\|$ and $ \lambda_a$ is the Aitchison measure introduced above.  
Then, we have
$$ K(\bu)=\widetilde{K}\left(\bu^*\right) =\frac{\itK\left(\left\|\bu^*\right\|\right)}{\int_{\real^{D-1}} \itK\left(\left\|\bu^*\right\|\right) d \bu^*}\,.$$
The kernel $K$ is a density on $\itS^D$ with respect to
the Aitchison measure, whereas $\widetilde{K}$ is a density on $\real^{D-1}$ with respect to the Lebesgue measure.

As mentioned in \citet{Marzio:etal:2015}, the kernel defined through $K(\bu)=\widetilde{K}\left(\bu^*\right)$ is invariant when changing the orthonormal
basis for $\itS^D$, for that reason, henceforth,  we will consider   the following explicit expression of the $ilr-$transformation
\begin{equation}
	\label{eq:ilr}
	u_j^*=\sqrt{\frac{j}{j+1}}\log\left\{\frac{g_j(u_1, \dots, u_j)}{u_{j+1}}\right\}\,,
\end{equation}
where $\bu^*=ilr(\bu)=(u^*_1, \dots, u^*_{D-1})\trasp$ and $g_j(u_1, \dots, u_j)= \left(\prod_{\ell=1}^j u_\ell\right)^{1/j}$ is the geometric mean of components of $(u_1, \dots, u_j)$. Note that this transformation is the one considered in \citet{Egozcue:etal:2003} and \citet{Chacon:etal:2011}.

Given a symmetric positive definite matrix $\bH\in \real^{(D-1)\times (D-1)}$ and $\bu \in \itS^D$, a simplicial kernel centered at
$\bx\in \itS^D$   and rescaled by $\bH$, is defined as 
\begin{equation}
	\label{eq:nucleoK}
	K_{\bH} (\bu\ominus \bx)= \frac{1}{\mbox{det}(\bH)}\widetilde{K}\left(\bH^{-1} (\bu^*-\bx^* )\right)=\widetilde{K}_{\bH} \left(  \bu^*-\bx^* \right)\,.
\end{equation}
\citet{Marzio:etal:2015} introduced a local constant estimator for $m(\bx)$ as the minimizer of \linebreak  $\sum_{i=1}^n \left(y_i-b\right)^2 K_{\bH} (\bx_i\ominus \bx)$,
which leads to the following expression for the resulting estimator $\wm(\bx) = \sum_{i=1}^n w_i(\bx) y_i  $,
where 
\begin{equation}
	\label{eq:nucleo}
	w_i(\bx)= K_{\bH} (\bx_i\ominus \bx)\; \left\{\sum_{j=1}^n    K_{\bH} (\bx_j\ominus \bx)\right\}^{-1}\;.
\end{equation}
The local linear version of this estimator is defined as $\wm(\bx)=\wb_0$, where
$$(\wb_0, \wbb_1)=\argmin_{(b_0, \bb_1): \bb_1\trasp \buno_D=0}\;\sum_{i=1}^n \left\{y_i-b_0-\bb_1\trasp \log(\bx_i\ominus \bx)\right\}^2 K_{\bH} (\bx_i\ominus \bx)=\argmin_{(b_0, \bb_1): \bb_1\trasp \buno_D=0}L_n(b_0, \bb_1, \bx)\,.$$
Recall that we have defined the hyperplane  $\itV$ as  $\itV= \{\bz\in \real^D: \sum_{j=1}^D z_j=0\}$. Since $\bb_1\in \itV$,  we have that $L_n(b_0, \bb_1,  \bx)=\sum_{i=1}^n \left\{y_i-b_0-\langle inv.clr(\bb_1), \bx_i\ominus \bx\rangle_a\right\}^2 K_{\bH} (\bx_i\ominus \bx)$. Hence the estimator can be viewed as a local linear smoother with respect to the natural Aitchison geometry in the simplex.

As mentioned in \citet{Marzio:etal:2015}, the loss $L_n(b_0, \bb_1, \bx)$ may be re--written using the $ilr-$transformation as
$$\sum_{i=1}^n \left\{y_i-b_0-\bb_1^{*\tras}  (\bx_i^*- \bx^*)\right\}^2 \widetilde{K}_{\bH} \left(  \bx^*_i - \bx^*\right)\,,$$
which leads to an explicit expression for the local linear estimator as
\begin{equation}
	\label{eq:wmclasico}
	\wm(\bx)= \bi_1\trasp \left( \newX\trasp \newK \newX\right)^{-1} \newX\trasp \newK \newY\,,
\end{equation}
where $\bi_1$ is the first canonical vector in $\real^D$, that is, the  vector having   its first component equal to 1 and the remaining ones equal to 0,  $\newK=\diag\left(\widetilde{K}_{\bH} \left(  \bx^*_1 - \bx^*\right), \dots, \widetilde{K}_{\bH} \left(  \bx^*_n - \bx^*\right)\right)$,   
\begin{equation}
	\label{eq:matrixY-X}
	\newY=(y_1, \dots, y_n)\trasp \qquad \mbox{and} \qquad \newX=\left(\begin{array}{cc}
		1 & (  \bx^*_1 - \bx^*)\trasp\\
		\vdots & \vdots \\
		1 & (  \bx^*_n - \bx^*)\trasp
	\end{array}\right)\,.
\end{equation}
As mentioned, equation \eqref{eq:wmclasico} shows that the resulting estimators are a weighted mean of the responses and from this fact it becomes evident that the estimator may be very sensitive to outliers among $y_i$. In this sense, one could make the estimator to take an arbitrarily large value just by choosing an appropriate atypical response. In the next section we propose a robust alternative that will be much more stable in presence of anomalous data among the responses.

\subsection{The robust estimators}{\label{sec:robust-Simplicial-Real}}

To robustify the local constant and local linear kernel estimators under the simplicial--real regression model \eqref{eq:modelosimpl-real} one may consider an $M-$smoother with the kernel weights defined above.

More precisely, as in the Euclidean case, define  the empirical conditional distribution function $\wF(y|\bX=\bx)$ as
\begin{equation} 
	\wF(y|\bX=\bx) = \sum_{i=1}^n w_{i}(\bx) \indica_{(-\infty,y]}(y_i)  \, ,
	\label{distri0}
\end{equation}
with $w_i(\bx)$ the kernel weights defined in \eqref{eq:nucleo}. Note that, as above, this estimator can be computed using the $ilr-$transformation as 
$$\wF(y|\bX=\bx) = \sum_{i=1}^n \wtw_i(\bx^*) \indica_{(-\infty,y]}(y_i)\,,$$
where 
\begin{equation}
	\label{eq:nucleostar}
	\wtw_i(\bx^*)= \widetilde{K}_{\bH} \left(  \bx^*_i - \bx^*\right)\; \left\{\sum_{j=1}^n   \widetilde{K}_{\bH} \left(  \bx^*_j - \bx^*\right)\right\}^{-1}\;,
\end{equation}
since \eqref{eq:nucleoK} implies that $ w_{i}(\bx)= \wtw_i(\bx^*)$, where $\bx^* = ilr(\bx)$.

As it is well known, to decide which observations may be considered as atypical, the size of the residuals should be measured with respect to some appropriate scale or dispersion measure, see for instance, Sections 2.6 and  6.9.1 in  \citet{Maronna:etal:2019}. For that reason, the first step to define  robust estimators of the regression function is to introduce a robust scale.

Define $\wm_{\ini}(\bx)$ the local median, that is, the median of the empirical conditional distribution function $\wF(y|\bX=\bx)$. Take 
a preliminary robust scale estimator $\wsigma(\bx)$, such as  the local \textsc{mad}, that is,  the  median of the empirical conditional distribution of the absolute value of the residuals $\werre_{i,\bx}=y_i-\wm_{\ini}(\bx)$, which is computed as
$$\wG(r|\bX=\bx) =\sum_{i=1}^n w_{i}(\bx) \indica_{(-\infty,r]}\left(\left|\werre_{i,\bx}\right|\right)\,.$$
Other possible choices for $\wsigma(\bx)$ are local $S-$scale estimators. To define them, as in \citet{Maronna:etal:2019}, let
$\rho : \real \to \real_+$ be a bounded $\rho-$function, that is, an even function, non--decreasing on $|r|$, increasing for $r>0$ when $\rho(r)<\lim_{r\to +\infty}\rho(r)$ and such that $\rho(0) = 0$.  The local $S-$scale estimator  $\wsigma(\bx)$ is defined as the solution in $s$ of
\begin{equation*}\label{eq:Sescalalocal}
	\sum_{i=1}^n w_i(\bx) \rho_{c_0} \left(  \frac{ \werre_{i,\bx}}{s} \right) \ = b \,,
\end{equation*} 
where $\rho_c(r) = \rho(r/c)$, and $c_0 > 0$ is a user--chosen tuning constant.
When $\rho$ is the Tukey's biweight function, $\rho(r) = \min\left(3 r ^2 - 3 r ^4 +  r ^6, 1\right)$, the choice  
$c = 1.54764$ and $b=1/2$ ensures that the estimator is Fisher--consistent for normal errors.  Note that the local \textsc{mad} corresponds to the choice $\rho(r)=\indica_{(1,\infty)}(|r|)$, $c=1$ and $b=0.5$.  

When the model is homocedastic, a global robust scale estimator $\wsigma$ may be computed as  the solution in $s$ of
\begin{equation}\label{eq:Sescala}
	\sum_{i=1}^n   \rho_{c_0} \left(  \frac{ \werre_{i}}{s} \right) \ = b \,,
\end{equation} 
where $\werre_{i}=y_i-\wm_{\ini}(\bx_i)$. Consistency of these scale estimators for Euclidean covariates were given in \citet{boente:martinez:2017}. 
Moreover, for Eculidean covariates, local $M-$smoothers were studied, among others in  \citet{hardle:tsybakov:1988} and \citet{boente:fraiman:1989}, while robust local linear smoothers were considered in \citet{cleveland:1979} and \citet{jiang:mack:2001}.

In the definitions to be given below, it should be taking into account that if the model is homocedastic, a robust scale estimator $\wsigma$ should be used instead of the local one  $\wsigma(\bx)$.

Inspired in these approaches, we propose the following estimators:
\begin{itemize}
	\item Local $M-$smoothers may be defined using a loss function $\rho_{c_1}$ where $c_1>c_0$ as the minimizer $ \wm(\bx) $ of $\wgamma (\bx,a,\wsigma(\bx))$, where 
	\begin{equation}
		\wgamma_j(\bx,a,\varsigma)= \sum_{i=1}^{n}  w_i(\bx) \, \rho_{c_1}\left( \frac{ y_{i}-a}{\varsigma}\right)\,.
		\label{eq:funcionwgama}
	\end{equation}
	It is worth mentioning that  if $\rho$ is differentiable, $\psi=\rho^{\prime}$ and $\psi_{c}(t)=\psi(t/c)$, then  $\wm(\bx) $ is  the solution   of 
	$ \wlam(\bx,\wm(\bx),\wsigma(\bx))=0$,  with
	\begin{equation}
		\wlam(\bx,a,\varsigma)= \sum_{i=1}^{n}  w_i(\bx) \, \psi_{c_1}\left( \frac{ y_{i}-a}{\varsigma}\right)\,.
		\label{eq:funcionwlamda}
	\end{equation}
	Furthermore, if $\psi$ is an increasing function \eqref{eq:funcionwlamda} has a unique solution. In contrast, when considering a bounded $\rho-$function, the algorithm provided below  provides a procedure to obtain a consistent estimator.
	
	\item Local linear $M-$smoothers may be defined as follows. Let  $\rho_{c_1}$ be a $\rho-$function such that  $c_1>c_0$ and define 
	$$(\wb_0, \wbb_1)=\argmin_{(b_0, \bb_1): \bb_1\trasp \buno_D=0}\sum_{i=1}^n   w_i(\bx)\, \rho_{c_1}\left(\frac{y_i-b_0-\bb_1\trasp \log(\bx_i\ominus \bx)}{\wsigma(\bx)}\right)   \,.$$
	Then, $\wm(\bx)=\wb_0$. As in the classical case, using the $ilr-$transformation, we have that  $\wm(\bx)=\wb_0$, where
	$$(\wb_0, \wbb_1^{*})=\argmin_{(b_0, \bb_1^{*})}\sum_{i=1}^n \wtw_i(\bx^*)\, \rho_{c_1}\left(\frac{y_i-b_0-\bb_1^{*\tras}  (\bx_i^*- \bx^*)}{\wsigma(\bx)}\right)  \,,$$
	with  $\bx^*=ilr(\bx)$. Note that $(\wb_0, \wbb_1^{*})$ satisfy
	$$\sum_{i=1}^n \wtw_i(\bx^*)\, \psi_{c_1}\left(\frac{y_i-\wb_0-\wbb_1^{*\tras}  (\bx_i^*- \bx^*)}{\wsigma(\bx)}\right)  \left(\begin{array}{c}
		1\\
		\bx^*_i - \bx^*
	\end{array}\right)=\bcero\,,$$
	which suggests a reweighted version of the algorithm used to compute $M-$estimators, that will be described below. Indeed, defining $W_{c_1}(t)=\psi_{c_1}(t)/t$ and the standardized residuals 
	$$\werre_{i,\bx}=\frac{y_i-\wb_0-\wbb_1^{*\tras}  (\bx_i^*- \bx^*)}{\wsigma(\bx)}\,,$$
	and the matrix
	$$ \wbB_{\bx}=\sum_{i=1}^n\wtw_i(\bx^*)\,  W_{c_1}\left( \werre_{i,\bx} \right)    \left(\begin{array}{c}
		1\\
		\bx^*_i - \bx^*
	\end{array}\right) \left(\begin{array}{c}
		1\\
		\bx^*_i - \bx^*
	\end{array}\right)\trasp \,,$$
	we have that $$\left(\begin{array}{c}
		\wb_0\\
		\wbb_1^{*\tras} 
	\end{array}\right)= \wbB_{\bx}^{-1}\sum_{i=1}^n \wtw_i(\bx^*)\, W_{c_1}\left( \werre_{i,\bx}\right)    \left(\begin{array}{c}
		1\\
		\bx^*_i - \bx^*
	\end{array}\right) \, y_i\,.$$
	Note that, if we define the matrix $\wnewK= \diag\left(\wtw_1(\bx^*)\, W_{c_1}\left( {\werre_{1,\bx}} \right)  , \dots,  \wtw_n(\bx^*)\, W_{c_1}\left( {\werre_{n,\bx}} \right) \right) $, 
	we have that $\wm(\bx)$ satisfies
	$\wm(\bx)=   \bi_1\trasp \left( \newX\trasp \wnewK \newX\right)^{-1} \newX\trasp \wnewK \newY $,
	an expression closely related to \eqref{eq:wmclasico}, except that the matrix $\wnewK$ involves the estimators $(\wb_0, \wbb_1^{*})$. This is the basis for the algorithm described in Algorithm \ref{alg: Mlocallineal}.
\end{itemize}

Taking into account that  $ w_{i}(\bx)= \wtw_i(ilr(\bx))$, local $M-$smoothers may be numerically computed using the following algorithm.

\begin{algorithm}[H]
	\caption{The robust local $M-$smoother}
	\label{alg: Mlocal}
	\begin{algorithmic}[1]
		
		\STATE Let $\ell = 0$ and $\wm^{(0)} =  \wm_{\ini}(\bx)$ be the local median
		and let $\wsigma(\bx)$ be a robust local scale estimator, as the one defined in \eqref{eq:Sescala}. Furthermore, let $\wtw_i(\bx^*)$ be defined as in \eqref{eq:nucleostar}, with $\bx^*=ilr(\bx)$
		\REPEAT
		\STATE $\ell  \leftarrow \ell + 1$
		\FOR{$i=1$ \TO $n$}
		\STATE Let   $\werre_{i}(\bx)$ be the standardized residuals  $\werre_{i}(\bx)= (y_i- \wm^{(\ell)}(\bx))/\wsigma(\bx)$ and 
		$W_{i}^{(\ell)}= \wtw_i(\bx^*)\, W_{c_1}\left(\werre_{i}(\bx)\right)$.
		\ENDFOR
		\STATE Let $\wm^{(\ell+1)}(\bx) = \sum_{i=1}^n W_{i}^{(\ell)} y_i /\sum_{i=1}^n W_{i}^{(\ell)}$.
		
		\UNTIL convergence
	\end{algorithmic}
\end{algorithm}

Similarly, $M-$local linear estimators may be computed through

\begin{algorithm}[H]
	\caption{The robust local linear $M-$smoother}
	\label{alg: Mlocallineal}
	\begin{algorithmic}[1]
		\STATE Define the matrices $\newY$ and $\newX  $ as in \eqref{eq:matrixY-X}.
		
		\STATE Let $\ell = 0$ and take $\wb_0^{(0)} =  \wm_{\ini}(\bx)$ the local median, $\wbb_1^{*\,(0)}=\bcero_{D-1}$
		and let $\wsigma(\bx)$ be a robust local scale estimator, as the one defined in \eqref{eq:Sescala}. Furthermore, let $\wtw_i(\bx^*)$ be defined as in \eqref{eq:nucleostar}, with $\bx^*=ilr(\bx)$
		\REPEAT 
		\STATE $\ell  \leftarrow \ell + 1$
		\FOR{$i=1$ \TO $n$}
		\STATE Let   $\werre_{i}(\bx)$ be the standardized residuals  
		$$\werre_{i}(\bx)= \frac{y_i-\wb_0^{(\ell)}-\wbb_1^{*\,{(\ell)}\tras}  (\bx_i^*- \bx^*)}{\wsigma(\bx)} $$ and 
		$W_{i}^{(\ell)}=\wtw_i(\bx^*)\, W_{c_1}\left(\werre_{i}(\bx)\right) $.
		\ENDFOR
		\STATE Define the matrix 
		$$\wnewK^{(\ell)}= \diag\left(W_{1}^{(\ell)}, \dots,W_{n}^{(\ell)}\right)\,,$$ 
		\STATE Let $(\wb_0^{(\ell+1)}, \wbb_1^{*\,(\ell+1) \tras}) \trasp=  \left( \newX\trasp \wnewK^{(\ell)} \newX\right)^{-1} \newX\trasp \wnewK \newY$ and $\wm^{(\ell+1)}(\bx)=\wb_0^{(\ell+1)}$
		
		\UNTIL convergence
		
	\end{algorithmic}
\end{algorithm}

\subsection{Selection of the bandwidth parameter}{\label{sec:band-selec}}
As with other non--parametric procedures, the selection of the smoothing parameter is an
important practical issue when fitting additive models. The importance of using a robust
criterion for selecting smoothing parameters, even when one uses robust estimators, has
been extensively described, see for instance  \citet{Leung:etal:1993}, \citet{Wang:Scott:1994}, \citet{boente:etal:1997}, \citet{Cantoni:Ronchetti:2001} and \citet{Leung:2005}. 
Several proposals have been made in the literature, including $L^1-$cross--validation \citep{Wang:Scott:1994},   robust
versions of $C_p$ and cross--validation  \citep{Cantoni:Ronchetti:2001,  Bianco:Boente:2007} and a robust plug--in procedure discussed in \citet{boente:etal:1997}.

As in \citet{Marzio:etal:2015}, we will consider the case where the bandwidth matrix $\bH$ equals $h\,\identidad_{D-1}$. In such a case, the parameter $h$ may be chosen using a robust leave--one--out cross-validation criterion, see \citet{Bianco:Boente:2007} and \citet{boente:rodriguez:2008} or more generally a $K-$fold cross--validation procedure. To define the robust selection method, given the sample  $(y_i, \bx_i)$, $1\le i\le n$, denote $\wm_h^{(-i)}(\bx)$ the regression estimator computed with bandwidth $h$ using all the data except $(y_i, \bx_i)$ and define the cros--validation residual  $\weps_i(h)=y_i-\wm_h^{(-i)}(\bx_i)$. The classical cross-validation criterion considered in \citet{Marzio:etal:2015} obtains   adaptive data--driven bandwidths, by minimizing
\begin{equation}
CV(h)=\frac 1n \sum_{i=1}^n \weps_i(h)^2=\frac 1n \sum_{i=1}^n \left(\weps_i(h)-\frac 1n \sum_{\ell=1}^n  \weps_{\ell}(h)\right)^2 + \left(\frac 1n \sum_{\ell=1}^n  \weps_{\ell}(h)\right)^{\,2}\,.
\label{classCV}
\end{equation}
\citet{Bianco:Boente:2007}, following \citet{Leung:etal:1993},   suggests to replace the square function by a $\rho-$function after scaling the residuals to control the effect of large responses in partly linear autoregression models. These authors notice that this procedure  lacks of robustness when using the  the Huber's   $\rho$ functions  due to its unboundness, while the performance of the criterion based on Tukey's $\rho-$function, is similar   to that attained  by a method which replaces  the square function by the square of Huber's score function.  The robust cross--validation criterion to be defined  below is based on the idea that in \eqref{classCV}  the right hand side is decomposed into the sum of the squared bias and the variance, leading to an objective function that tries  to measure both bias and variance. Based on this property, \citet{Bianco:Boente:2007} suggested a   new procedure that establishes a trade--off between robust measures of bias and   dispersion instead and that, according to their simulation study, turn out to be   preferable than those based on bounding the squared residuals. In the actual framework we  will follow these ideas to provide a robust data--driven bandwidth in regression models with compositional data. For that purpose, let  $\mu_n(z_1,\dots, z_n)$ and $s_n(z_1,\dots, z_n)$ denote robust estimators of location and 
dispersion based on the 
sample $\{ z_1,\dots, z_n \}$, respectively.  Possible  location estimator, $\mu_n$, may be the median, while $s_n$ can be taken as the \textsc{mad}, the bisquare a-scale estimator or the Huber $\tau-$scale estimator. A robust cross--validation
criterion can be defined by minimizing on $h$ the quantity
\begin{equation}
RCV(h) = \mu_n^2 \left(\weps_1(h), \ldots, \weps_n(h) \right)  \, + \, s_n^2 \left(\weps_1(h), \ldots, \weps_n(h) \right)\,.
\label{RCV}
\end{equation}
Another possibility is to consider a $K-$fold cross-validation method. For that purpose, first randomly partition the data set into $K$ 
disjoint subsets of approximately equal sizes, with indices $\itC_j$, $1 \le j \le K$, so that $\bigcup_{j=1}^K \itC_j
= \{ 1, \ldots, n \}$. 
Let $\itH \subset \real$ be the set of bandwidth combinations to be considered, and let 
$\wm_{h}^{(-j)}(\bx)$  be the robust estimator of $m(\bx)$  computed with the smoothing parameter $h\in \itH$  and 
without using the observations with indices in $\itC_j$.
For each $i=1,\dots, n$, define as above the  residuals $\weps_i(h)$ as
$$
\weps_i(h) \, = \, y_i -  \wm_{h}^{(-j)}(\bx_i) \, , \quad i \in \itC_j \, , \  j = 1, \, \ldots, K \, . 
$$
The robust $K-$fold cross-validation bandwidth parameter $h$ is selected by minimizing over 
$h \in \itH$ the following criterion:
\begin{equation}
RCV_{K}(h) \, = \, \mu_n^2 \left(\weps_1(h), \ldots, \weps_n(h) \right)  
\, + \, 
s_n^2 \left(\weps_1(h), \ldots, \weps_n(h) \right)  \, . 
\label{RKCV}
\end{equation}
Note that when $K=n$ and ${\itC}_{j}= \{ j \}$, $1 \le j \le n$, $RCV_{K}(h)$ equals the leave--one--out version of the robust cross--validation criterion defined through \eqref{RCV}.

\section{Monte Carlo study}{\label{sec:monte}}

This section contains the results of a simulation study conducted to compare,  under different models and contamination schemes, the performance of the robust  estimators defined in Section \ref{sec:robust-Simplicial-Real} with their classical  competitors based on least squares.  All computations were carried out in  \texttt{R}.

The classical estimators are labelled \textsc{CL}$_{0}$ and \textsc{CL}$_{1}$ when using the local mean or the local linear estimator introduced in \citet{Marzio:etal:2015}.   For the robust estimators, we  first compute the scale estimator using an $S-$scale as described in Section  \ref{sec:robust-Simplicial-Real} using   the  Tukey's bisquare function with tuning constants $c_0= 1.54764$ and $b = 1/2$, since the model is homoscedastic. The local $M-$smoother, labelled  \textsc{ROB}$_{0}$, was computed using the Tukey's bisquare function with tuning constant    $c_1=4.685$. As mentioned above, the value $c_0 = 1.54764$ ensures Fisher--consistency of the scale estimator when the errors have a normal distribution.  

In all scenarios, we performed $N=500$ replications, the sample size was $n=100$.

The samples $\left\{(y_i, \bx_i ) \right\}_{i=1}^n$ are generated with the same distribution as $ (Y, \bX ) $, $\bX =(X_1,X_2, X_3)\trasp\in\itS^3$. In all cases, the response and the covariates are related through  the nonparametric model 
$Y=m(\bX)+\sigma \epsilon$, 
where  $\sigma =1$, $m(\bx)=\sin\left(\langle \bx, \bb \rangle_{a}\right)$, where $\bb= (0.05920067, 0.7193872, 0.2214121)$. The covariates were generated using a Dirichlet distribution with parameter  $\balfa=(5, 7, 1)\trasp$ or $\balfa=(5, 7, 4)\trasp$. For clean samples, denoted from now on as $C_0$,  the errors have a standardized normal distribution, $\epsilon\sim N(0,1)$.  For this numerical study we choose the bandwidth matrix $\bH$ as $\bH=2\, \identidad_{D-1}$ when $\balfa=(5, 7, 1)\trasp$ and   $\bH=  \identidad_{D-1}$ when $\balfa=(5, 7, 4)\trasp$.

To study the effect of atypical data on  the estimators,   we considered  a contamination scheme  denoted   $C_{1,\delta, \mu}$,   which is obtained generating the errors according to  $\epsilon_i\sim (1-\delta) N(0,1) + \delta N(\mu,0.1^2)$. We choose $\delta=0.10$ and $\mu=5, 10$. 
 
\begin{figure}[ht!]
		\renewcommand{\arraystretch}{0.1}
		\newcolumntype{G}{>{\centering\arraybackslash}m{\dimexpr.34\linewidth-1\tabcolsep}}
	\hskip-0.6in
				\begin{tabular}{GGG}
			Ternary Diagram &  $m(\bx)$ & ilr Plot \\ 
			\includegraphics[scale=0.35]{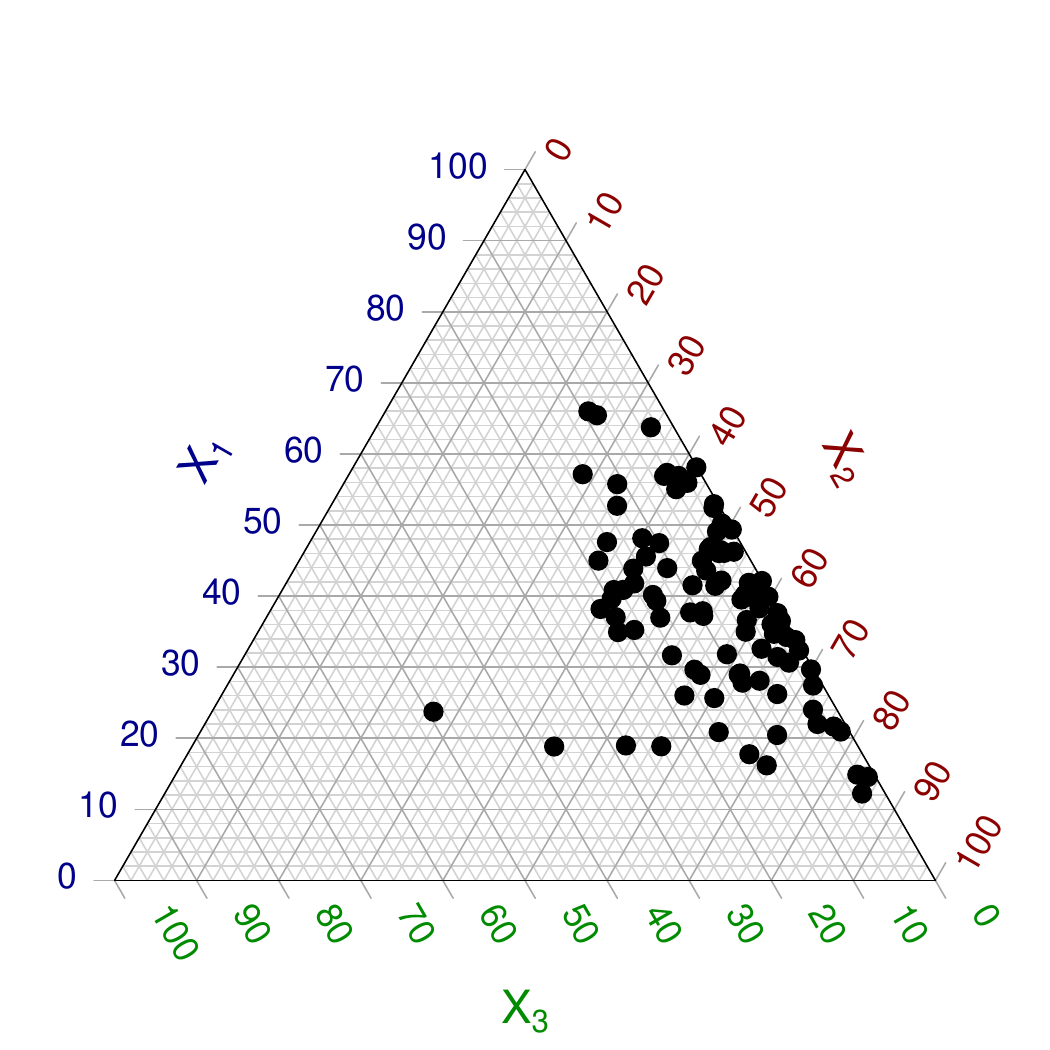} &
			\includegraphics[scale=0.35]{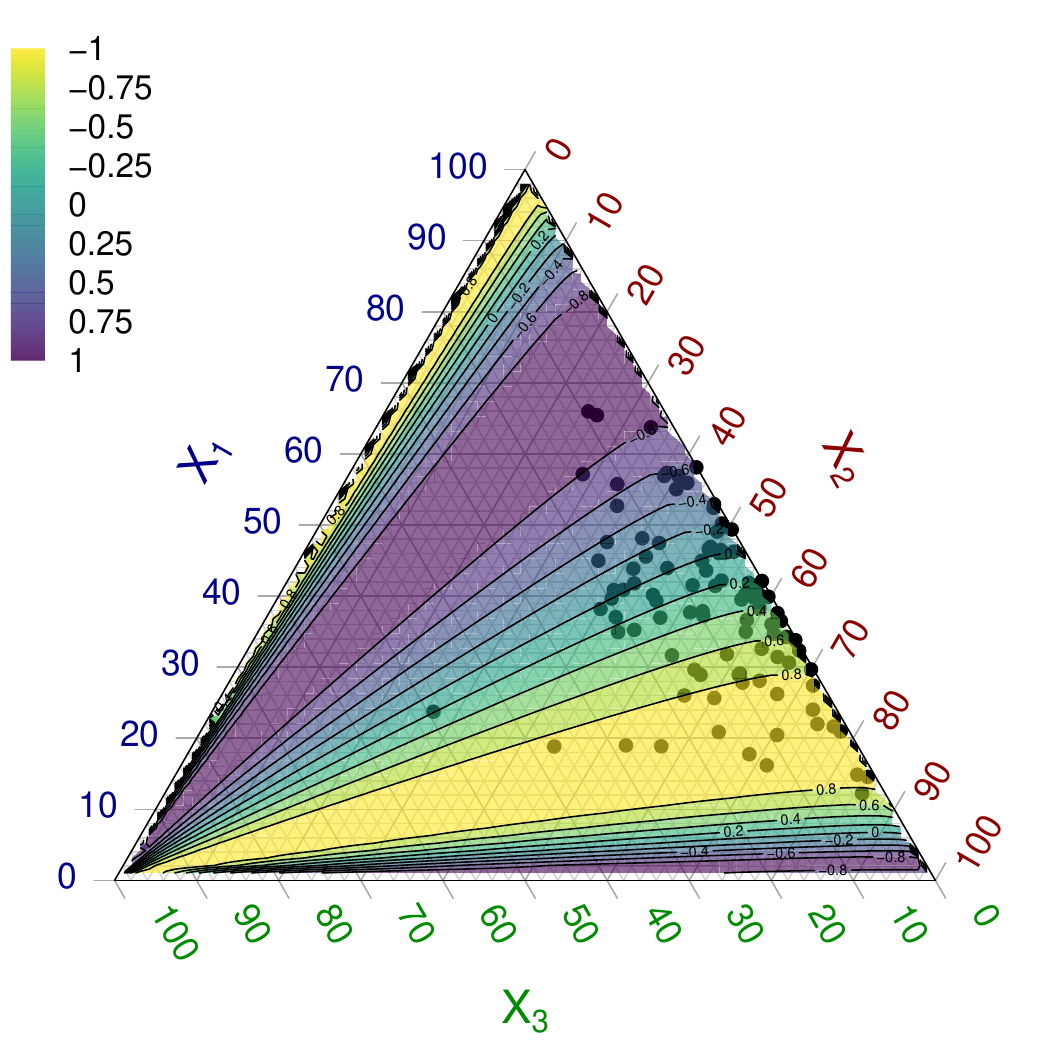} &
			\includegraphics[scale=0.32]{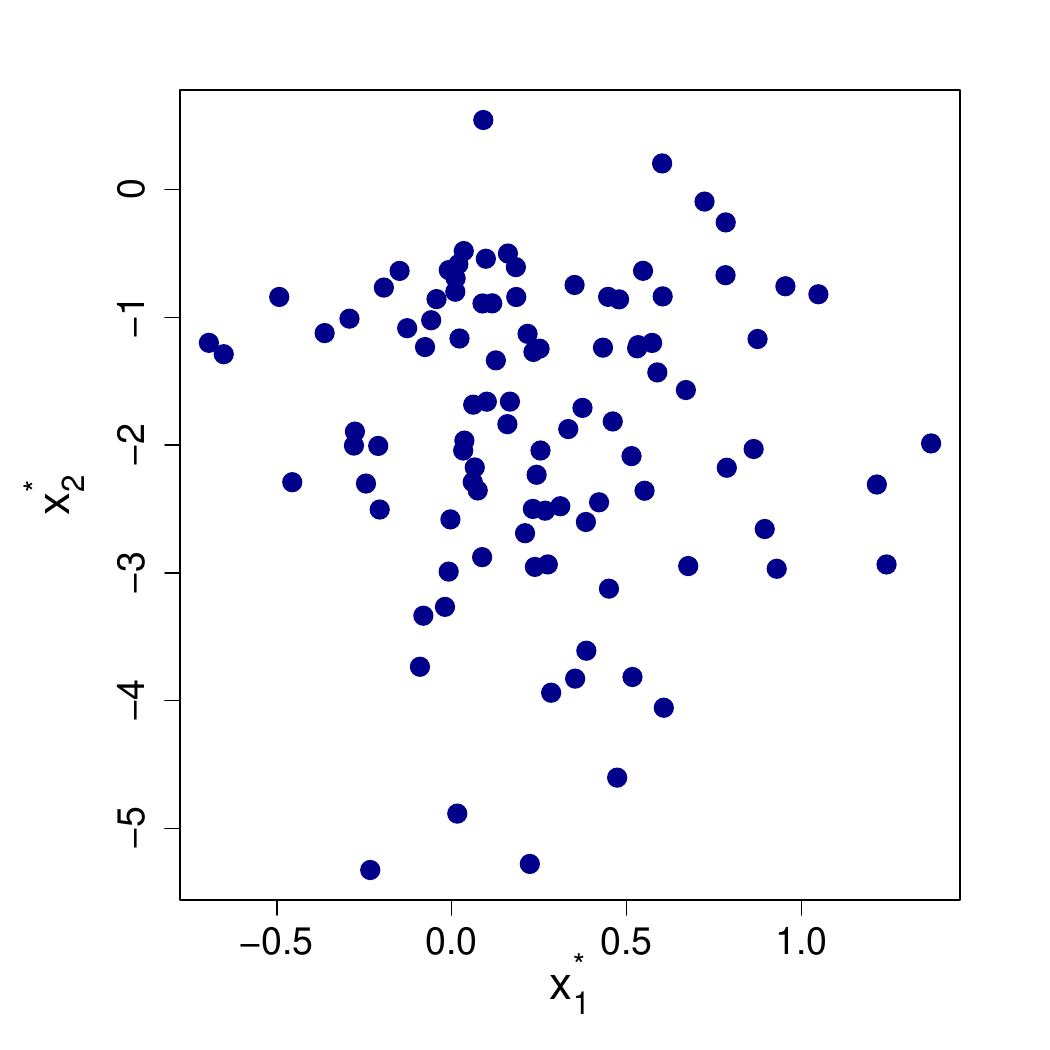} \\ 
			\includegraphics[scale=0.35]{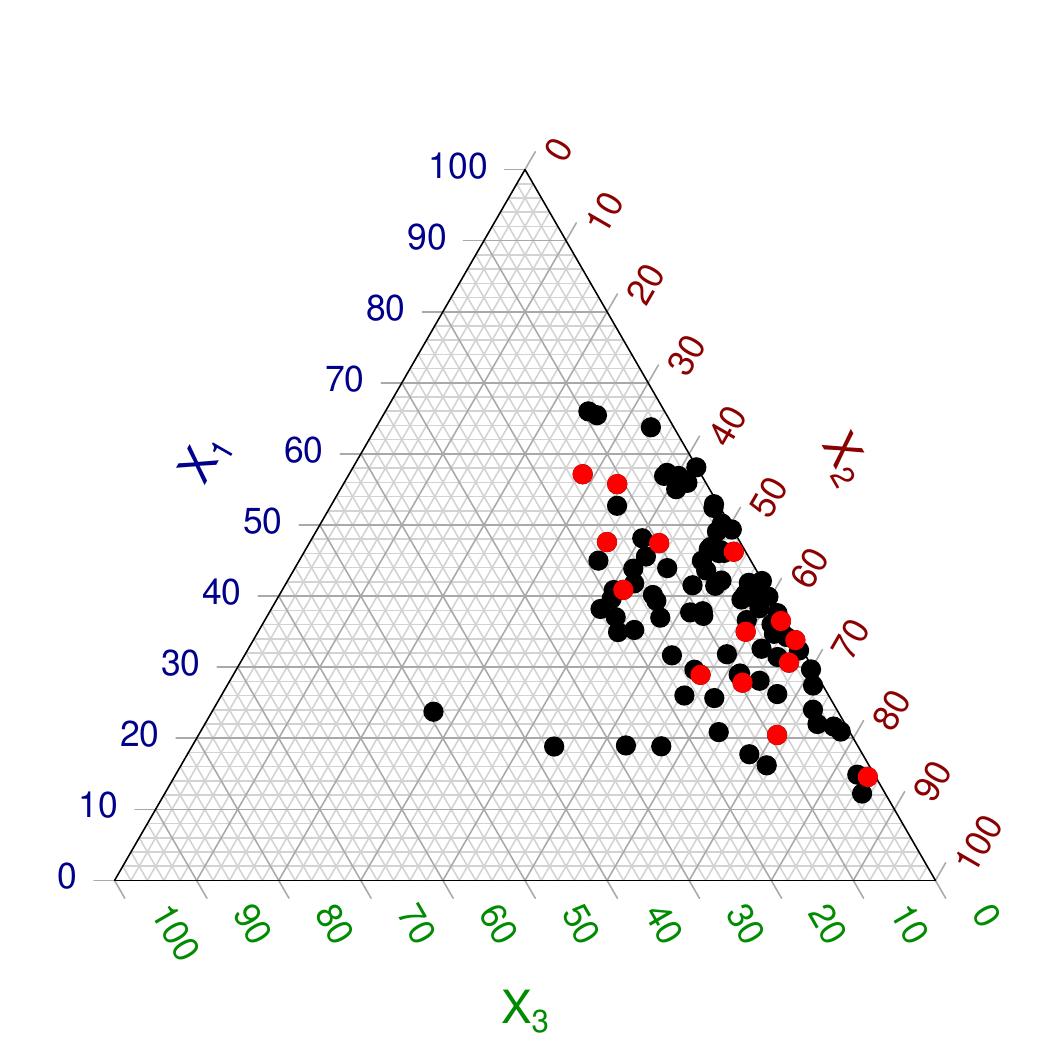} &
			\includegraphics[scale=0.35]{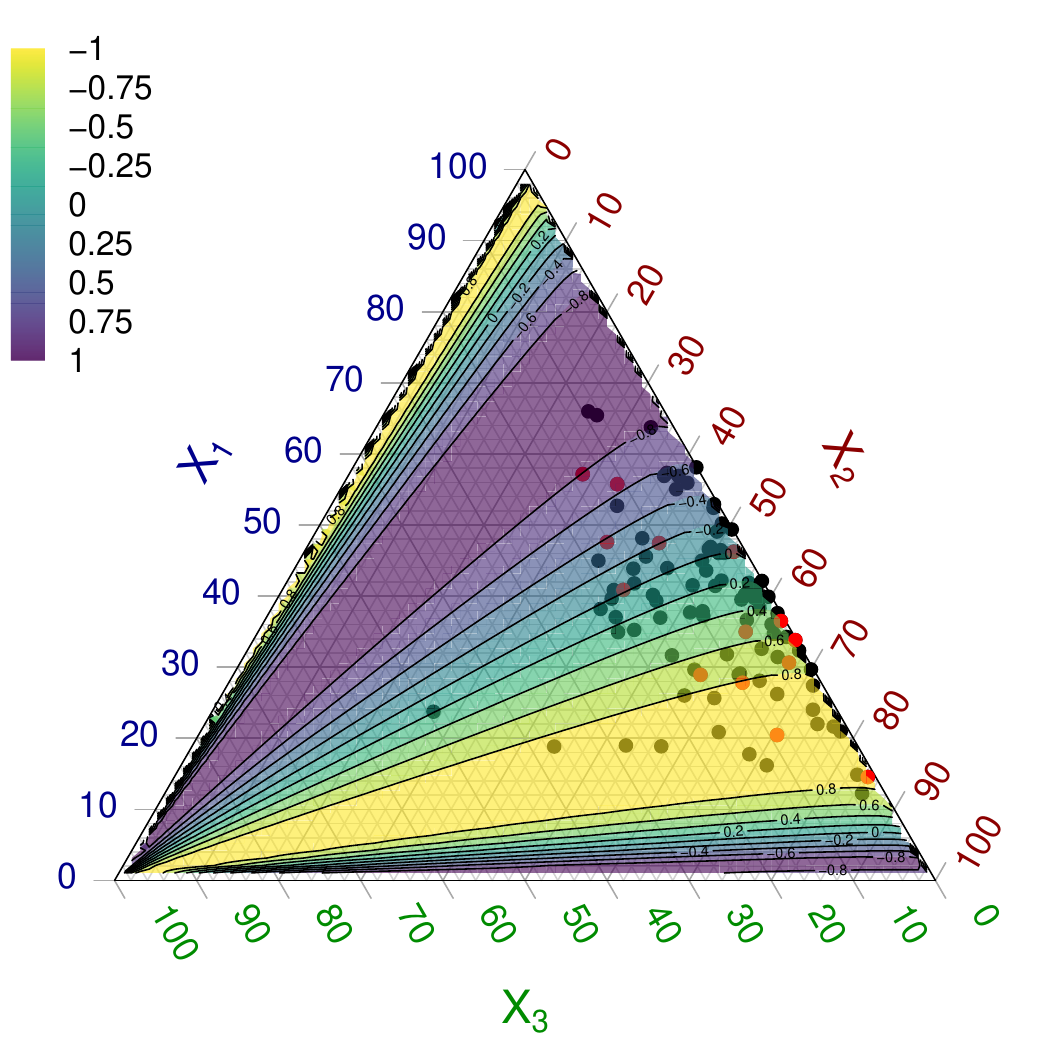} &
			\includegraphics[scale=0.32]{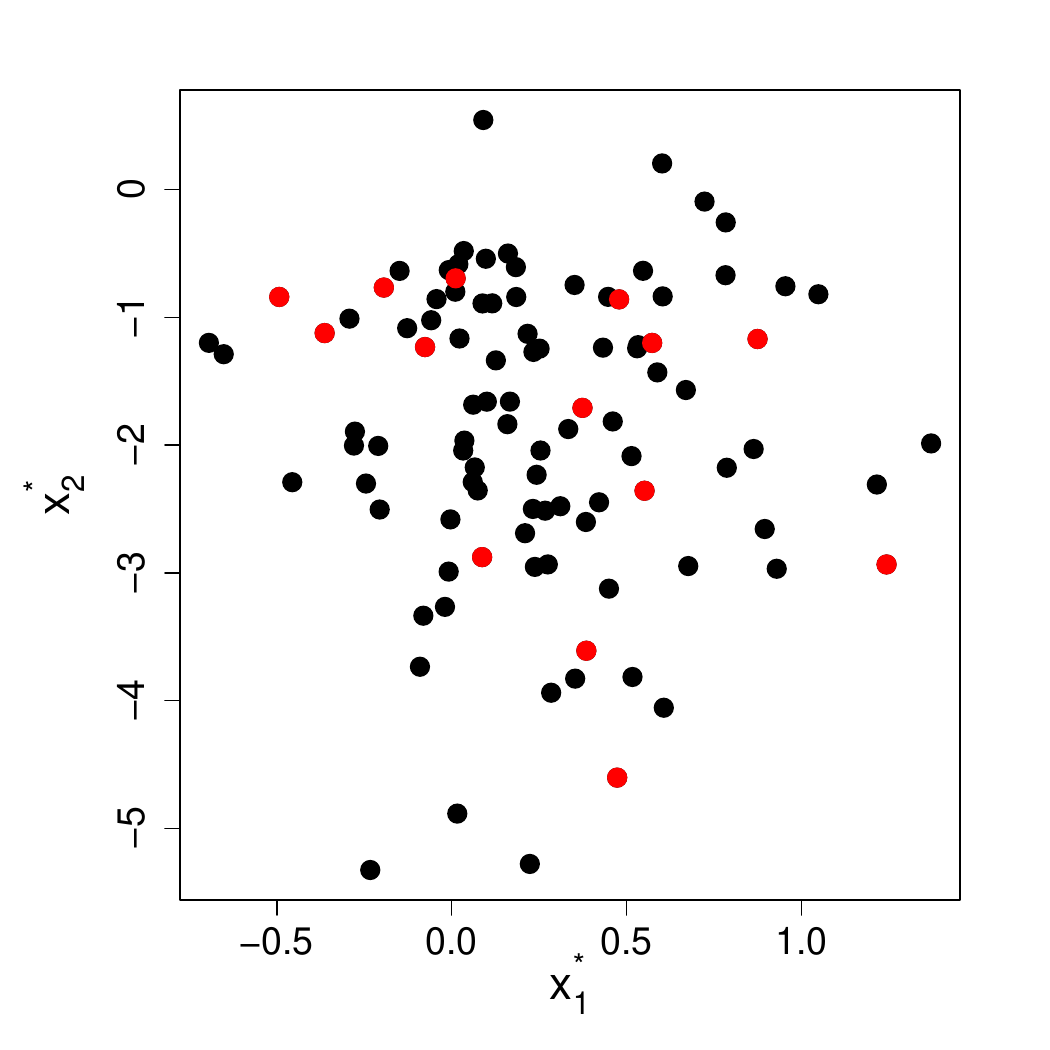} 
		\end{tabular}
		\vskip-0.2in \caption{ \small \label{fig:datos_simulados}  Synthetic data for one of the considered replications  when   $\balfa=(5, 7, 1)\trasp$: the left panel corresponds to the ternary diagram,   the middle one represents the contour plot of the regression function with the data over-imposed and the right one to the transformed data. The lower panels display  in red the covariates corresponding to the vertical outliers.}
\end{figure}

\begin{figure}[ht!]
		\renewcommand{\arraystretch}{0.1}
		\newcolumntype{G}{>{\centering\arraybackslash}m{\dimexpr.34\linewidth-1\tabcolsep}}
	\hskip-0.6in
				\begin{tabular}{GGG}
			Ternary Diagram &  $m(\bx)$ & ilr Plot \\ 
			\includegraphics[scale=0.35]{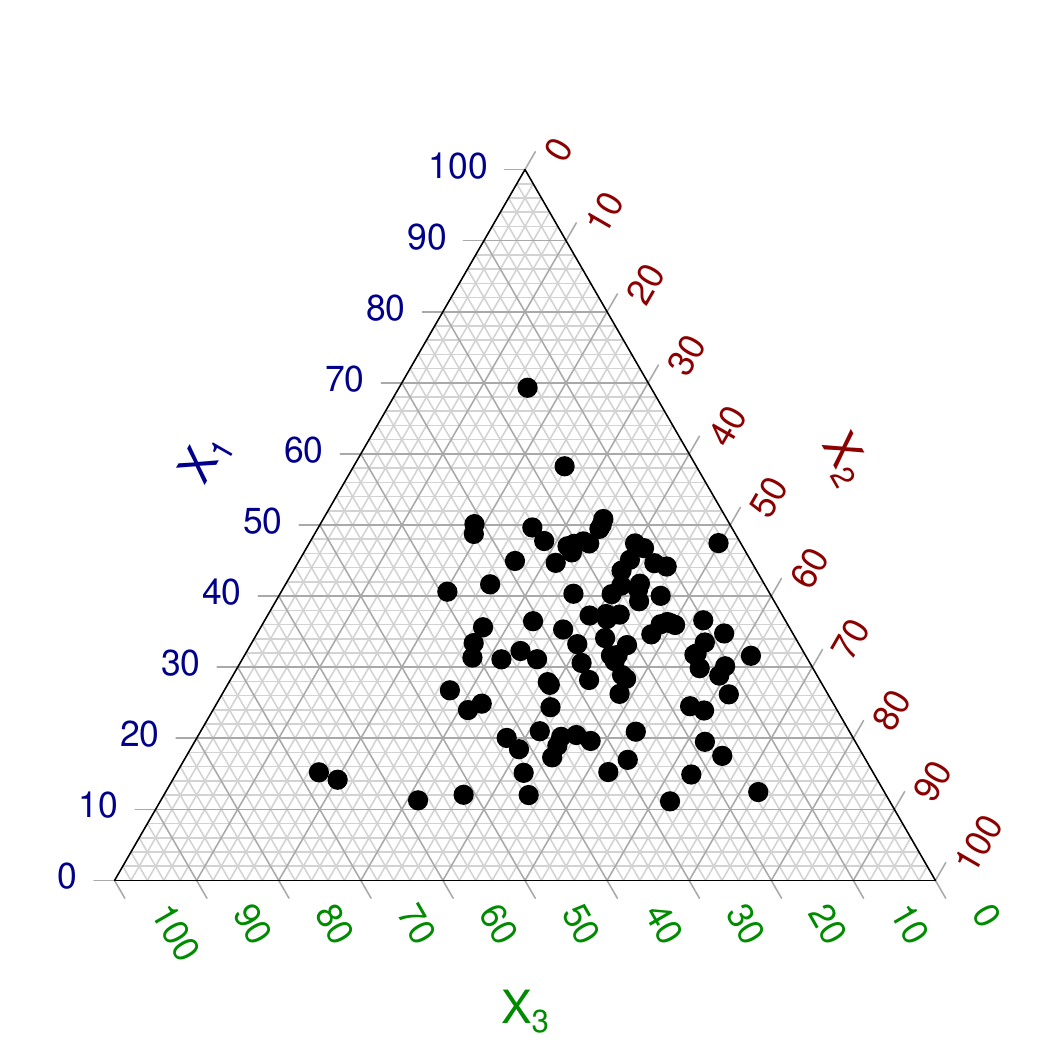} &
			\includegraphics[scale=0.35]{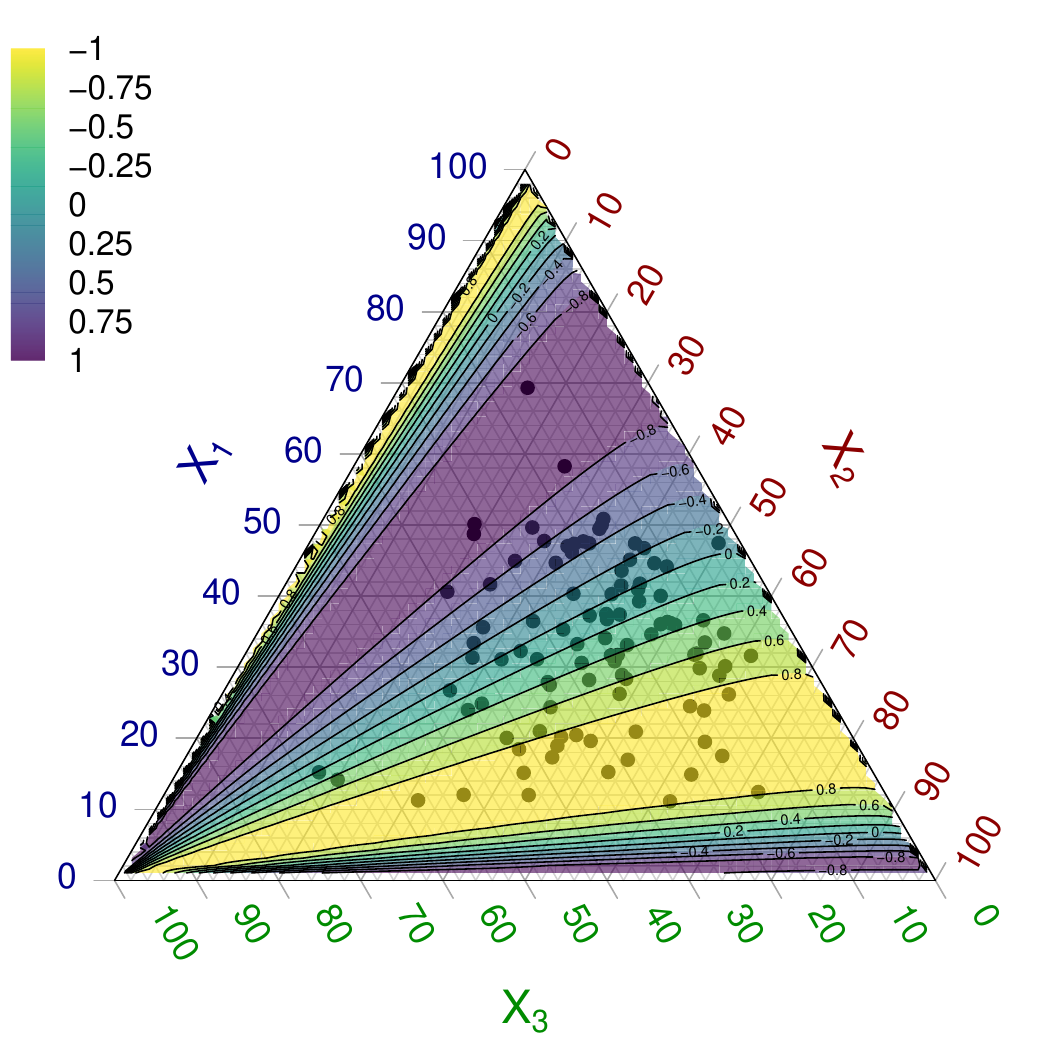} &
			\includegraphics[scale=0.32]{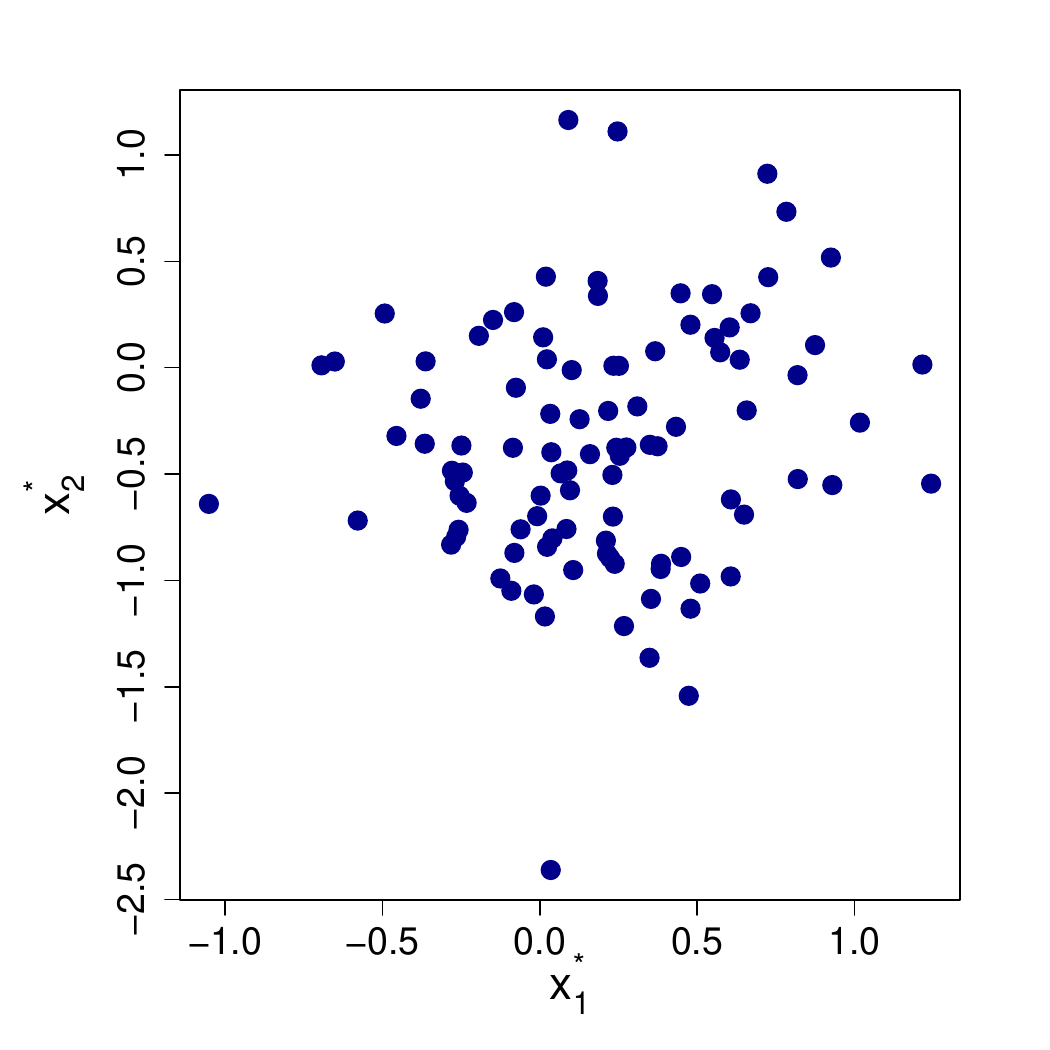} \\ 
			\includegraphics[scale=0.35]{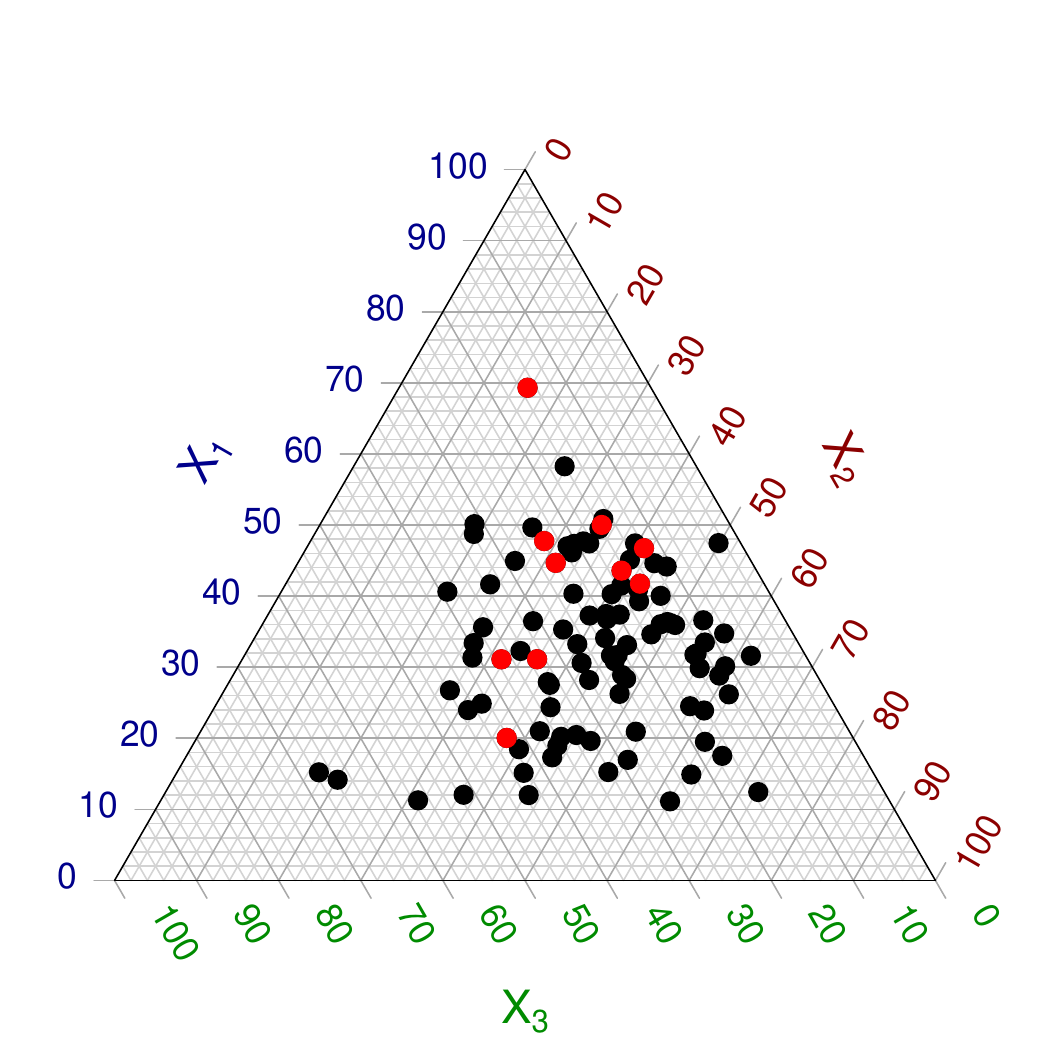} &
			\includegraphics[scale=0.35]{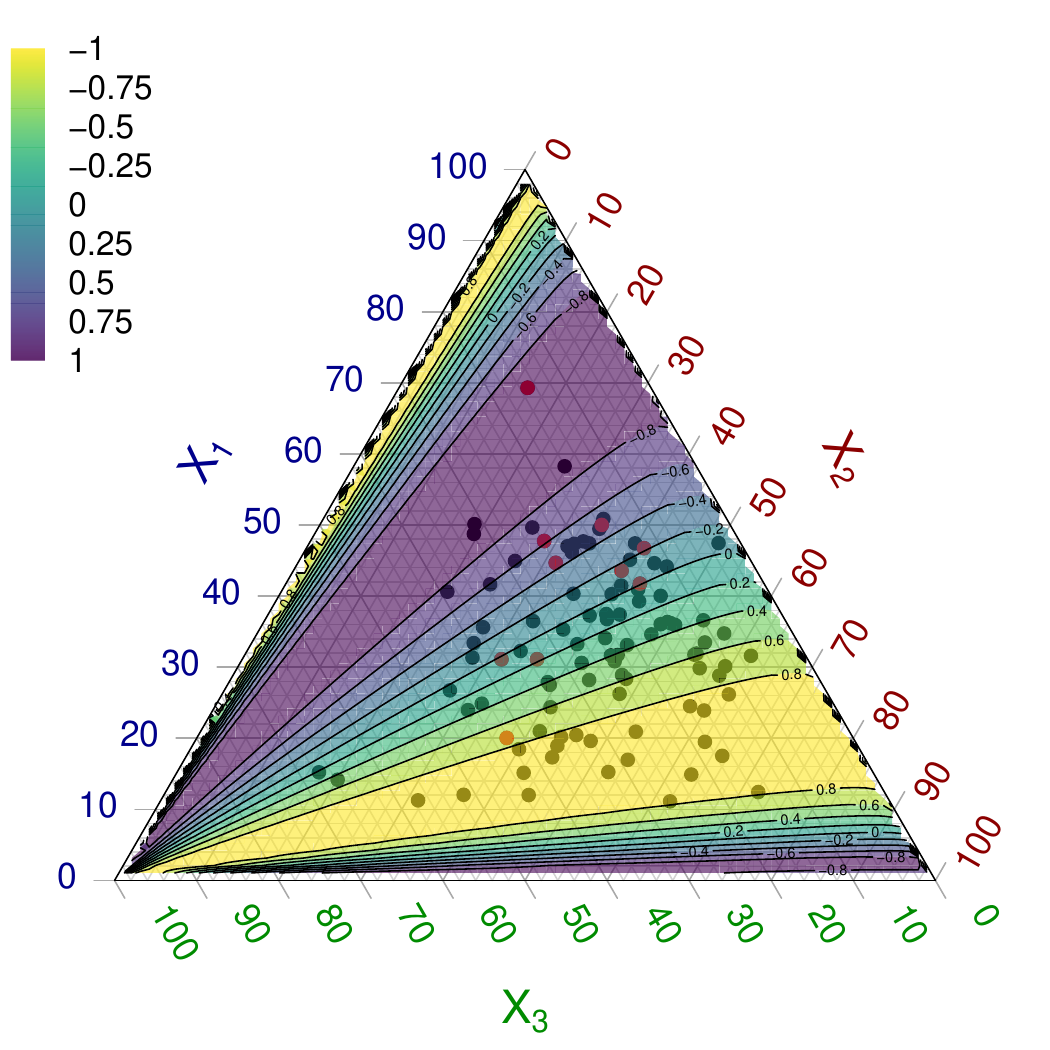} &
			\includegraphics[scale=0.32]{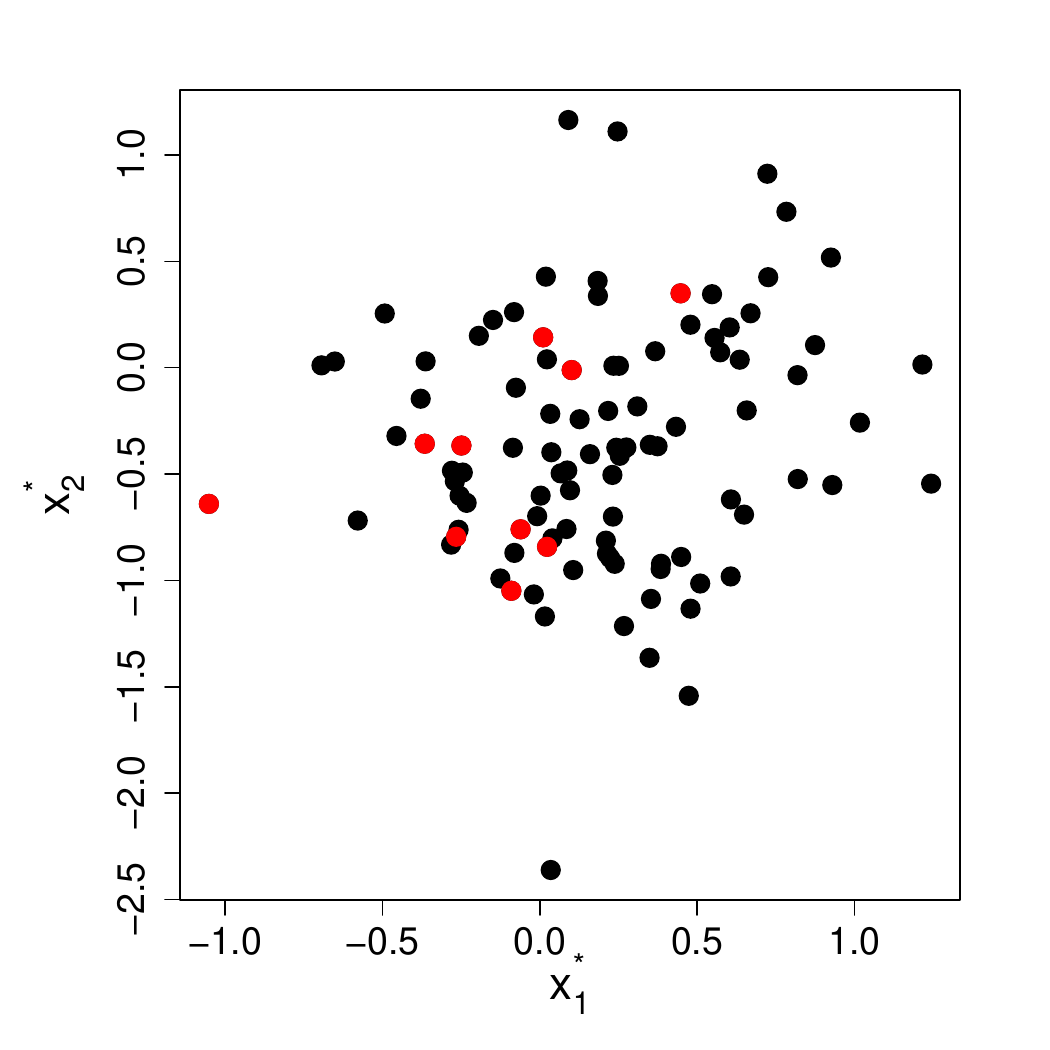} 
		\end{tabular}
		\vskip-0.2in \caption{ \small \label{fig:datos_simulados-5-7-4}  Synthetic data for one of the considered replications  when   $\balfa=(5, 7, 4)\trasp$: the left panel corresponds to the ternary diagram,   the middle one represents the contour plot of the regression function with the data over-imposed and the right one to the transformed data. The lower panels display  in red the covariates corresponding to the vertical outliers.}
\end{figure}

The panels in Figures \ref{fig:datos_simulados}  and \ref{fig:datos_simulados-5-7-4} illustrate the shape of the covariates in one generated sample when   $\balfa=(5, 7, 1)\trasp$ and $\balfa=(5, 7, 4)\trasp$, respectively. The left panel corresponds to the ternary diagram, the middle one represents the contour plot of the regression function and the right one to the ilr transformation of the data. The lower panel indicates in red   the  covariates corresponding to the vertical outliers generated by $C_{1,0.10, \mu}$.

To evaluate the behaviour of the regression estimators,   we measured the performance  of the estimator $\wm$     through the integrated squared error (\textsc{ise}). We approximate this measure   by evaluating the regression function and its estimators on points $\{\bx_{s}^{(0)}\}_{s=1}^M$,  generated with the same distribution as the covariates. We choose  $M=100$, that is, if  $\wm_{\ell}$ is the estimate of the function $m $ obtained with the $\ell$-th sample ($1 \le \ell \le N=1000$), we computed
$$\mbox{\textsc{ise}}=\frac{1}{M}\sum_{s=1}^M \left(\wm_{\ell}(\bx_{s})-m(\bx_{s})\right)^2\,.$$
We also provide a measure of the global squared bias given by
$$\mbox{Bias}^2     =    \frac 1M \sum_{s=1}^M
 \left( \frac{1}{N} \sum_{\ell=1}^{N} \wm_{\ell}(\bx_{s})-m(\bx_{s}) \right )^2 \,.$$

\begin{figure}[ht!]
	\begin{center}
		\renewcommand{\arraystretch}{0.1}
		\newcolumntype{G}{>{\centering\arraybackslash}m{\dimexpr.35\linewidth-1\tabcolsep}}
				\begin{tabular}{GGG}
			$C_0$  & $C_{1,0.10,5}$ & $C_{1,0.10,10}$ \\[-3ex]
			 \includegraphics[scale=0.3]{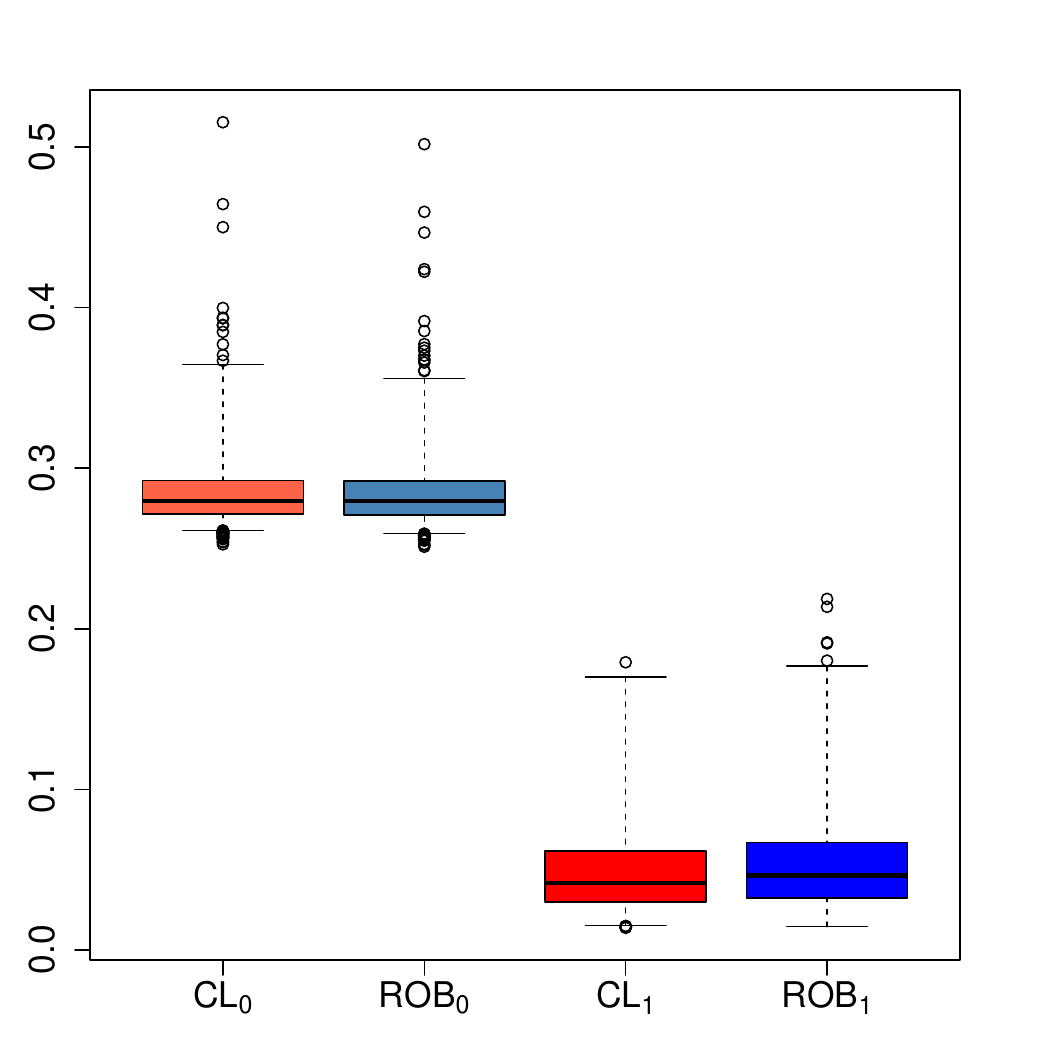}  &
	 		\includegraphics[scale=0.3]{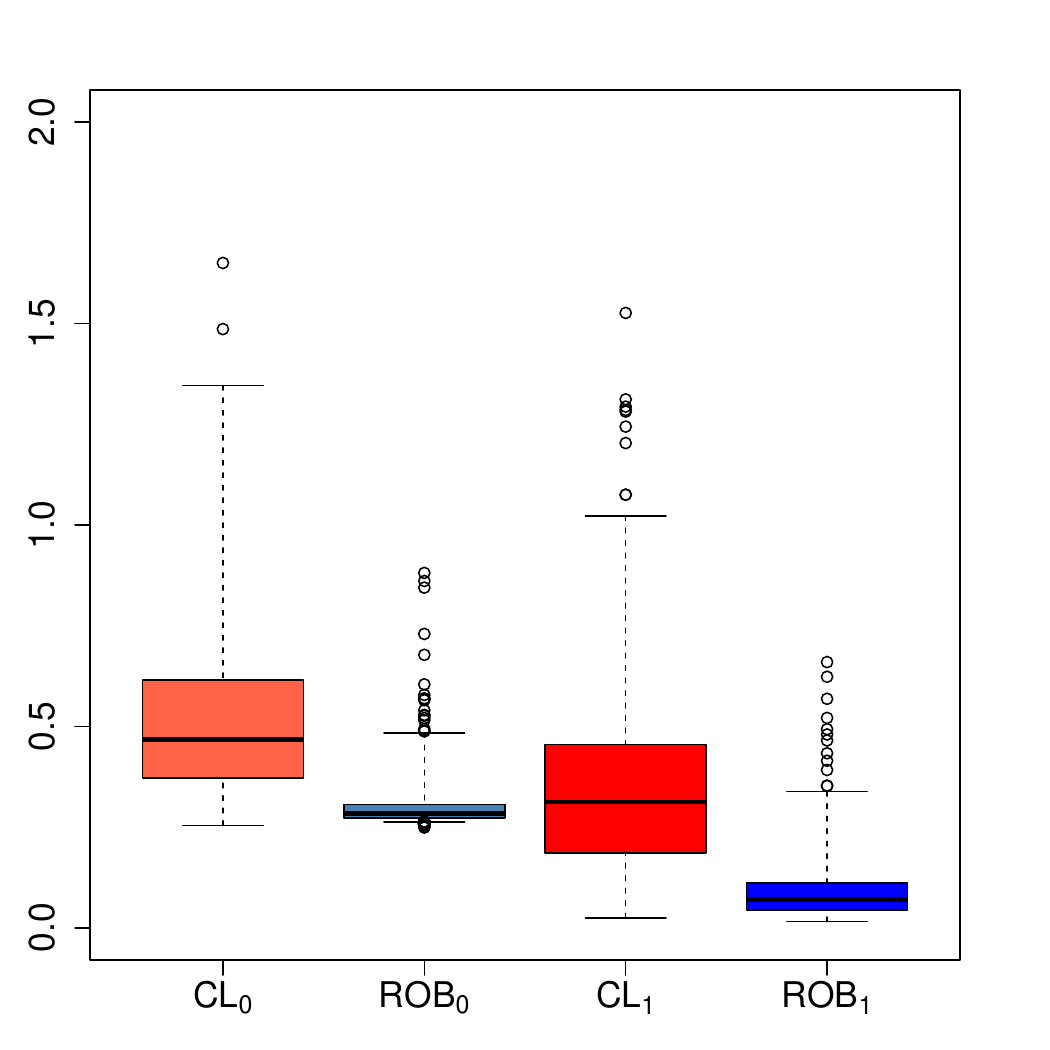} &
	 		\includegraphics[scale=0.3]{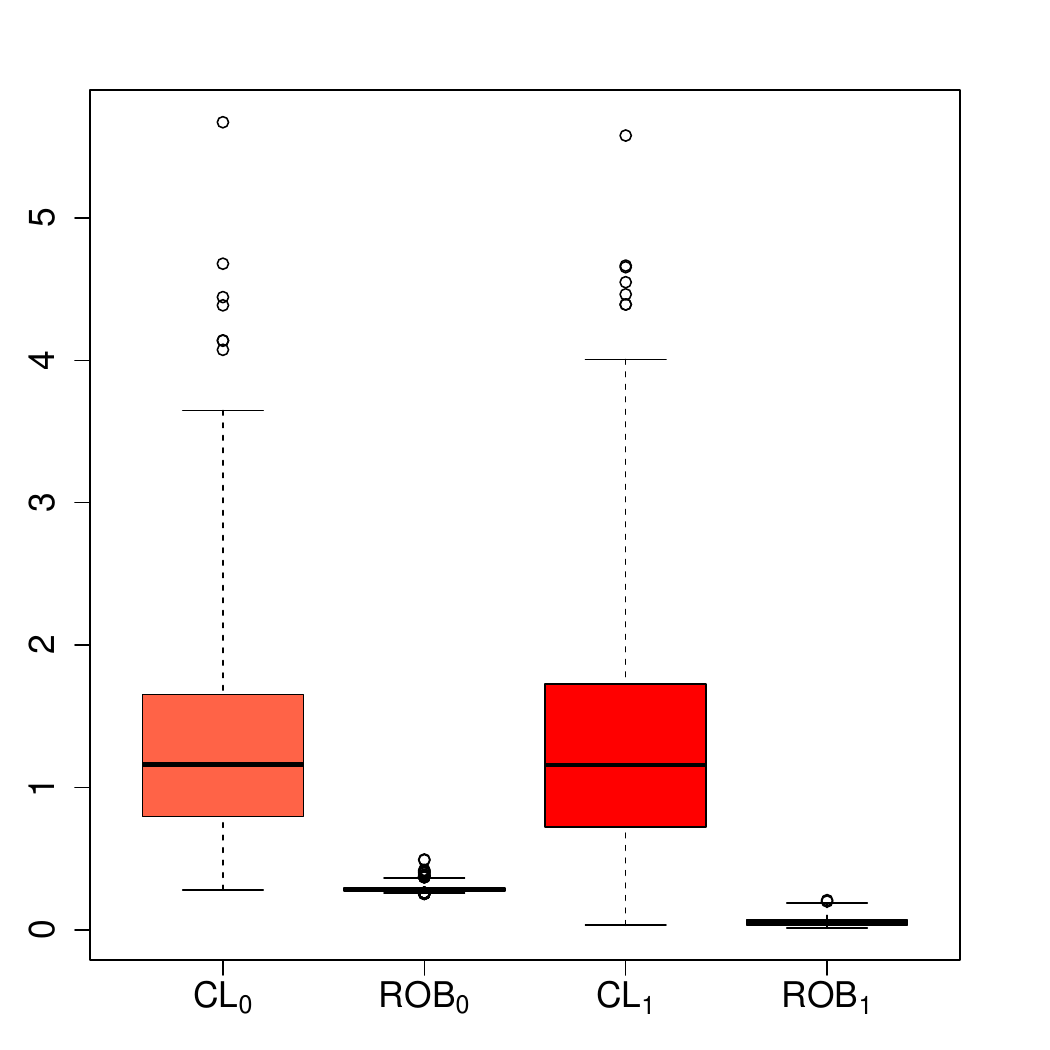}  \\
			$C_0$  & $C_{1,0.10,5}$ & $C_{1,0.10,10}$\\[-3ex]
	 		\includegraphics[scale=0.3]{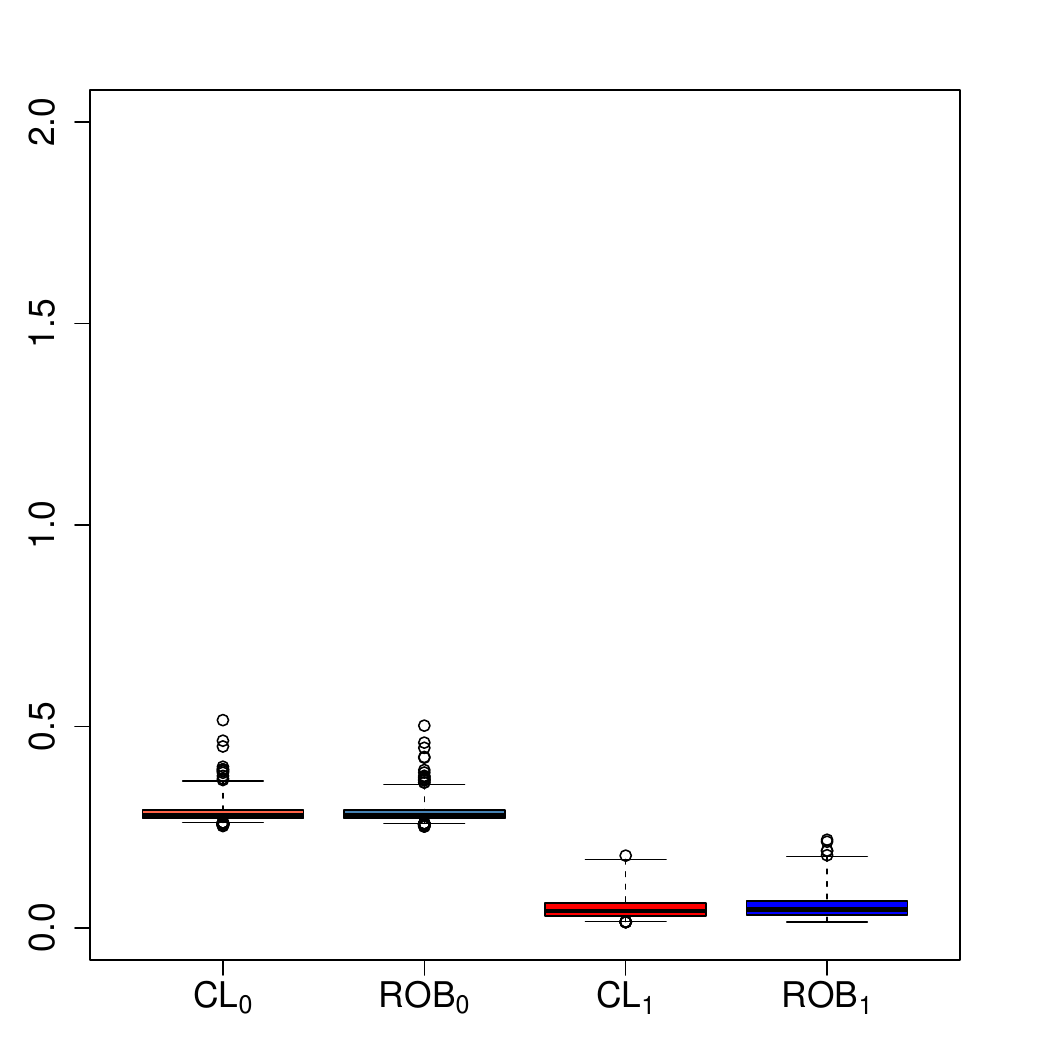} &
	 		\includegraphics[scale=0.3]{ADJboxplots_a-5-7-1_n100_cont_C1_delta_10_shift_5.pdf} &
	 		\includegraphics[scale=0.3]{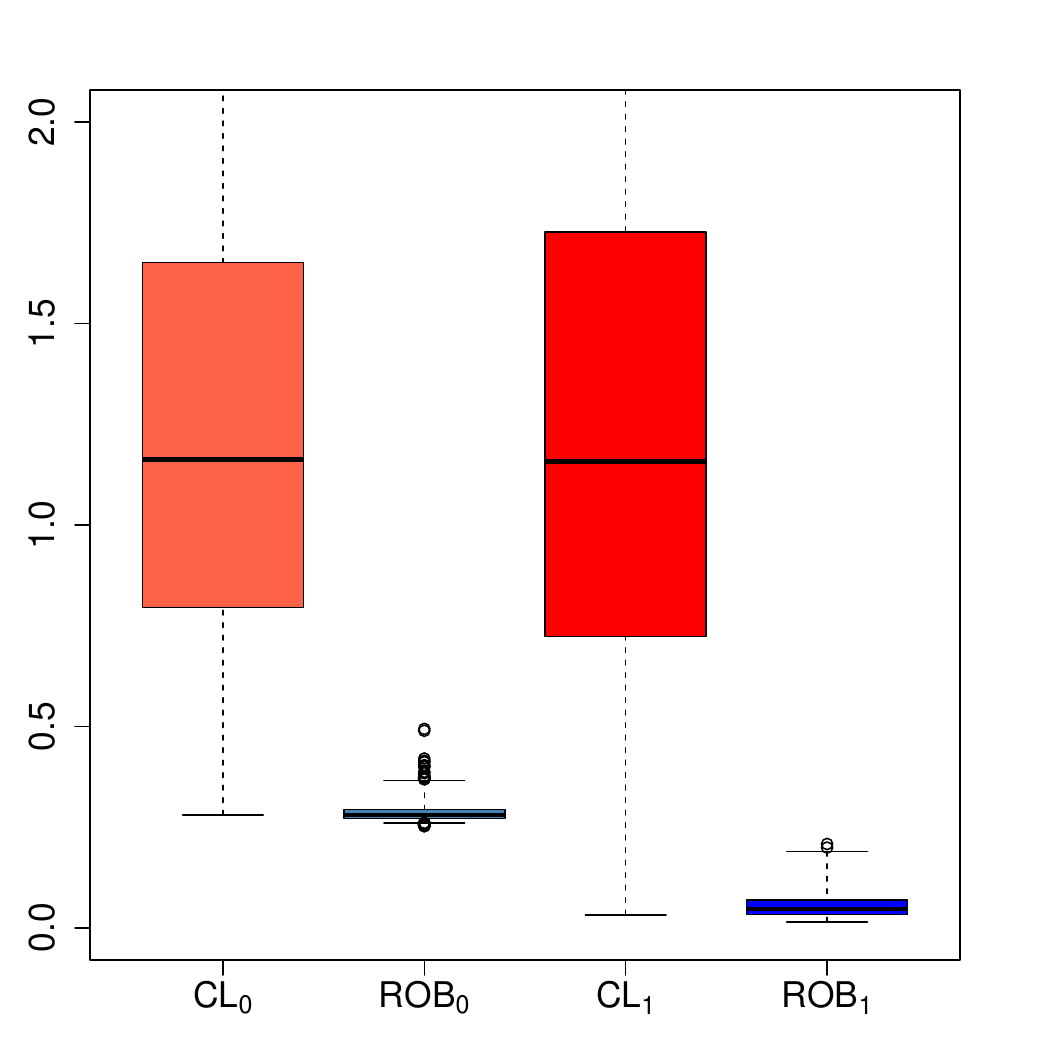} 
		\end{tabular}
		\vskip-0.1in \caption{ \small \label{fig:ECM}  Adjusted boxplots of the $\textsc{ise}$, when $\balfa=(5,7,1)\trasp$. The second row present the boxplots in the same scale.}
	\end{center} 
\end{figure}

\begin{figure}[ht!]
	\begin{center}
		\renewcommand{\arraystretch}{0.1}
		\newcolumntype{G}{>{\centering\arraybackslash}m{\dimexpr.35\linewidth-1\tabcolsep}}
				\begin{tabular}{GGG}
			$C_0$  & $C_{1,0.10,5}$ & $C_{1,0.10,10}$ \\[-3ex]
			 \includegraphics[scale=0.3]{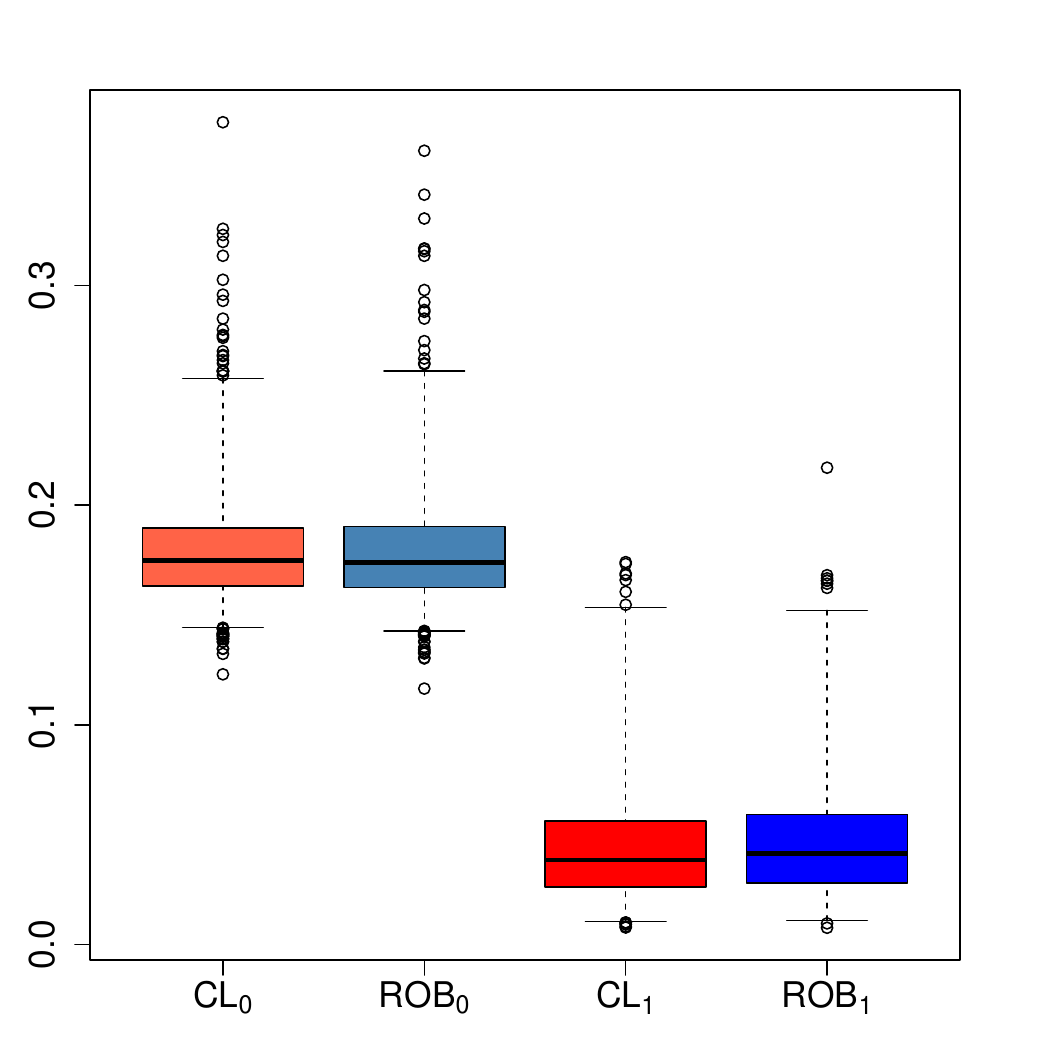}  &
	 		\includegraphics[scale=0.3]{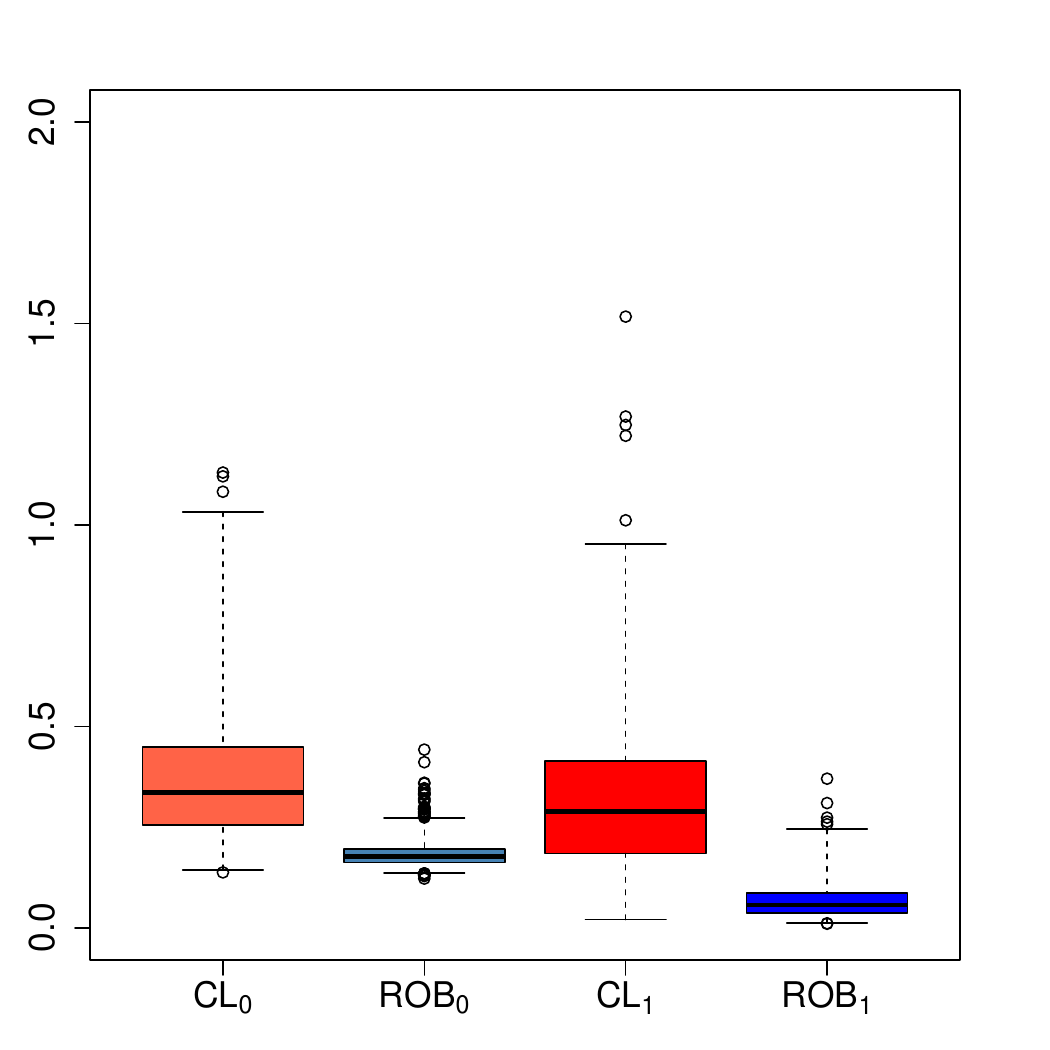} &
	 		\includegraphics[scale=0.3]{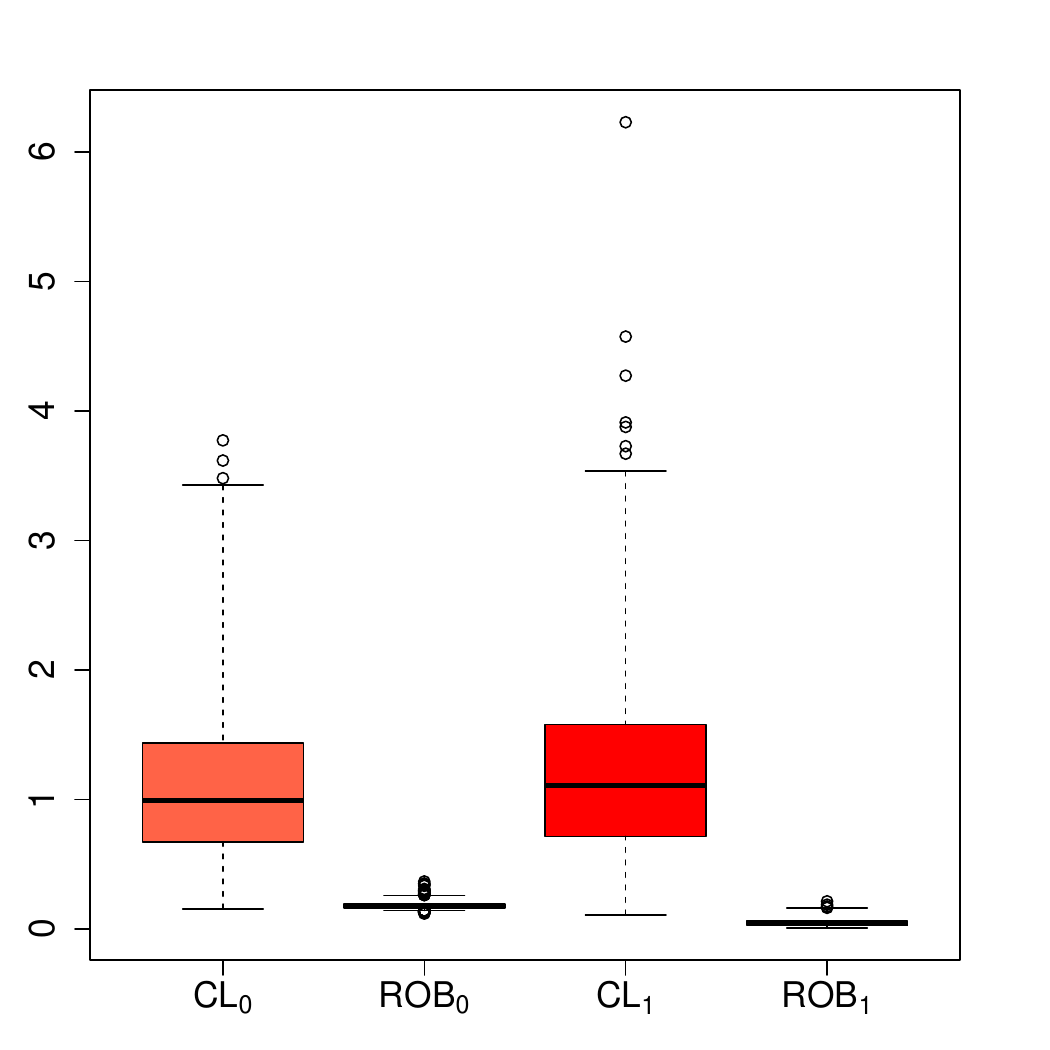}  \\
			$C_0$  & $C_{1,0.10,5}$ & $C_{1,0.10,10}$\\[-3ex]
	 		\includegraphics[scale=0.3]{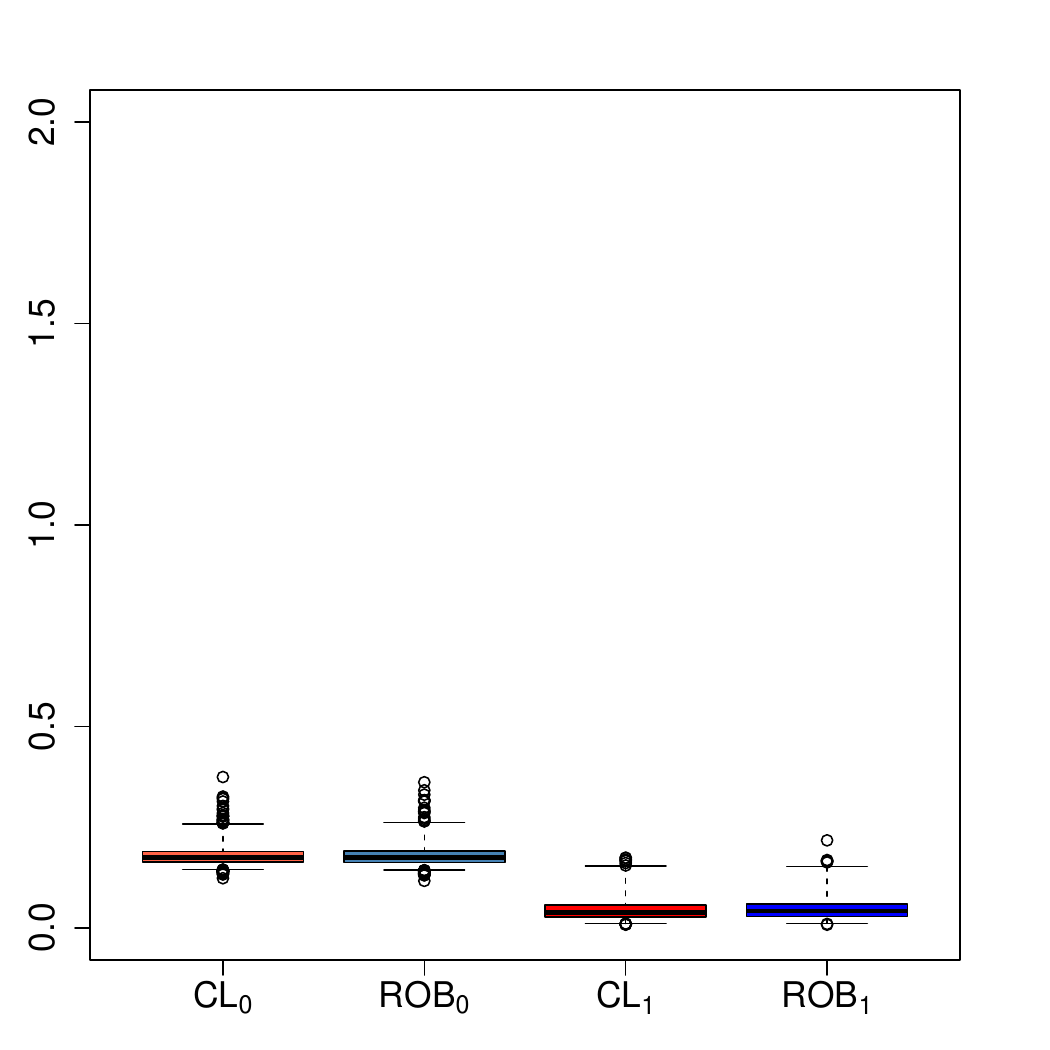} &
	 		\includegraphics[scale=0.3]{ADJboxplots_a-5-7-4_n100_cont_C1_delta_10_shift_5.pdf} &
	 		\includegraphics[scale=0.3]{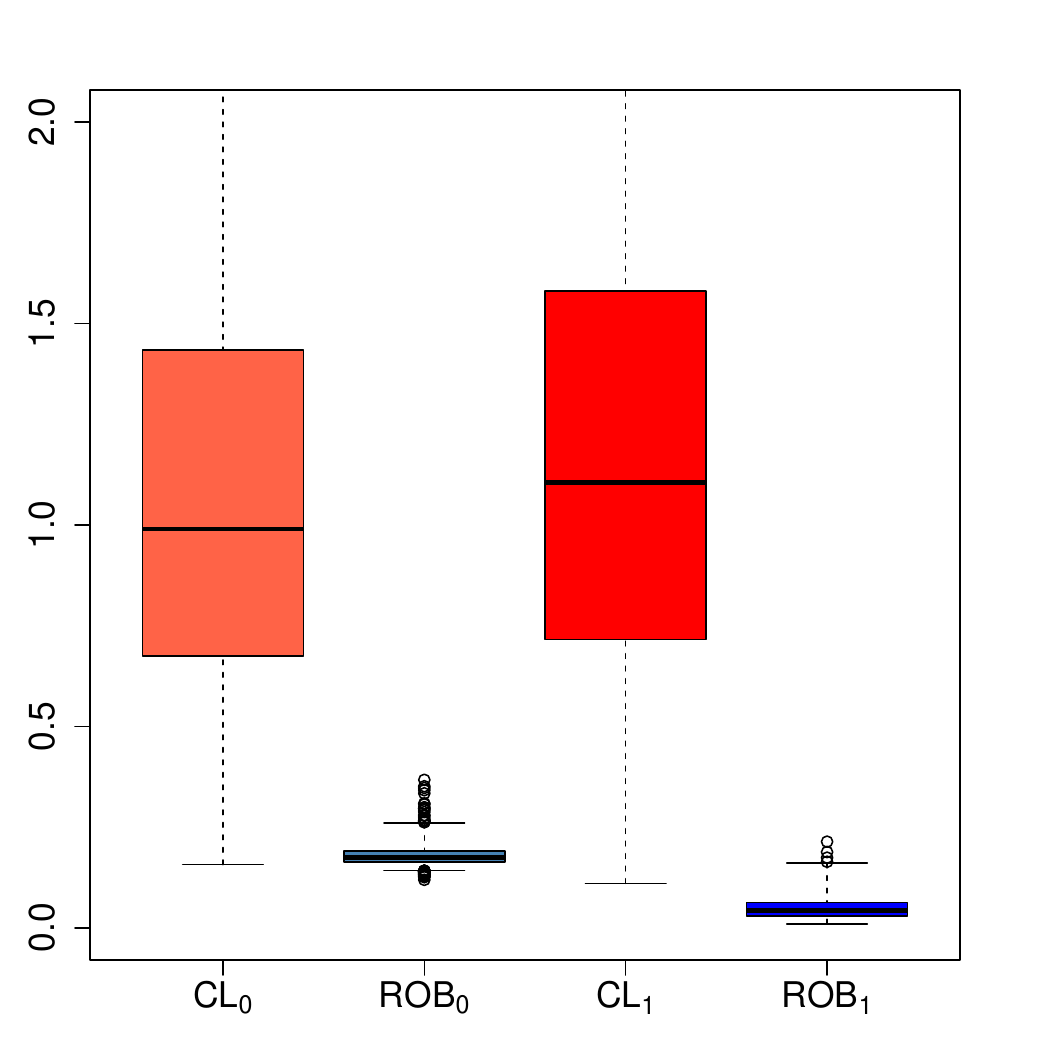} 
		\end{tabular}
		\vskip-0.1in \caption{ \small \label{fig:ECM-5-7-4}  Adjusted boxplots of the $\textsc{ise}$, when $\balfa=(5,7,4)\trasp$. The second row present the boxplots in the same scale.}
	\end{center} 
\end{figure}

Tables \ref{tab:MISE-BIAS} and \ref{tab:MISE-BIAS-5-7-4}  reports the mean over replications of the  \textsc{ise}, labelled \textsc{mise},  and  the squared bias, $\mbox{Bias}^2$, when   $\balfa=(5, 7, 1)\trasp$ and $\balfa=(5, 7, 4)\trasp$, respectively. The effect of the introduced vertical outliers is present on the classical procedures, in particular when $\mu=10$. The situation $\mu=5$ corresponds to a mild atypical data due to the considered central errors distribution and to the shape of the regression function explaining why the effect on the robust procedures is larger when $\mu=5$ than when $\mu=10$. However, in all cases, the robust methods described in this paper provide stable and reliable estimators. Figures \ref{fig:ECM} and \ref{fig:ECM-5-7-4} provide  the adjusted boxplots, see \citet{Hubert:Vandervieren:2008}, of the \textsc{ise}. These plots show the stability of the robust procedures across contaminations and the advantage of considering local linear estimators over local constants.

\begin{table}[ht!]
\begin{center}
\renewcommand{\arraystretch}{1.2}
    \setlength{\tabcolsep}{4pt} 
    \begin{tabular}{|c|c|c|c|c|c|c|c| }
	\hline
	& \multicolumn{3}{c|}{ \textsc{mise}} & &  \multicolumn{3 }{c|}{$\mbox{Bias}^2$}\\
	\hline
	& $C_{0}$ & $C_{1, 0.10, 5}$ & $C_{1, 0.10, 10}$ &   & $C_{0}$ & $C_{1, 0.10, 5}$ & $C_{1, 0.10, 10}$     \\
	\hline 
\textsc{CL}$_{0}$ &  0.2856 & 0.5150 & 1.3056 & & 0.2721 &  0.4762  & 1.1917  \\

\textsc{CL}$_{1}$ & 0.0489 & 0.3461 &  1.2975 & & 0.0156 & 0.2438 &  0.9801   \\
\textsc{ROB}$_{0}$ & 0.2853 & 0.3025 & 0.2869 & & 0.2711 & 0.2732 &  0.2713  \\
\textsc{ROB}$_{1}$ & 0.0539 & 0.0892 & 0.0549 & & 0.0232 & 0.0392 & 0.0198  \\
\hline
 \end{tabular}
\vskip-0.1in
\caption{ \small \label{tab:MISE-BIAS} MISE and squared bias of the regression estimators  when $\balfa=(5,7,1)\trasp$. }
\end{center}
\end{table}

\begin{table}[ht!]
\begin{center}
\renewcommand{\arraystretch}{1.2}
    \setlength{\tabcolsep}{4pt} 
    \begin{tabular}{|c|c|c|c|c|c|c|c| }
	\hline
	& \multicolumn{3}{c|}{ \textsc{mise}} & &  \multicolumn{3 }{c|}{$\mbox{Bias}^2$}\\
	\hline
	& $C_{0}$ & $C_{1, 0.10, 5}$ & $C_{1, 0.10, 10}$ &   & $C_{0}$ & $C_{1, 0.10, 5}$ & $C_{1, 0.10, 10}$     \\
	\hline 
\textsc{CL}$_{0}$ &  0.1801 &   0.3731 & 1.1058 & & 0.1669 & 0.3397  &  1.0054 \\
\textsc{CL}$_{1}$ &  0.0456 &   0.3207 & 1.2212 & & 0.0135 & 0.2291  &  0.9359 \\
\textsc{ROB}$_{0}$ & 0.1794 &   0.1838 & 0.1804 & & 0.1653 & 0.1616  &  0.1655 \\
\textsc{ROB}$_{1}$ & 0.0481 &   0.0680 & 0.0489 & & 0.0184 & 0.0262  &  0.0158 \\
\hline
 \end{tabular}
\vskip-0.1in
\caption{ \small \label{tab:MISE-BIAS-5-7-4} MISE and squared bias of the regression estimators when $\balfa=(5,7,4)\trasp$. }
\end{center}
\end{table}

\begin{figure}[ht!]
	\begin{center}
		\renewcommand{\arraystretch}{0.1}
		\newcolumntype{G}{>{\centering\arraybackslash}m{\dimexpr.35\linewidth-1\tabcolsep}}
				\begin{tabular}{GGG}
			$m$ & CL$_0$ & CL$_1$ \\[-3ex]
			 \includegraphics[scale=0.31]{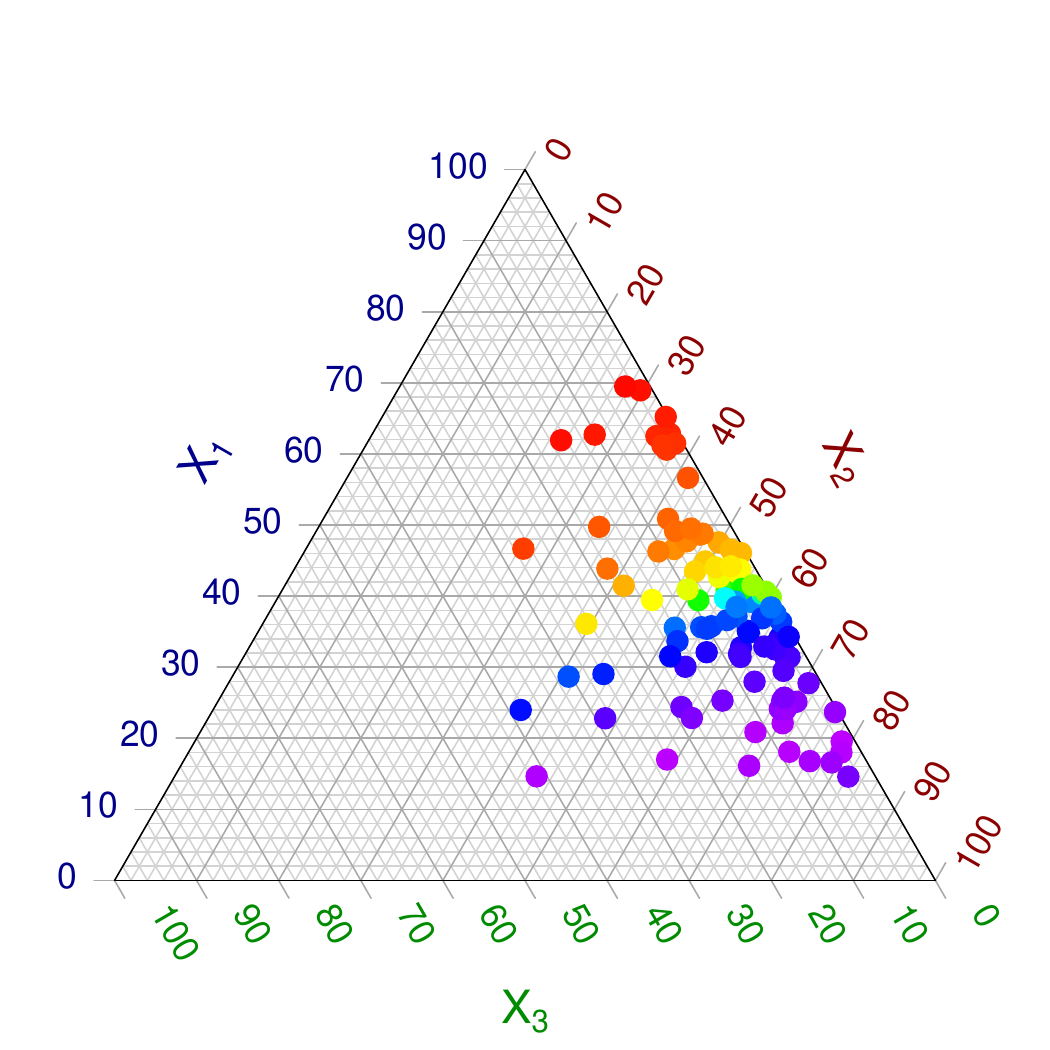}  &
	 		\includegraphics[scale=0.31]{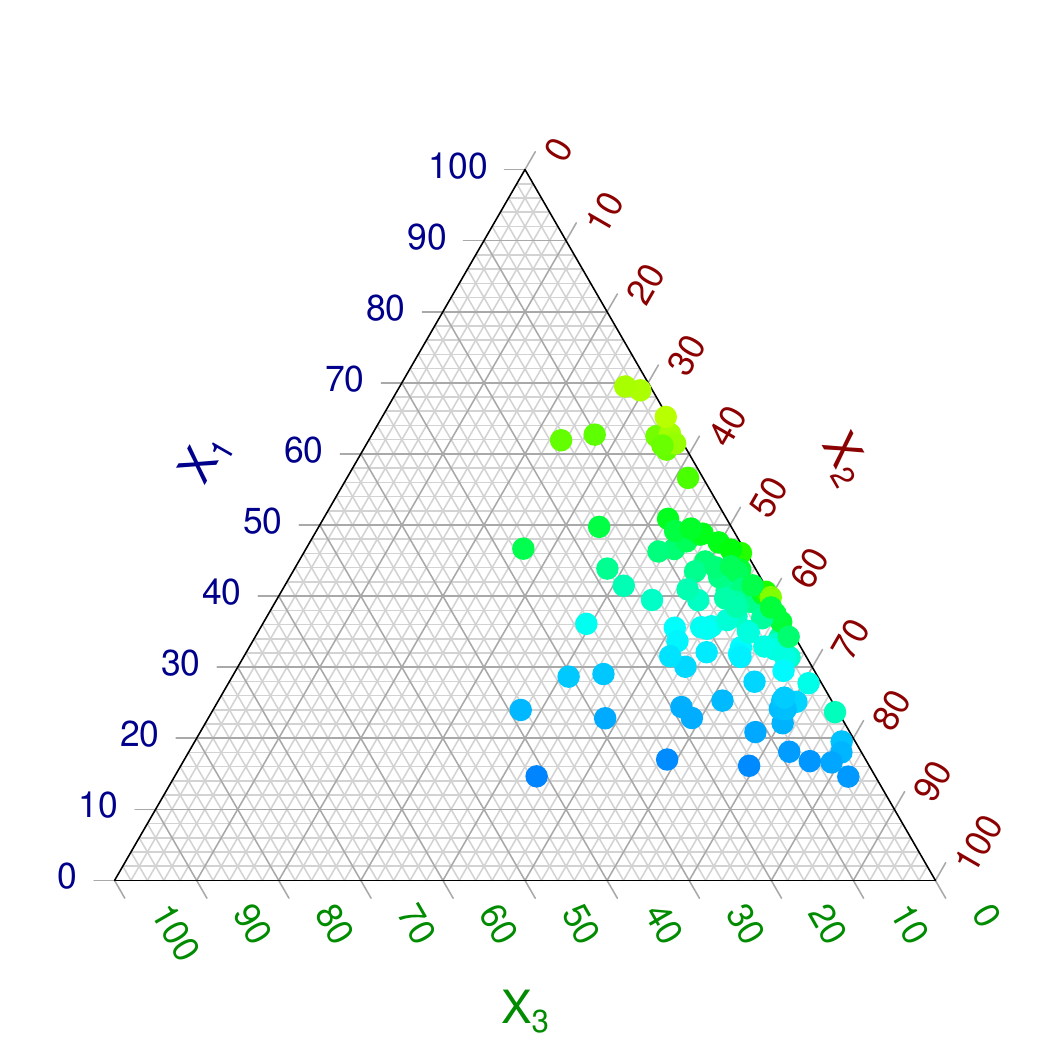} &
	 		\includegraphics[scale=0.31]{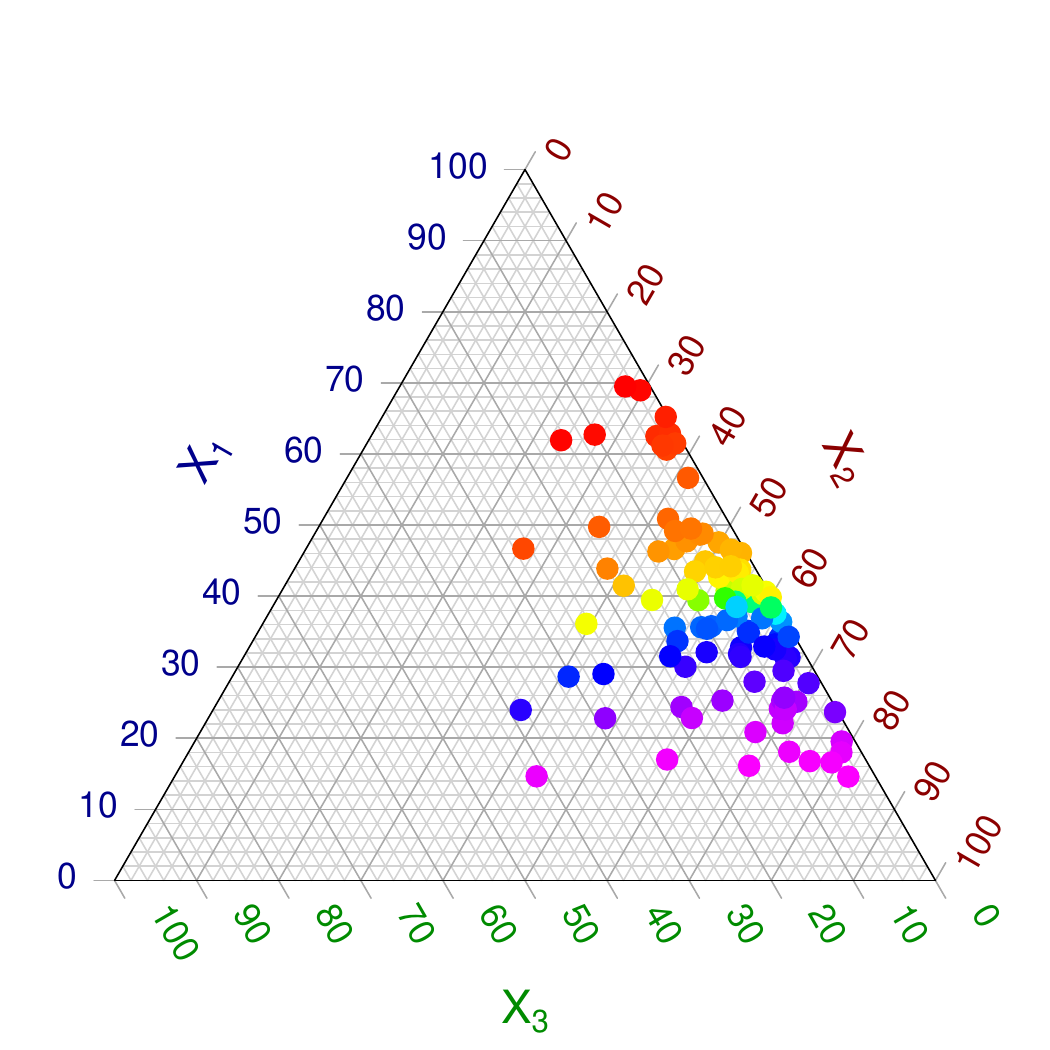}  \\
			$m$ & ROB$_0$ & ROB$_1$\\[-3ex]
	 		 \includegraphics[scale=0.31]{mhat-real-a5-7-1_n100_cont_C0-bis.pdf}  &
	 		\includegraphics[scale=0.31]{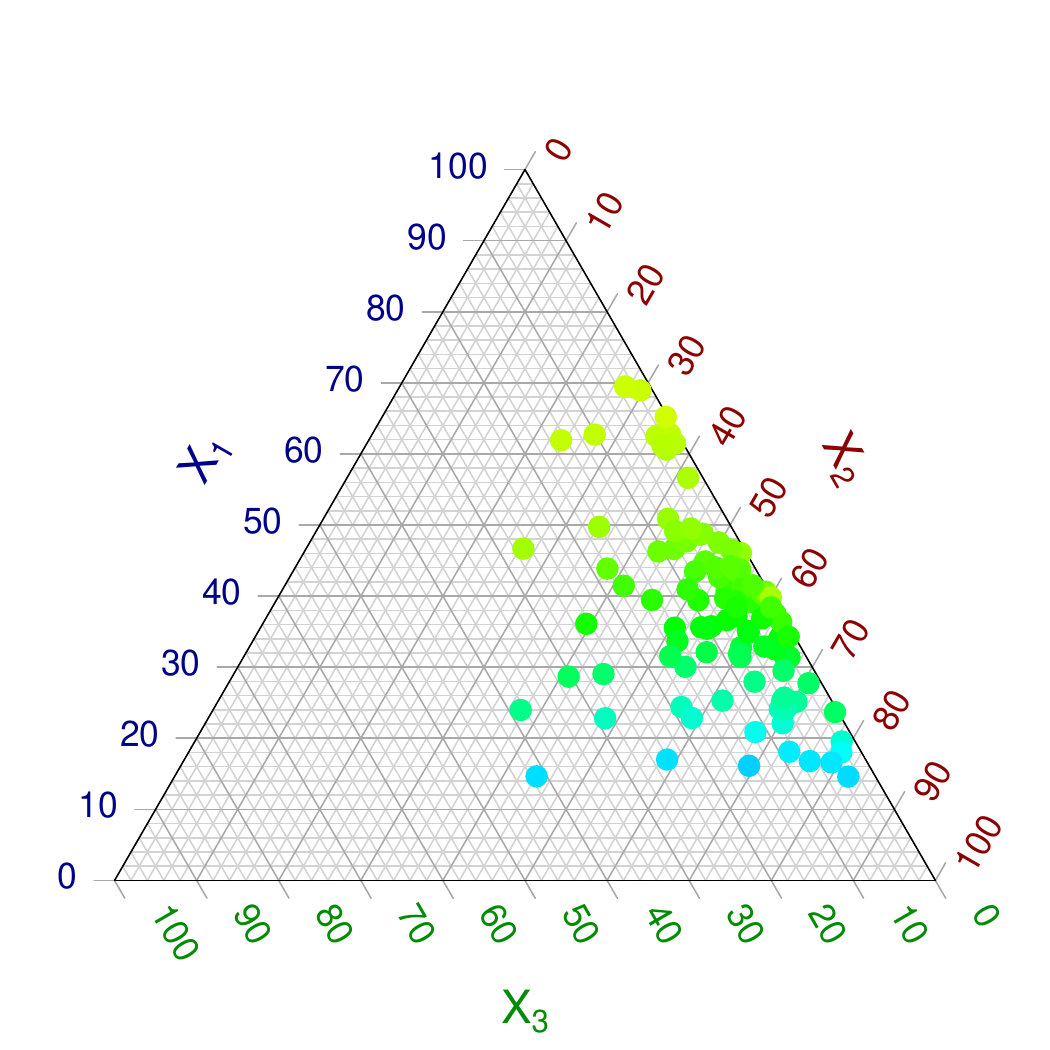} &
	 		\includegraphics[scale=0.31]{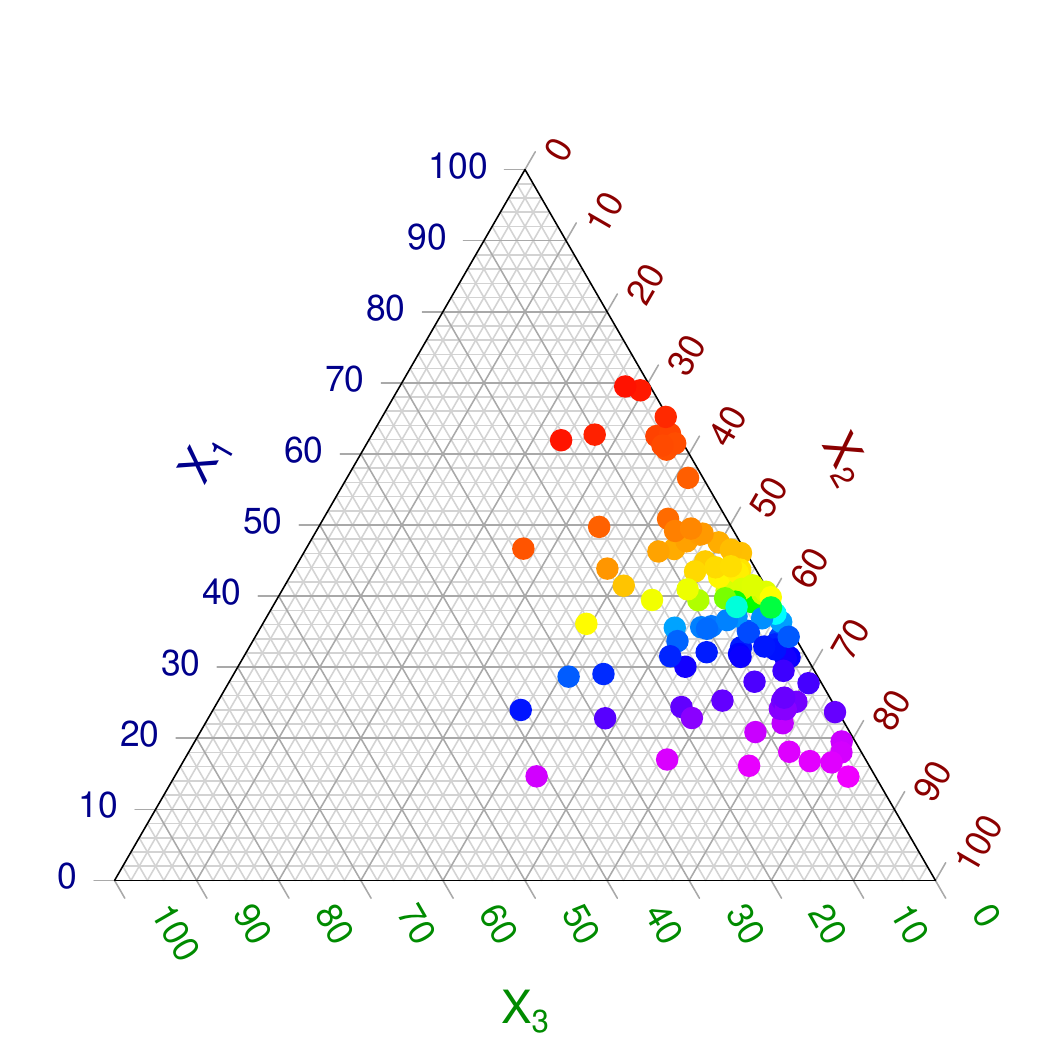}   
		\end{tabular}
		\vskip-0.1in \caption{ \small \label{fig:mhat.C0}  Ternary diagrams with the prediction points $\{\bx_{s}^{(0)}\}_{s=1}^M$ in different colours when $\balfa=(5,7,1)\trasp$. The colour palette indicates the value of $\wm(\bx_{s}^{(0)})$ under $C_{0}$ and $m(\bx_{s}^{(0)})$.}
	\end{center} 
\end{figure}

\begin{figure}[ht!]
	\begin{center}
		\renewcommand{\arraystretch}{0.1}
		\newcolumntype{G}{>{\centering\arraybackslash}m{\dimexpr.35\linewidth-1\tabcolsep}}
				\begin{tabular}{GGG}
			$m$ & CL$_0$ & CL$_1$ \\[-3ex]
			 \includegraphics[scale=0.31]{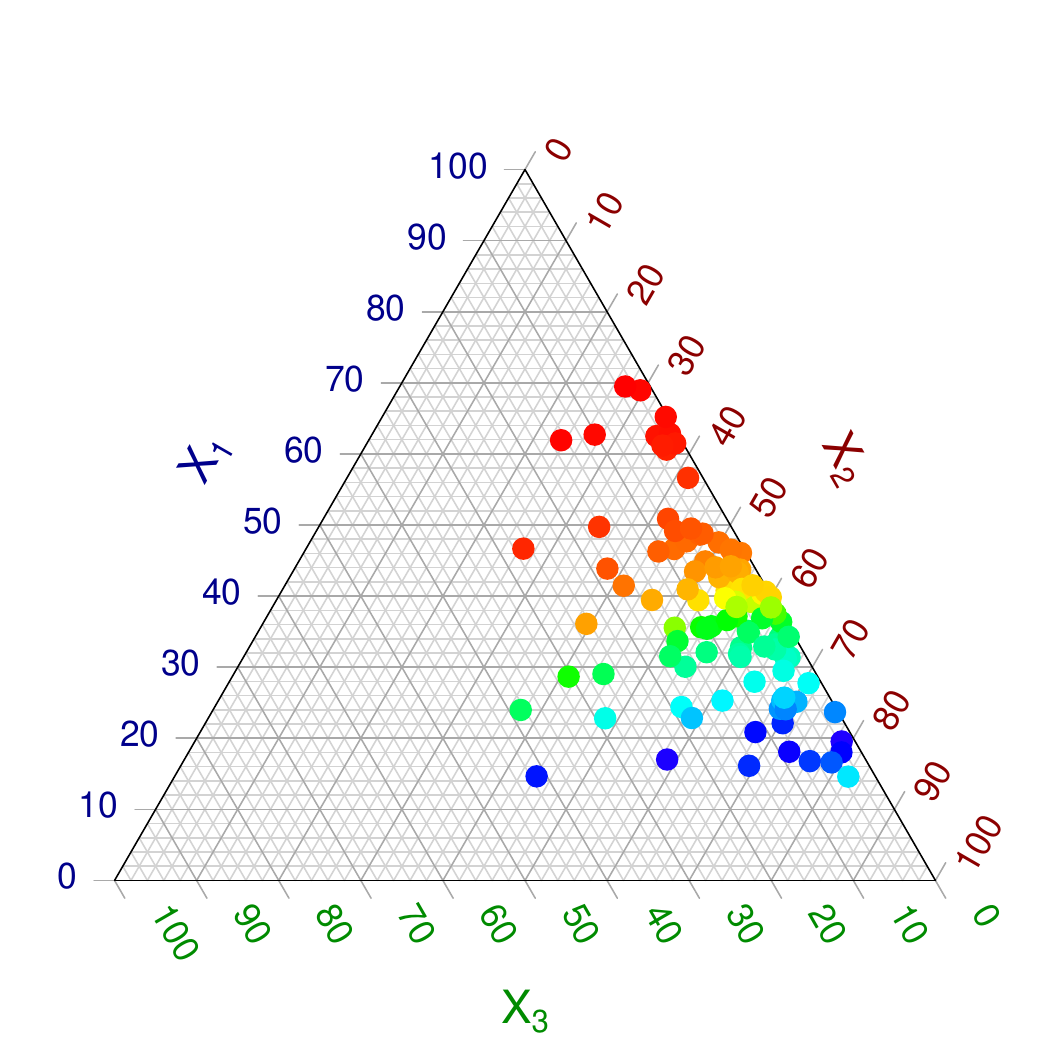}  &
	 		\includegraphics[scale=0.31]{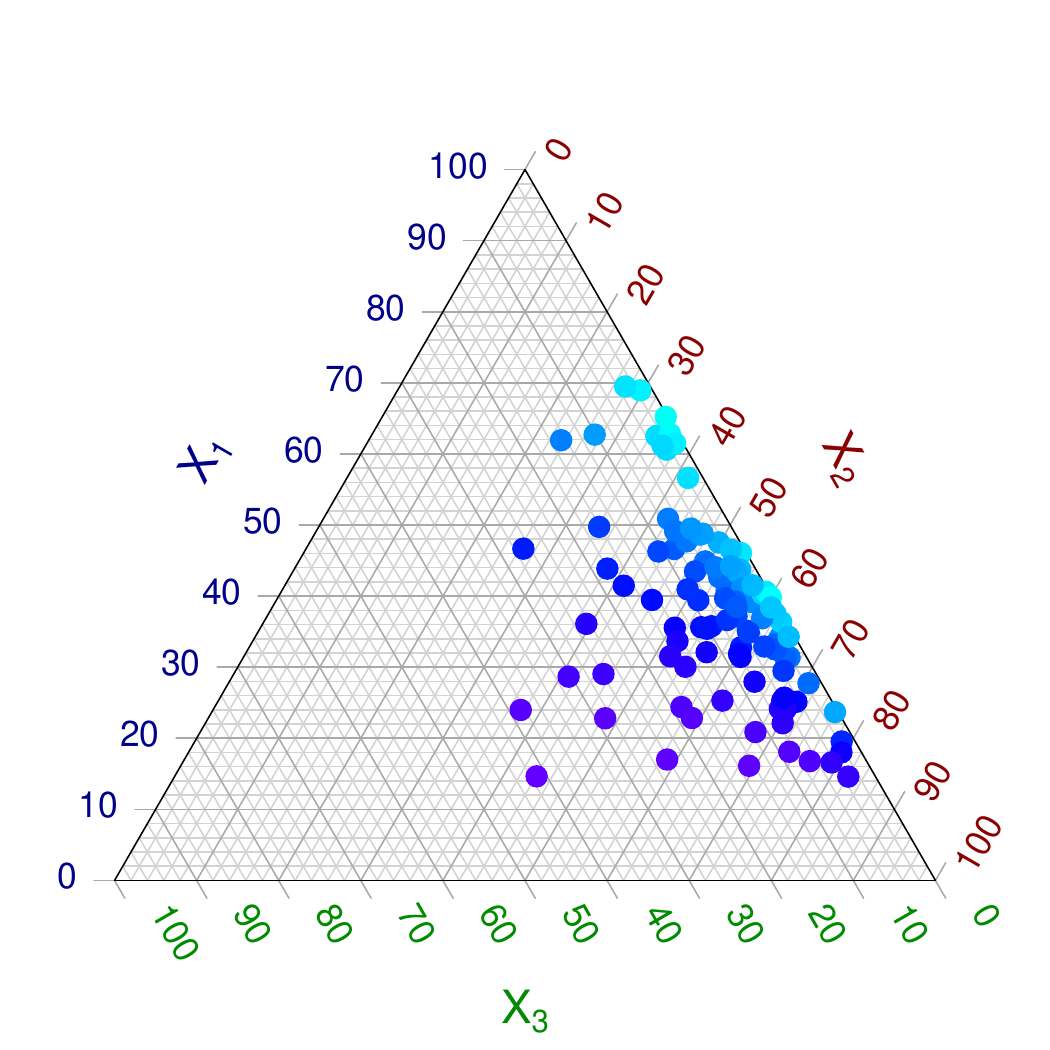} &
	 		\includegraphics[scale=0.31]{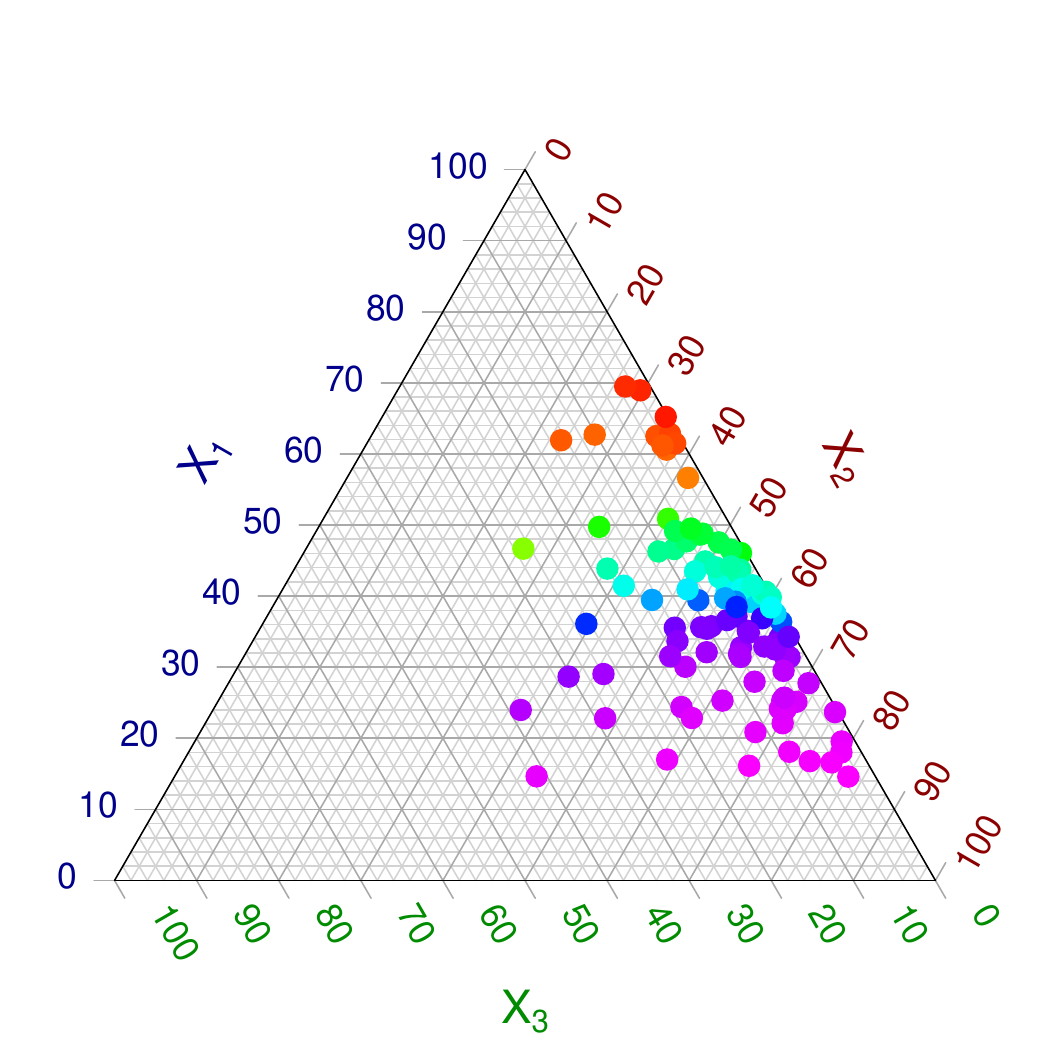}  \\
			$m$ & ROB$_0$ & ROB$_1$\\[-3ex]
	 		 \includegraphics[scale=0.31]{mhat-real-a5-7-1_n100_cont_C1_delta_10_shift_5-bis.pdf}  &
	 		\includegraphics[scale=0.31]{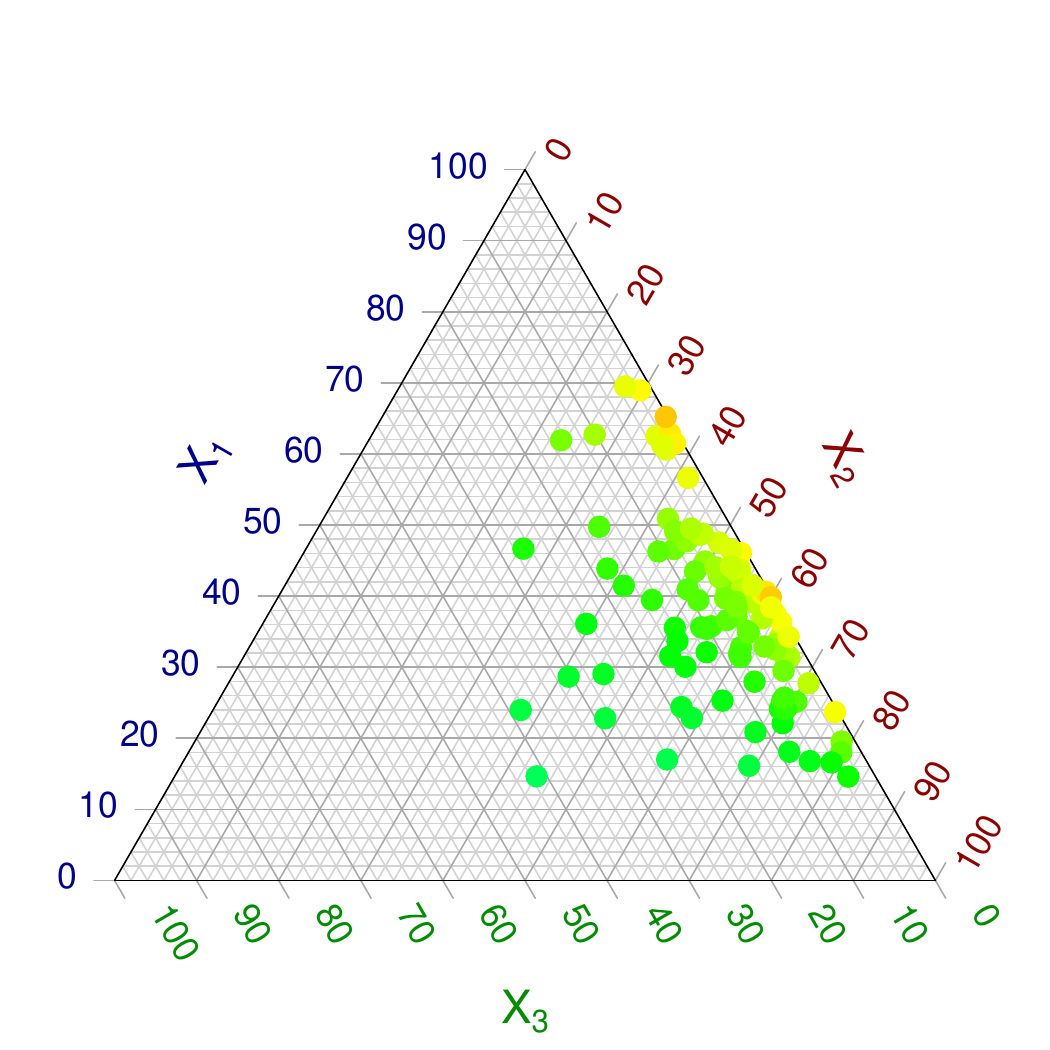} &
	 		\includegraphics[scale=0.31]{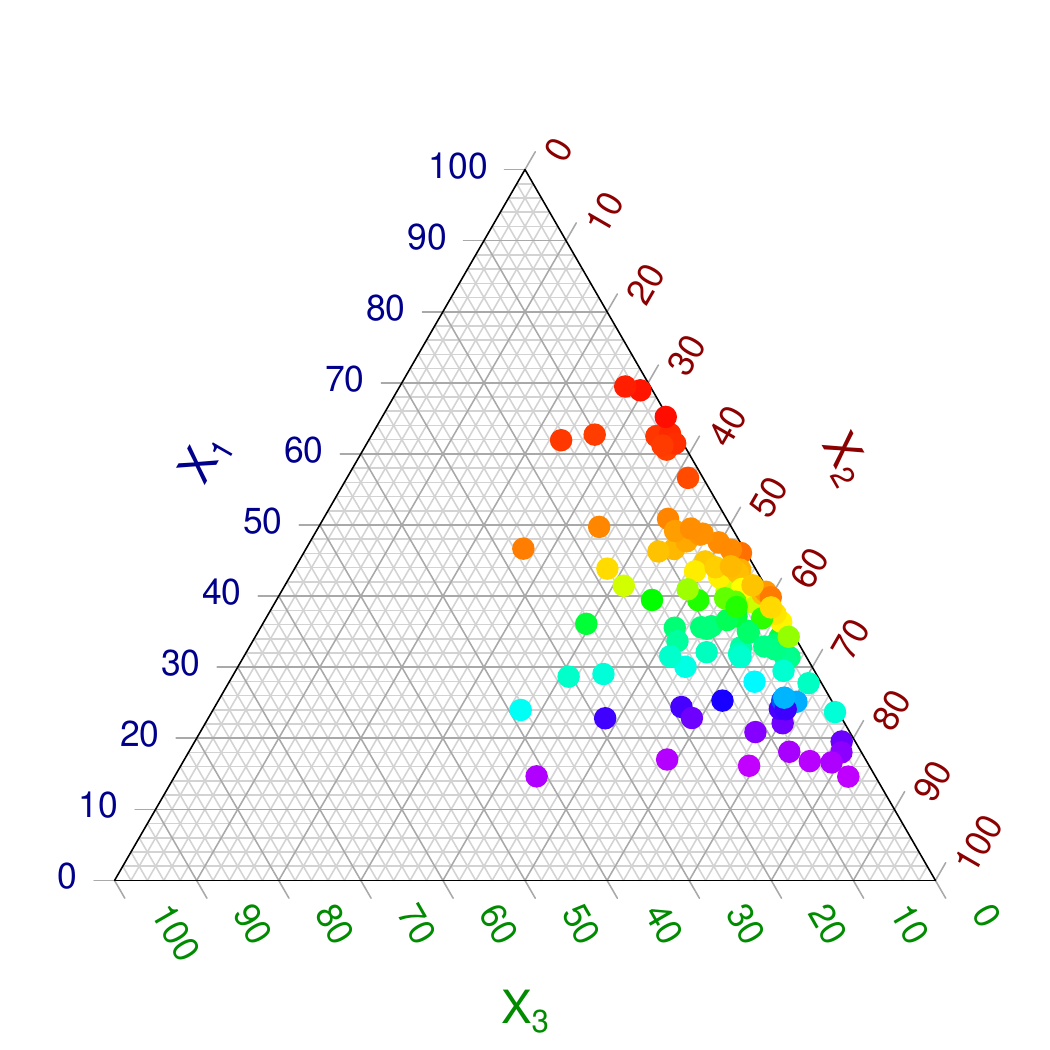}   
		\end{tabular}
		\vskip-0.1in \caption{ \small \label{fig:mhat.C1.5}  Ternary diagrams with the prediction points $\{\bx_{s}^{(0)}\}_{s=1}^M$ in different colours, when $\balfa=(5,7,1)\trasp$. The colour palette indicates the value of $\wm(\bx_{s}^{(0)})$ under $C_{1,0.10,5}$ and $m(\bx_{s}^{(0)})$.}
	\end{center} 
\end{figure}

\begin{figure}[ht!]
	\begin{center}
		\renewcommand{\arraystretch}{0.1}
		\newcolumntype{G}{>{\centering\arraybackslash}m{\dimexpr.35\linewidth-1\tabcolsep}}
				\begin{tabular}{GGG}
			$m$ & CL$_0$ & CL$_1$ \\[-3ex]
			 \includegraphics[scale=0.31]{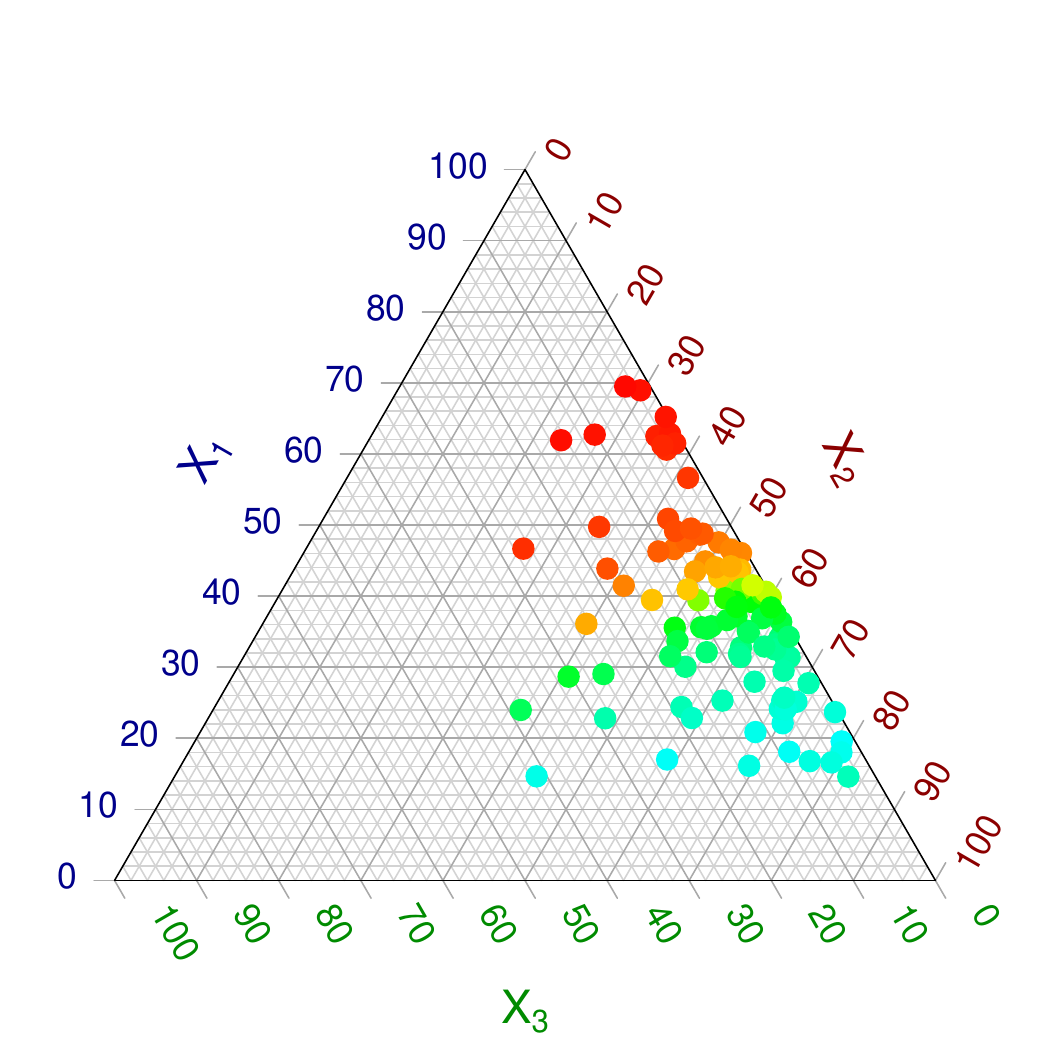}  &
	 		\includegraphics[scale=0.31]{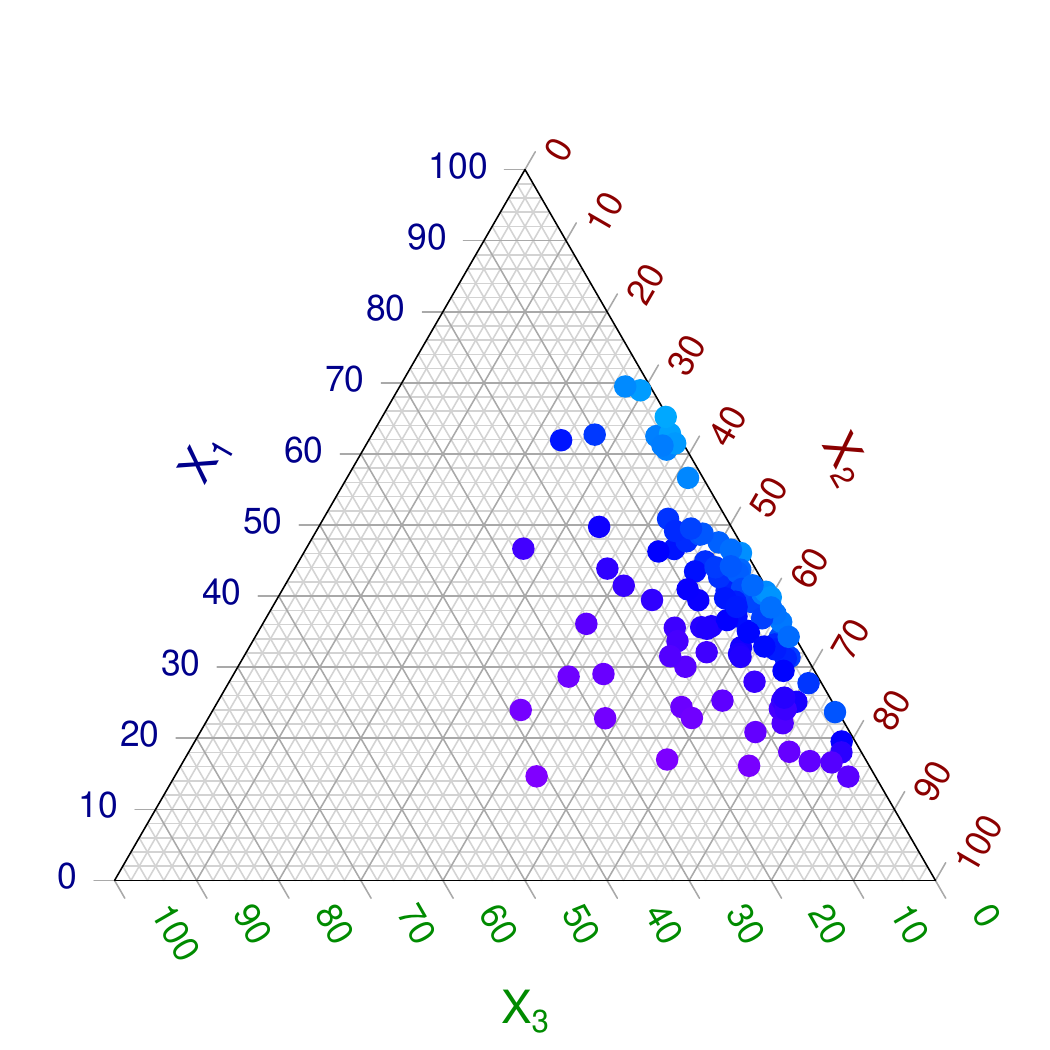} &
	 		\includegraphics[scale=0.31]{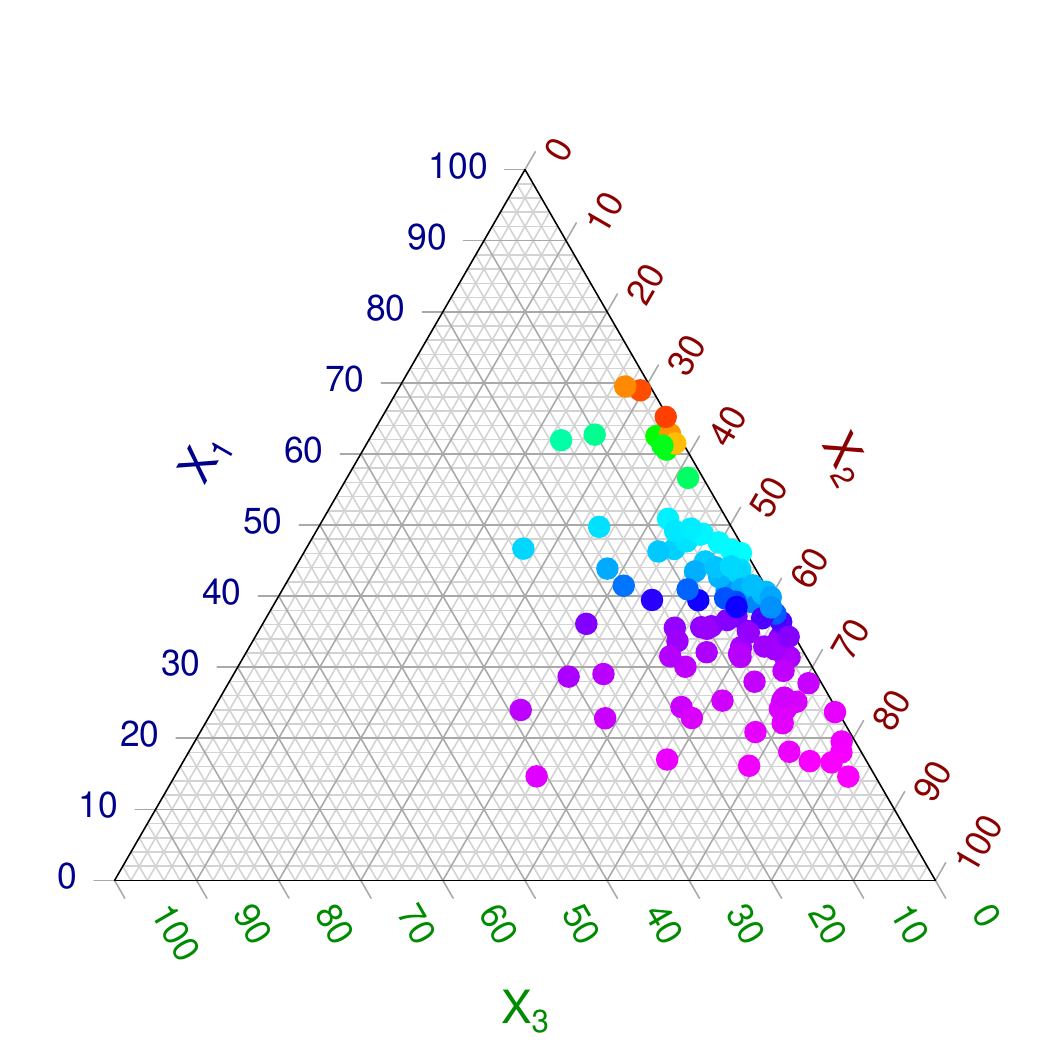}  \\
			$m$ & ROB$_0$ & ROB$_1$\\[-3ex]
	 		 \includegraphics[scale=0.31]{mhat-real-a5-7-1_n100_cont_C1_delta_10_shift_10-bis.pdf}  &
	 		\includegraphics[scale=0.31]{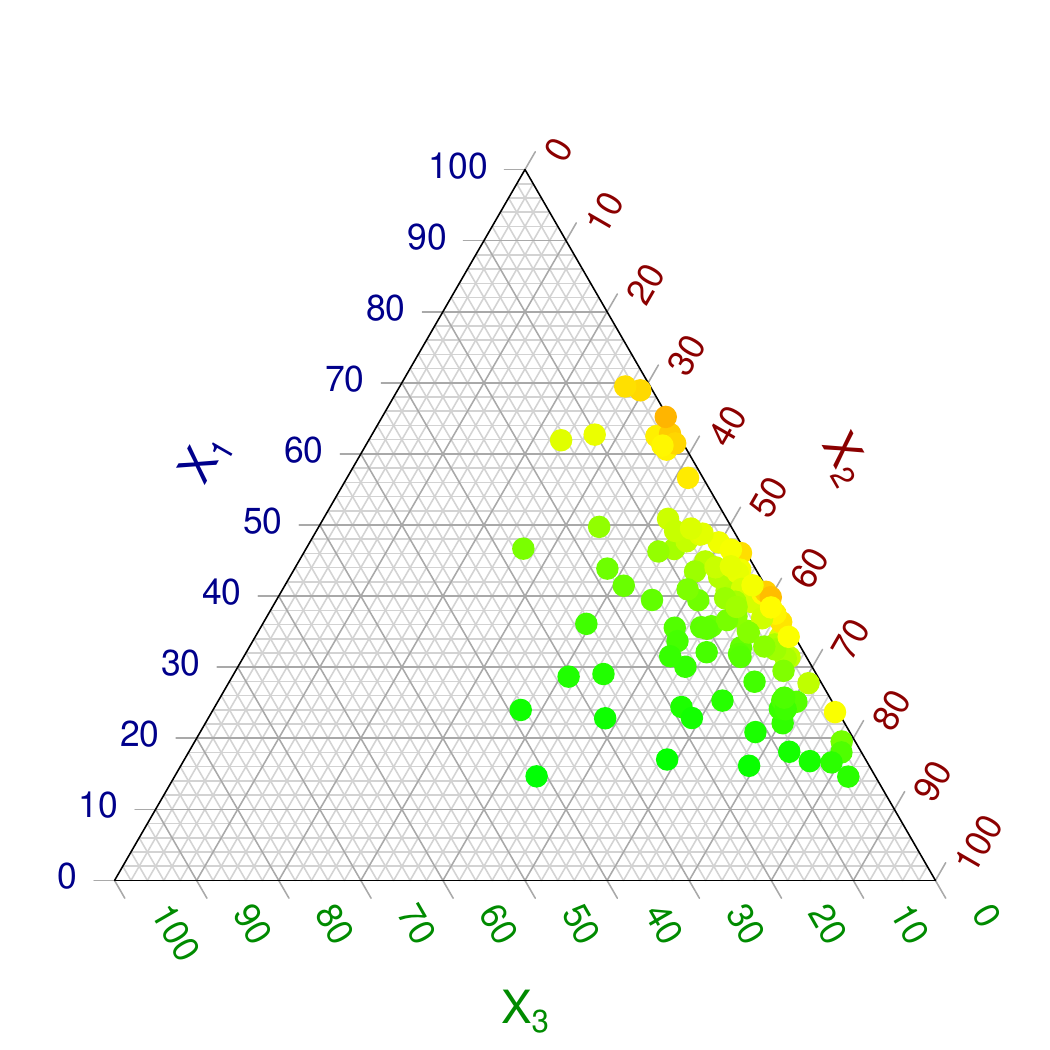} &
	 		\includegraphics[scale=0.31]{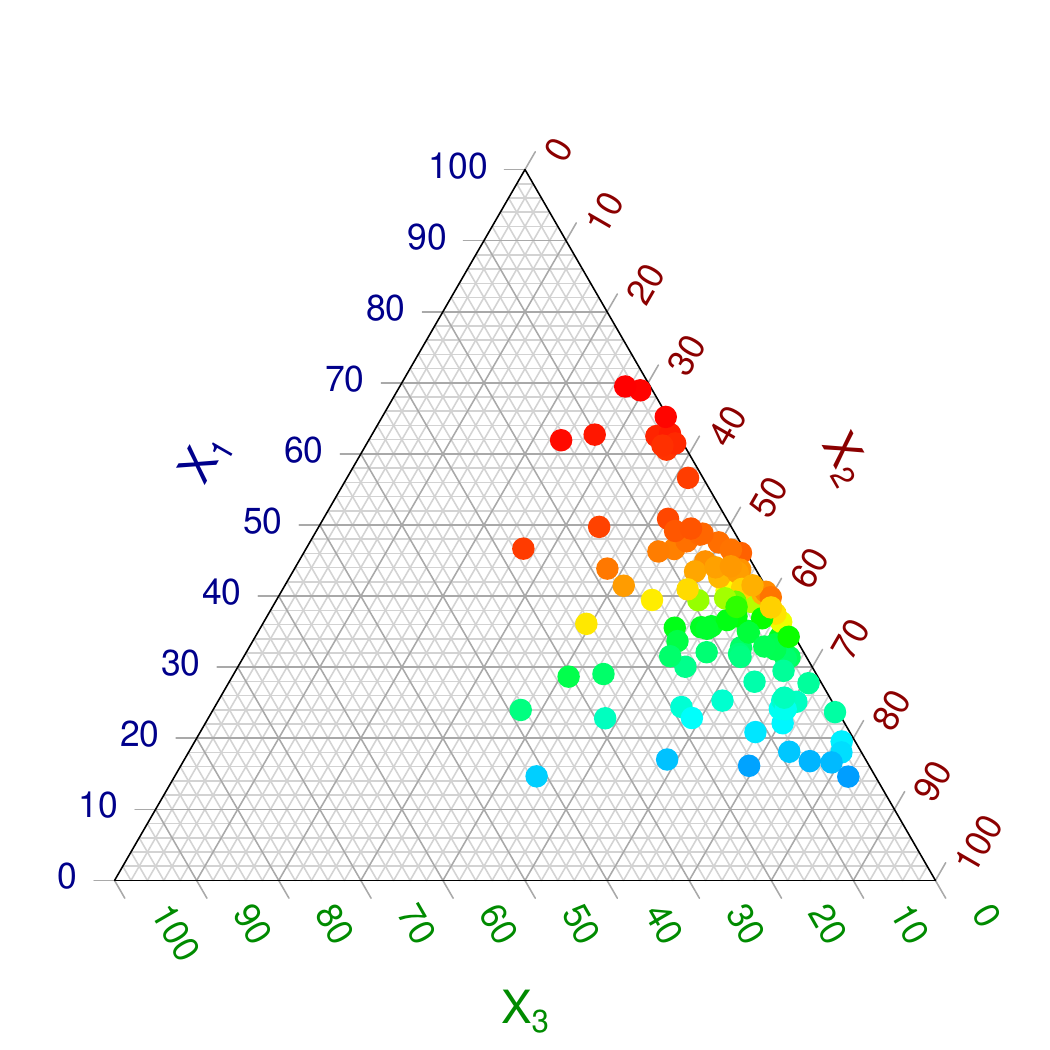}   
		\end{tabular}
		\vskip-0.1in \caption{ \small \label{fig:mhat.C1.10}  Ternary diagrams with the prediction points $\{\bx_{s}^{(0)}\}_{s=1}^M$ in different colours, when $\balfa=(5,7,1)\trasp$. The colour palette indicates the value of $\wm(\bx_{s}^{(0)})$ under $C_{1,0.10,10}$ and $m(\bx_{s}^{(0)})$.}
	\end{center} 
\end{figure}

\begin{figure}[ht!]
	\begin{center}
		\renewcommand{\arraystretch}{0.1}
		\newcolumntype{G}{>{\centering\arraybackslash}m{\dimexpr.35\linewidth-1\tabcolsep}}
				\begin{tabular}{GGG}
			$m$ & CL$_0$ & CL$_1$ \\[-3ex]
			 \includegraphics[scale=0.31]{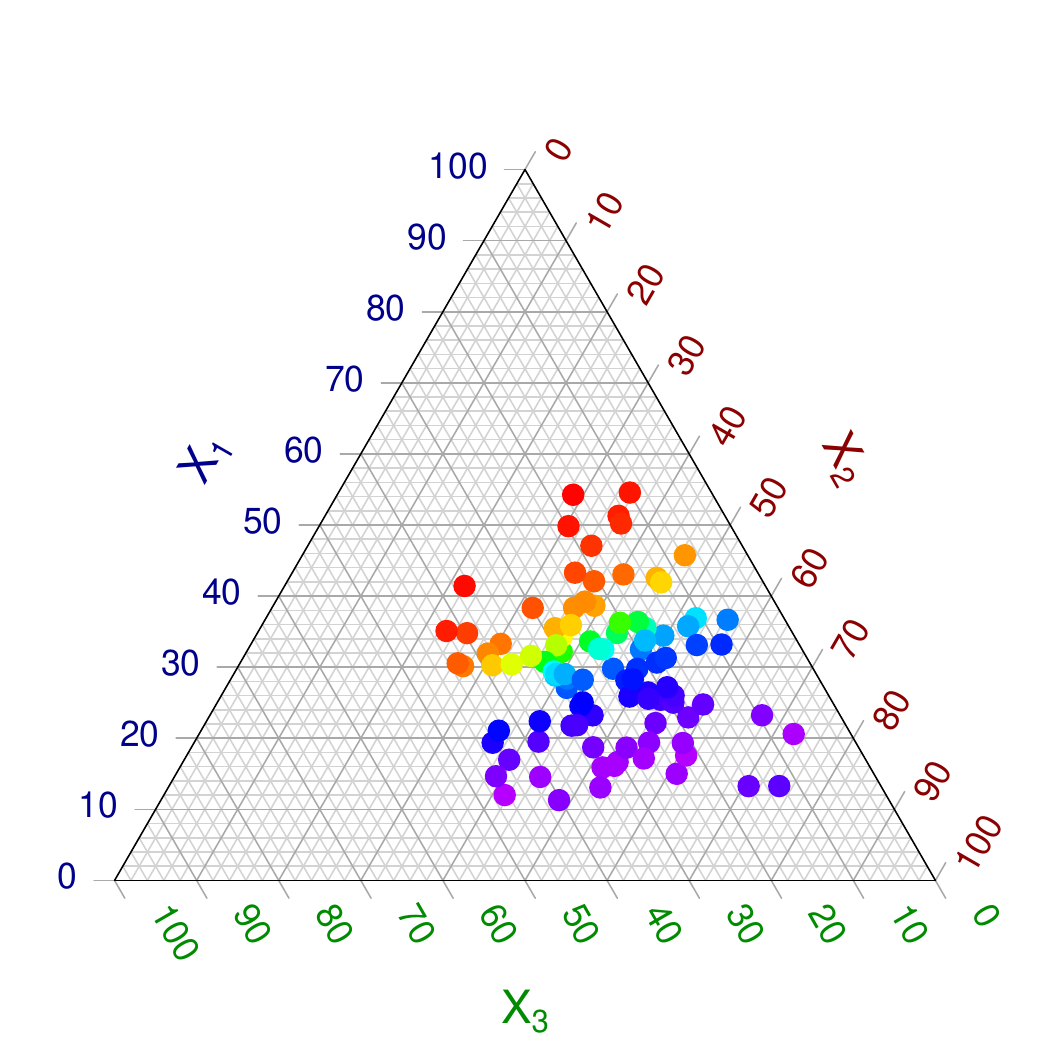}  &
	 		\includegraphics[scale=0.31]{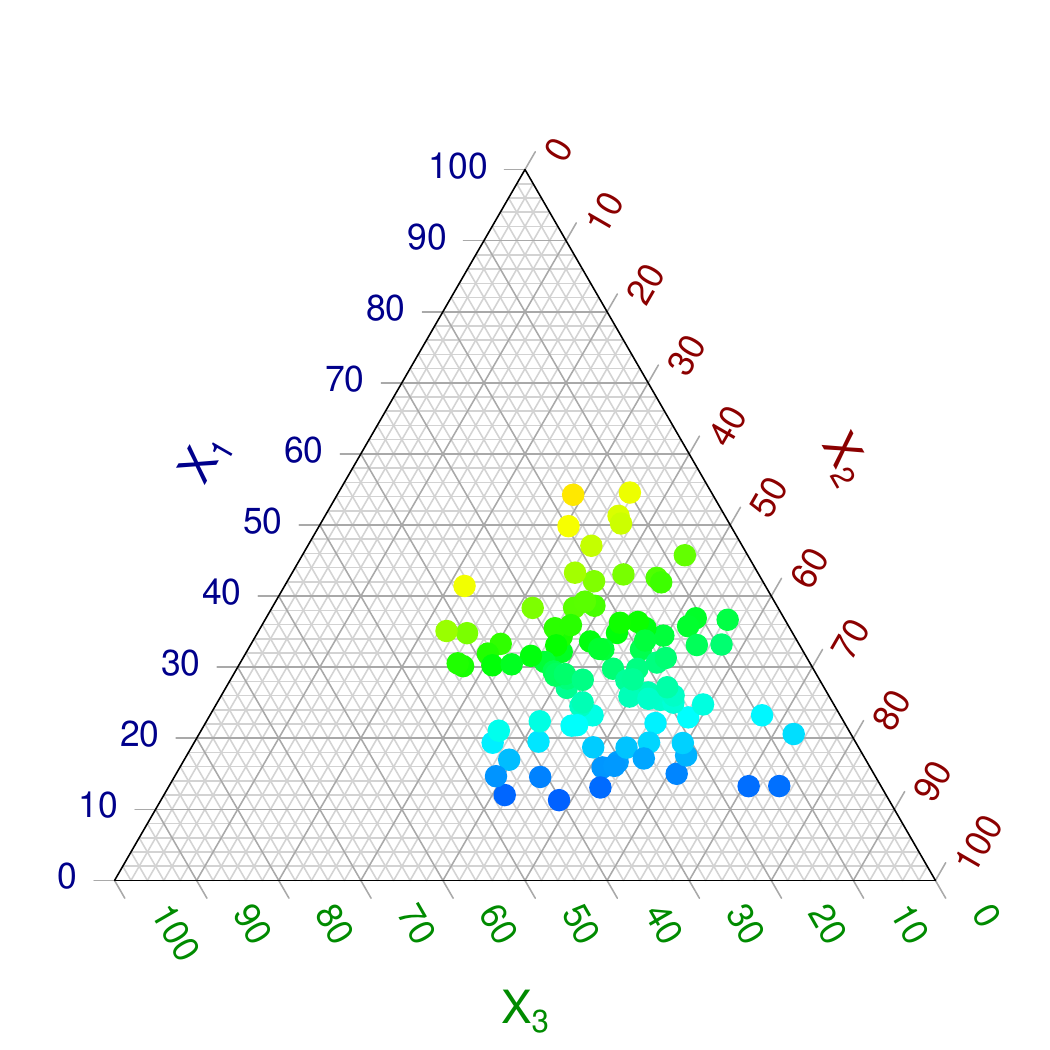} &
	 		\includegraphics[scale=0.31]{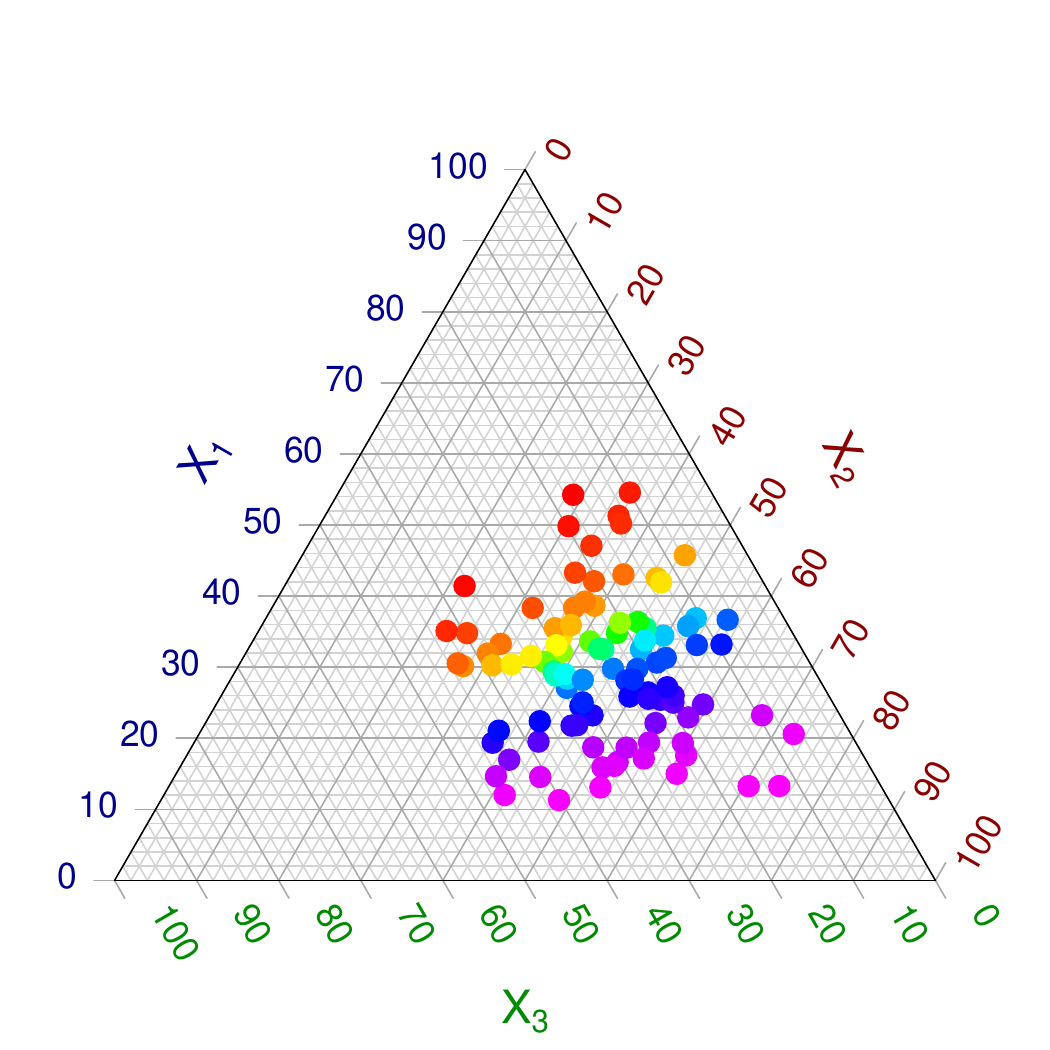}  \\
			$m$ & ROB$_0$ & ROB$_1$\\[-3ex]
	 		 \includegraphics[scale=0.31]{mhat-real-a5-7-4_n100_cont_C0-bis.pdf}  &
	 		\includegraphics[scale=0.31]{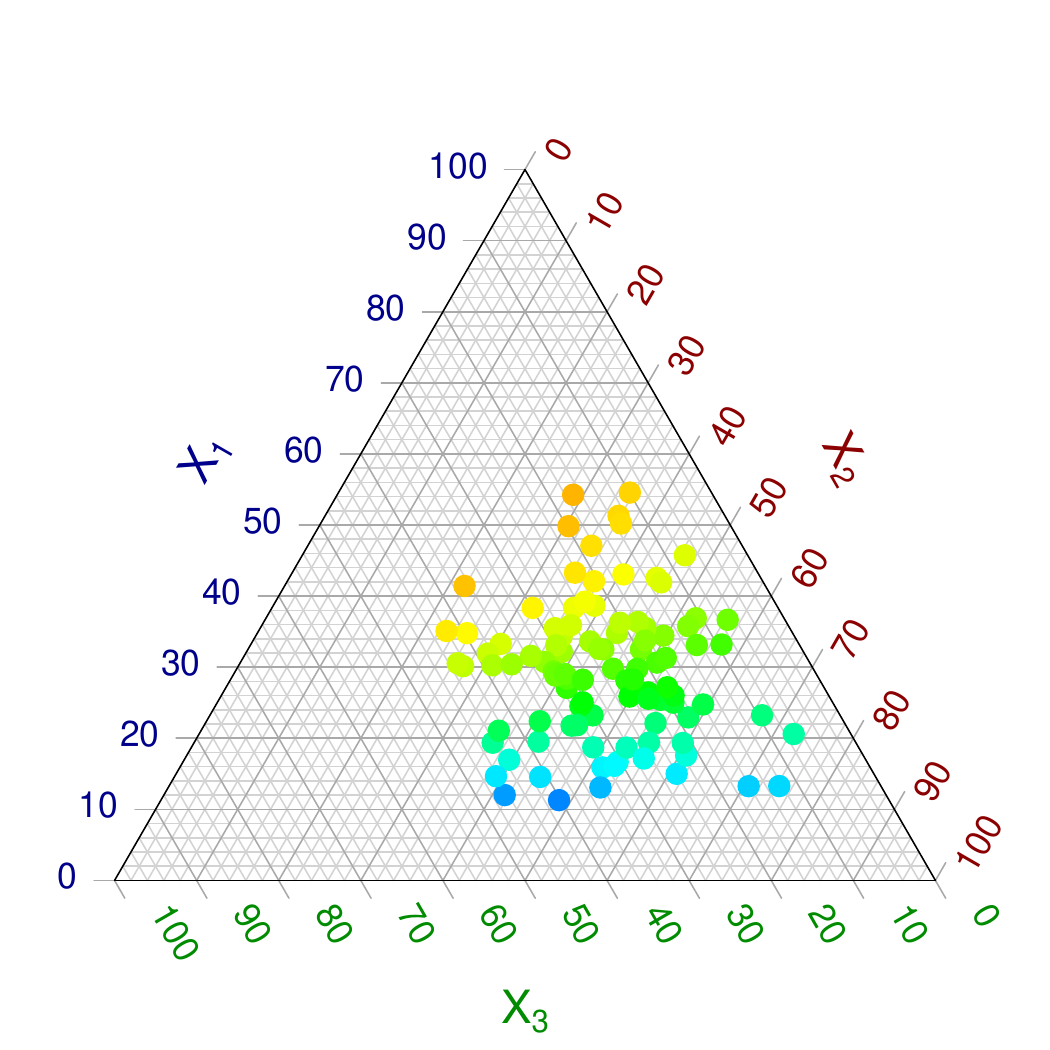} &
	 		\includegraphics[scale=0.31]{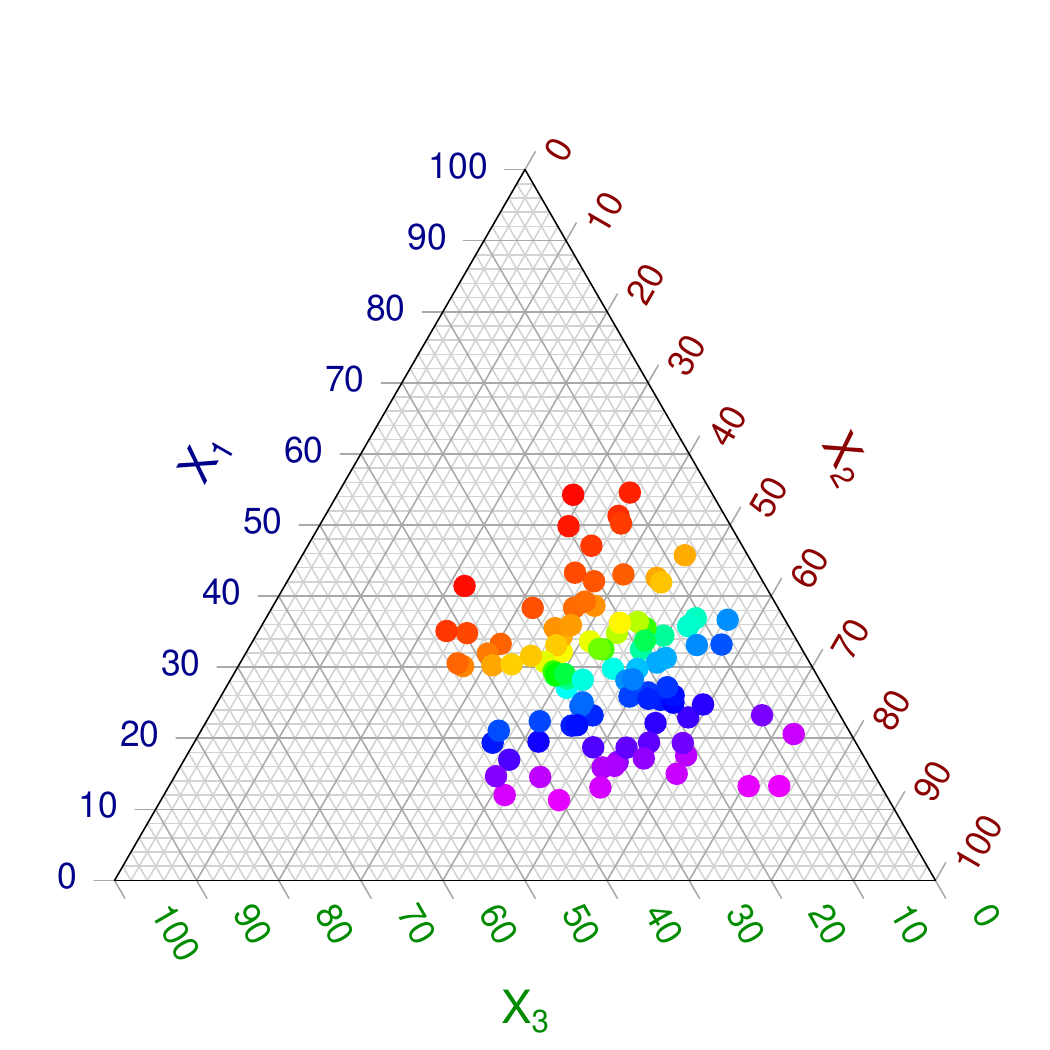}   
		\end{tabular}
		\vskip-0.1in \caption{ \small \label{fig:mhat.C0-5-7-4}  Ternary diagrams with the prediction points $\{\bx_{s}^{(0)}\}_{s=1}^M$ in different colours, when $\balfa=(5,7,4)\trasp$. The colour palette indicates the value of $\wm(\bx_{s}^{(0)})$ under $C_{0}$ and $m(\bx_{s}^{(0)})$.}
	\end{center} 
\end{figure}

\begin{figure}[ht!]
	\begin{center}
		\renewcommand{\arraystretch}{0.1}
		\newcolumntype{G}{>{\centering\arraybackslash}m{\dimexpr.35\linewidth-1\tabcolsep}}
				\begin{tabular}{GGG}
			$m$ & CL$_0$ & CL$_1$ \\[-3ex]
			 \includegraphics[scale=0.31]{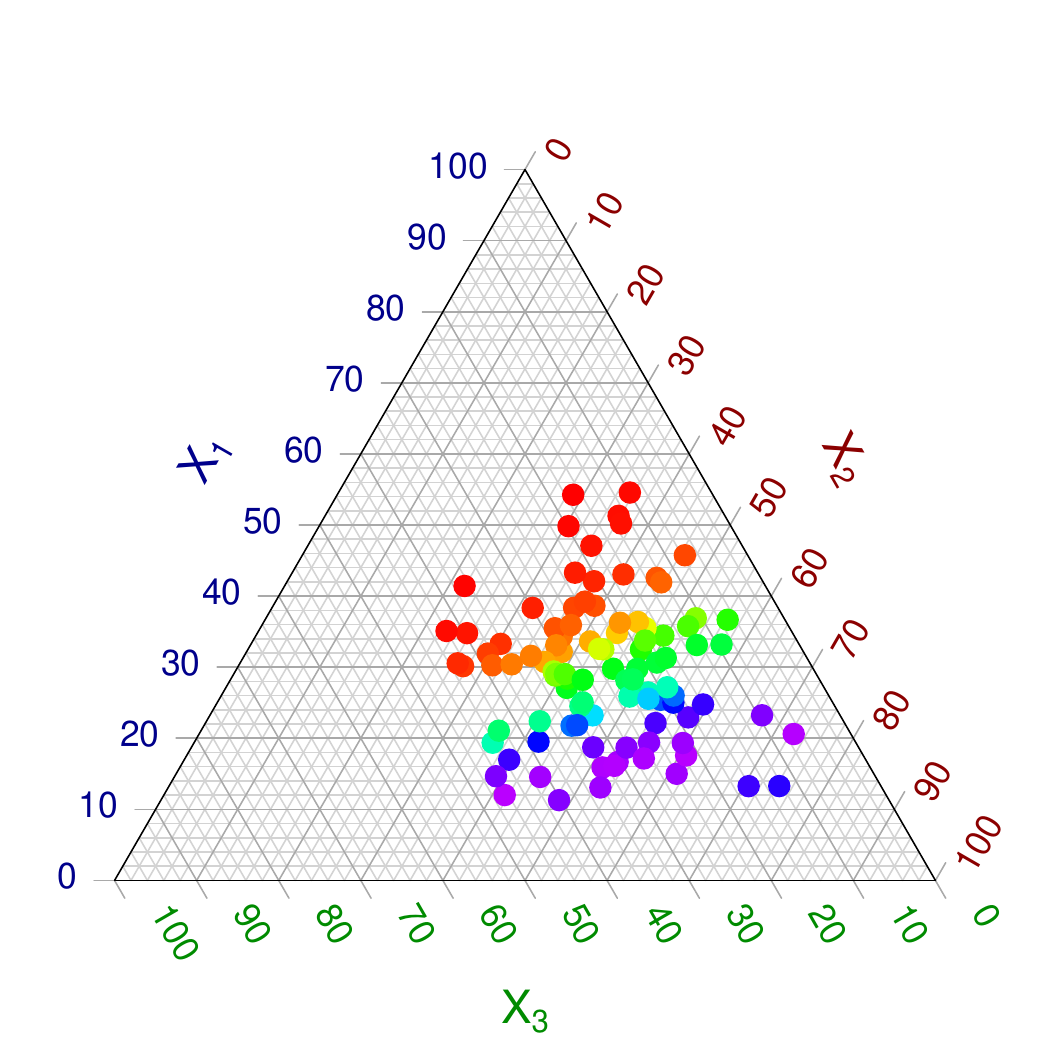}  &
	 		\includegraphics[scale=0.31]{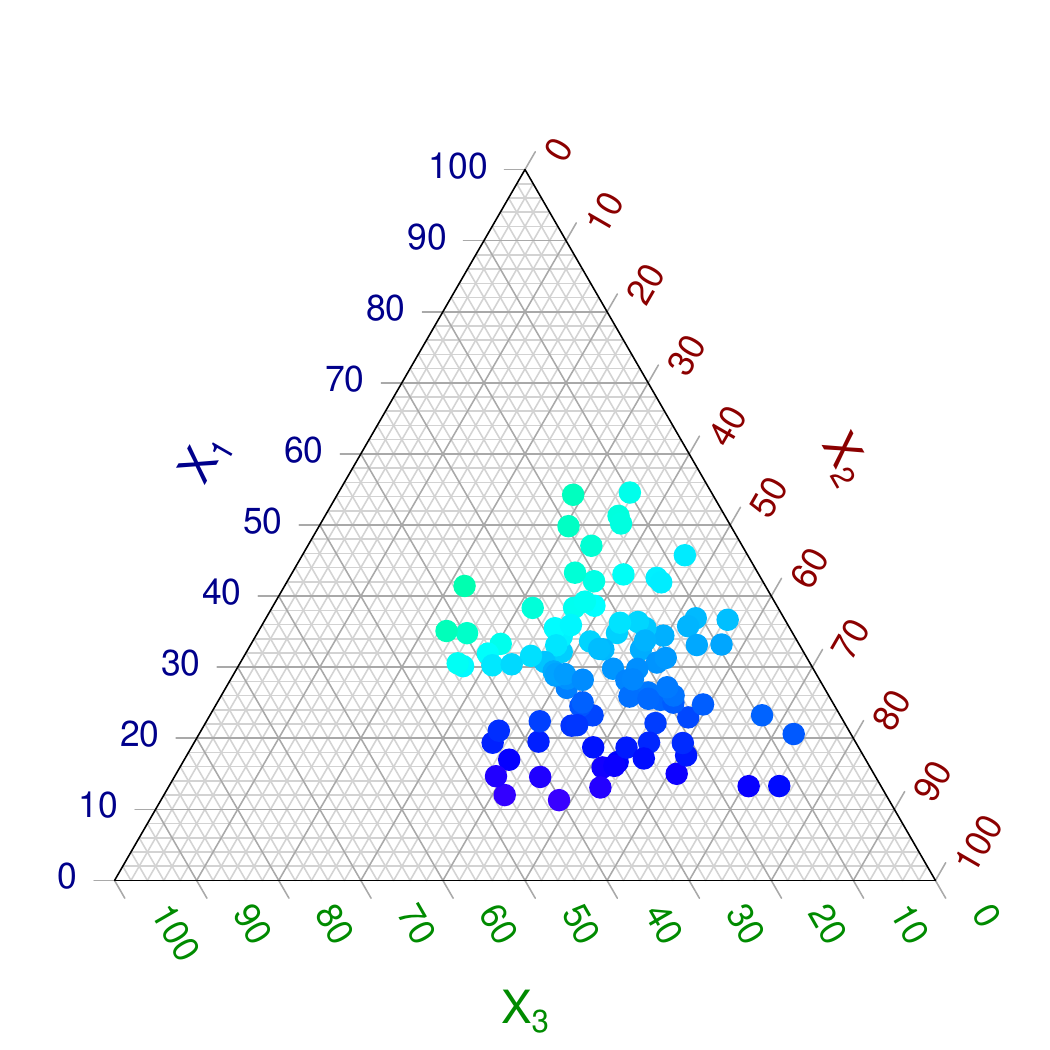} &
	 		\includegraphics[scale=0.31]{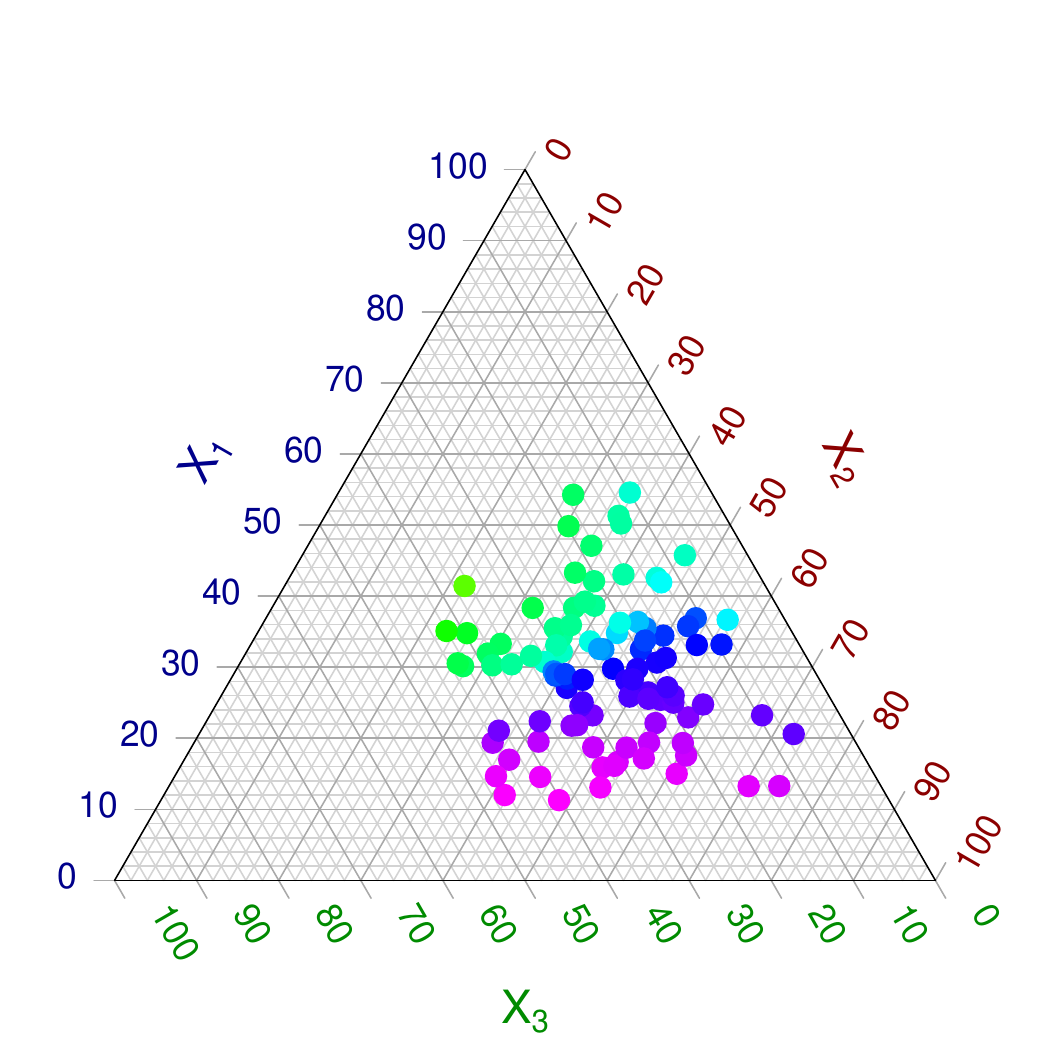}  \\
			$m$ & ROB$_0$ & ROB$_1$\\[-3ex]
	 		 \includegraphics[scale=0.31]{mhat-real-a5-7-4_n100_cont_C1_delta_10_shift_5-bis.pdf}  &
	 		\includegraphics[scale=0.31]{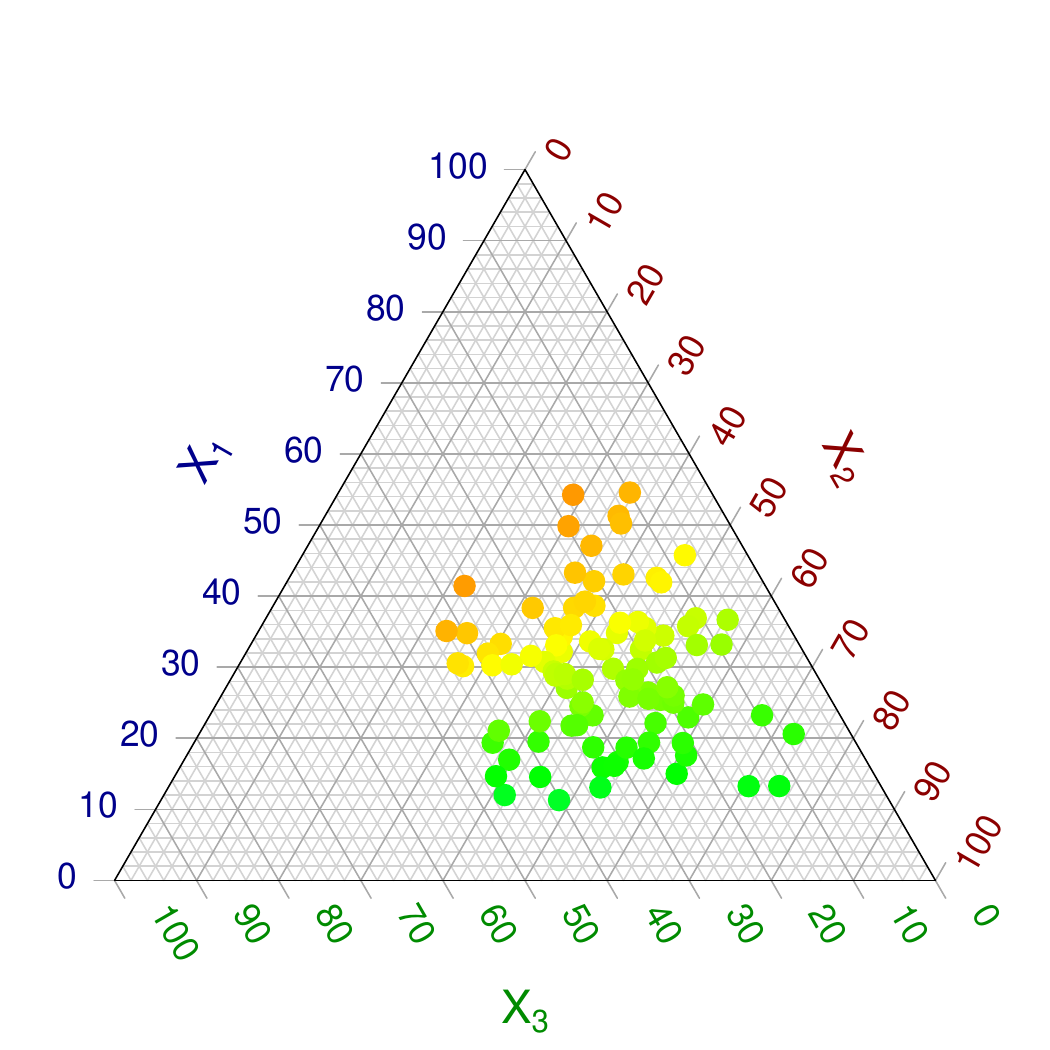} &
	 		\includegraphics[scale=0.31]{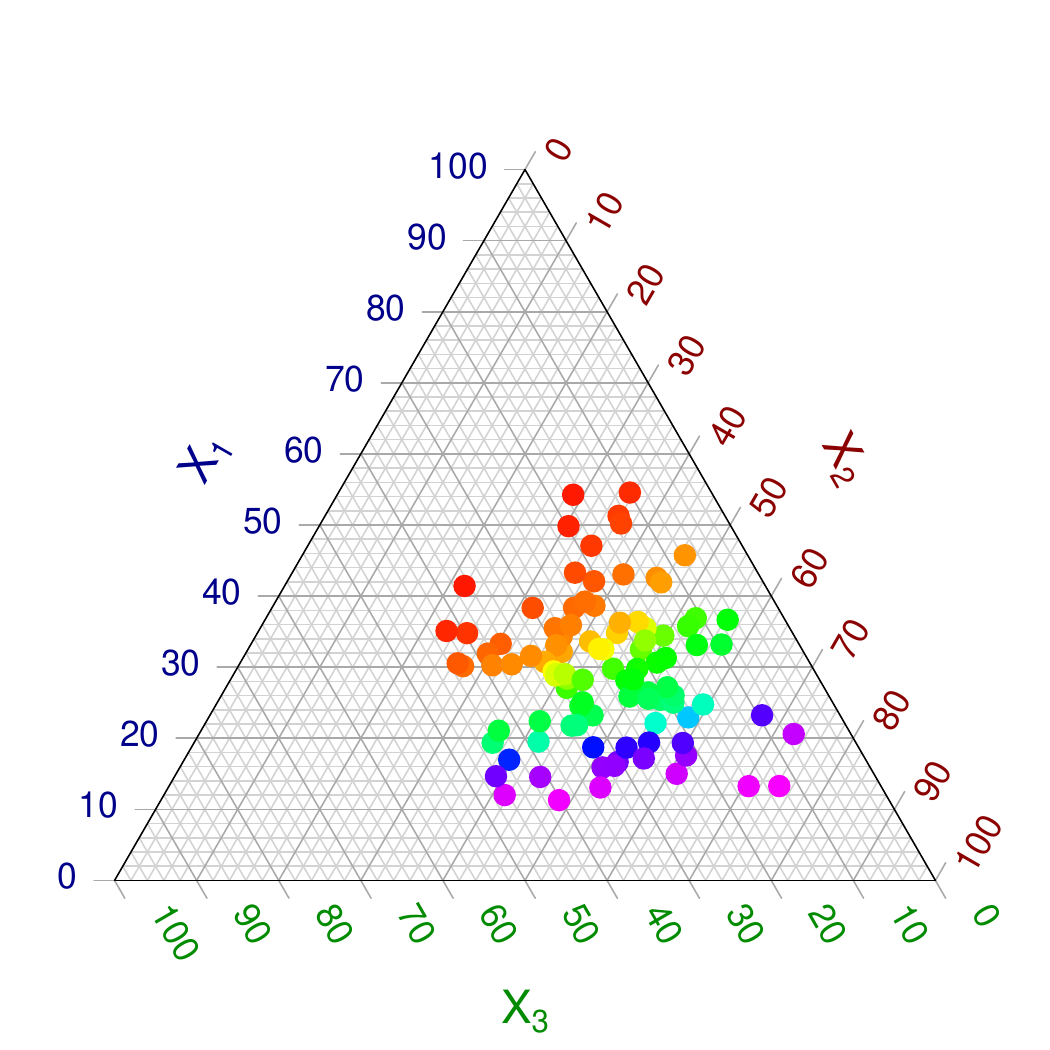}   
		\end{tabular}
		\vskip-0.1in \caption{ \small \label{fig:mhat.C1.5-5-7-4}  Ternary diagrams with the prediction points $\{\bx_{s}^{(0)}\}_{s=1}^M$ in different colours, when $\balfa=(5,7,4)\trasp$. The colour palette indicates the value of $\wm(\bx_{s}^{(0)})$ under $C_{1,0.10,5}$ and $m(\bx_{s}^{(0)})$.}
	\end{center} 
\end{figure}

\begin{figure}[ht!]
	\begin{center}
		\renewcommand{\arraystretch}{0.1}
		\newcolumntype{G}{>{\centering\arraybackslash}m{\dimexpr.35\linewidth-1\tabcolsep}}
				\begin{tabular}{GGG}
			$m$ & CL$_0$ & CL$_1$ \\[-3ex]
			 \includegraphics[scale=0.31]{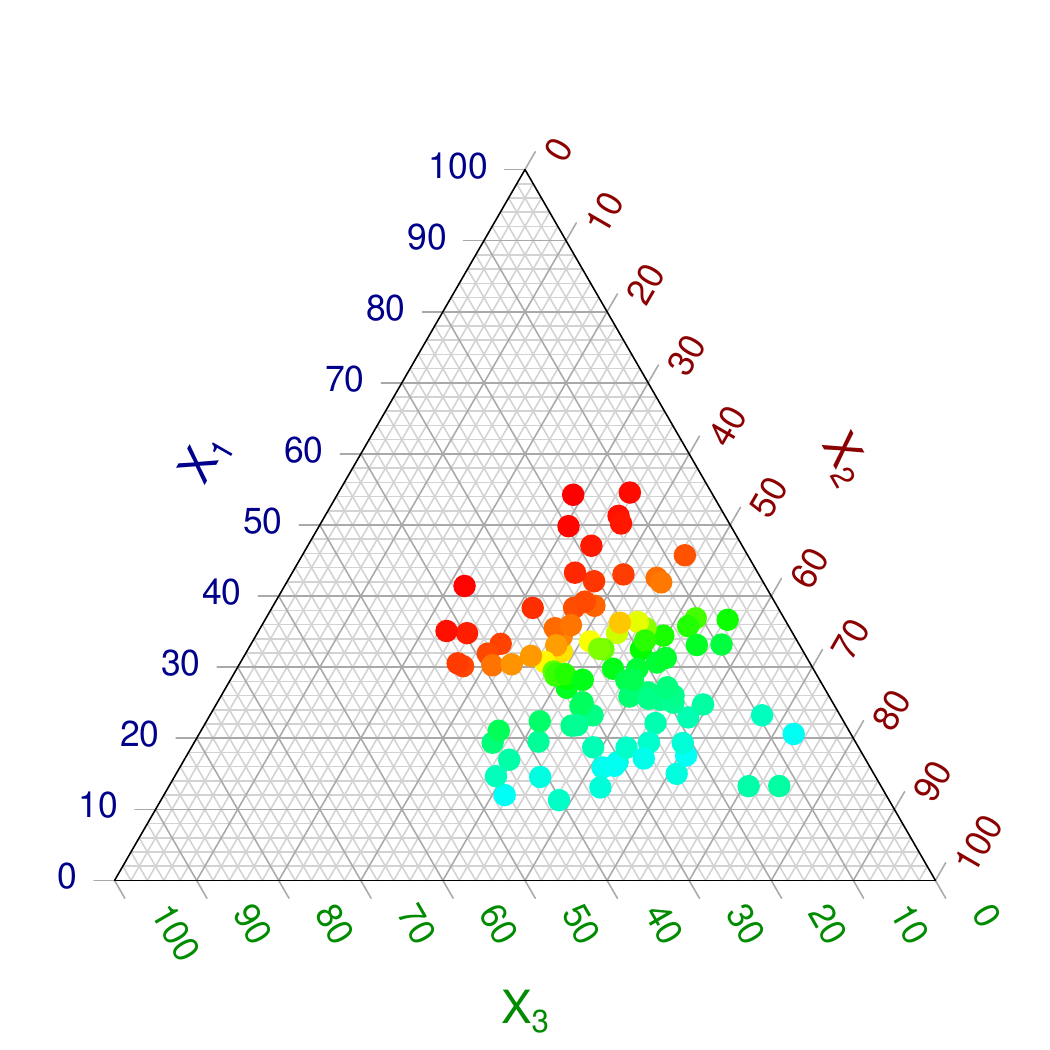}  &
	 		\includegraphics[scale=0.31]{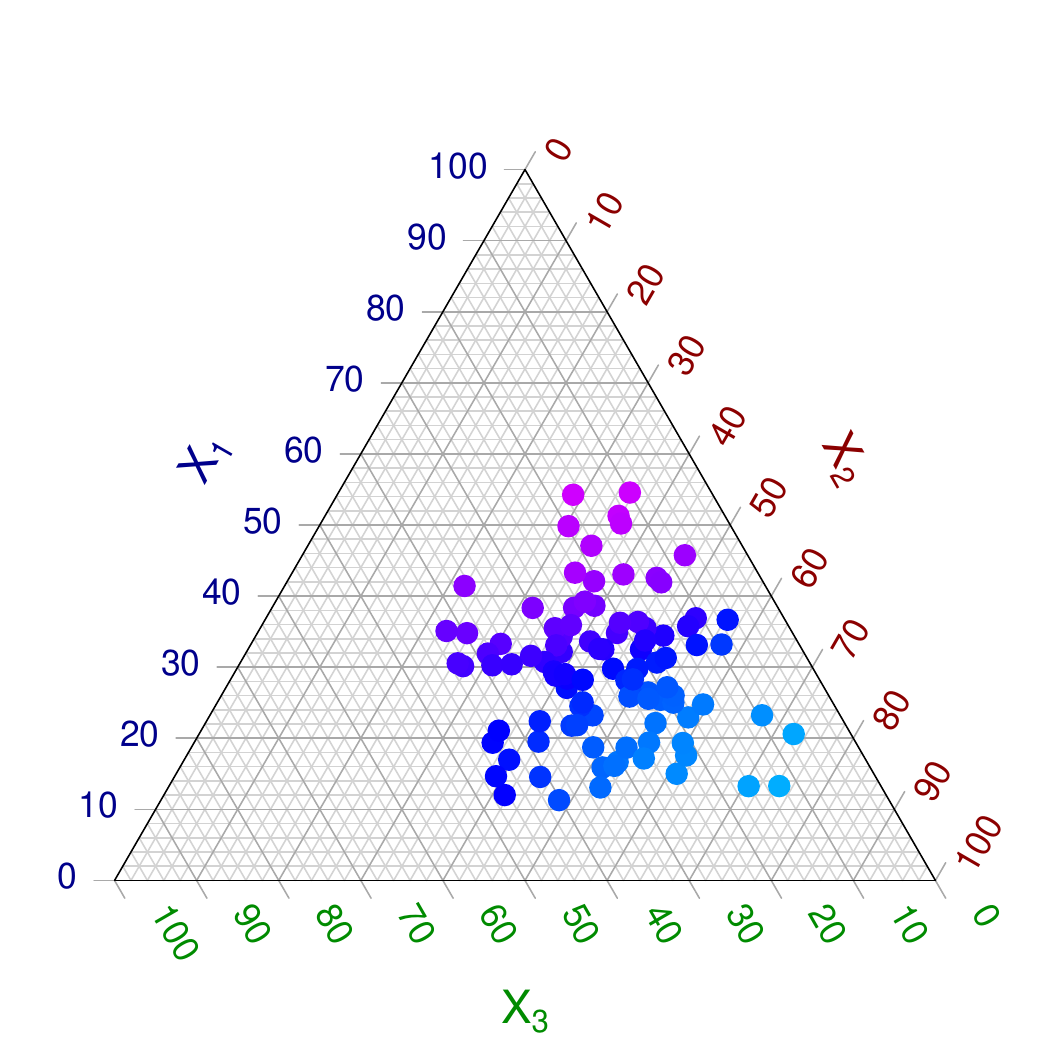} &
	 		\includegraphics[scale=0.31]{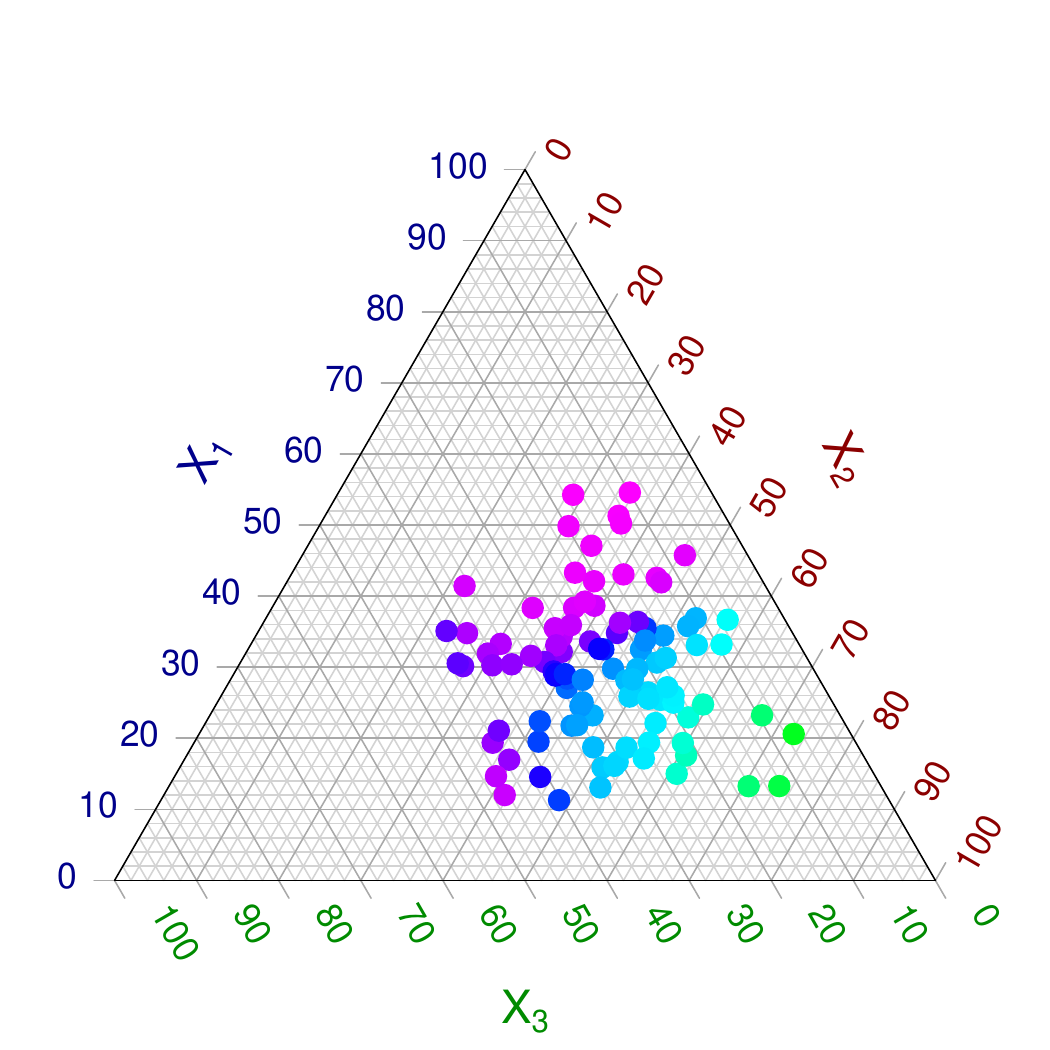}  \\
			$m$ & ROB$_0$ & ROB$_1$\\[-3ex]
	 		 \includegraphics[scale=0.31]{mhat-real-a5-7-4_n100_cont_C1_delta_10_shift_10-bis.pdf}  &
	 		\includegraphics[scale=0.31]{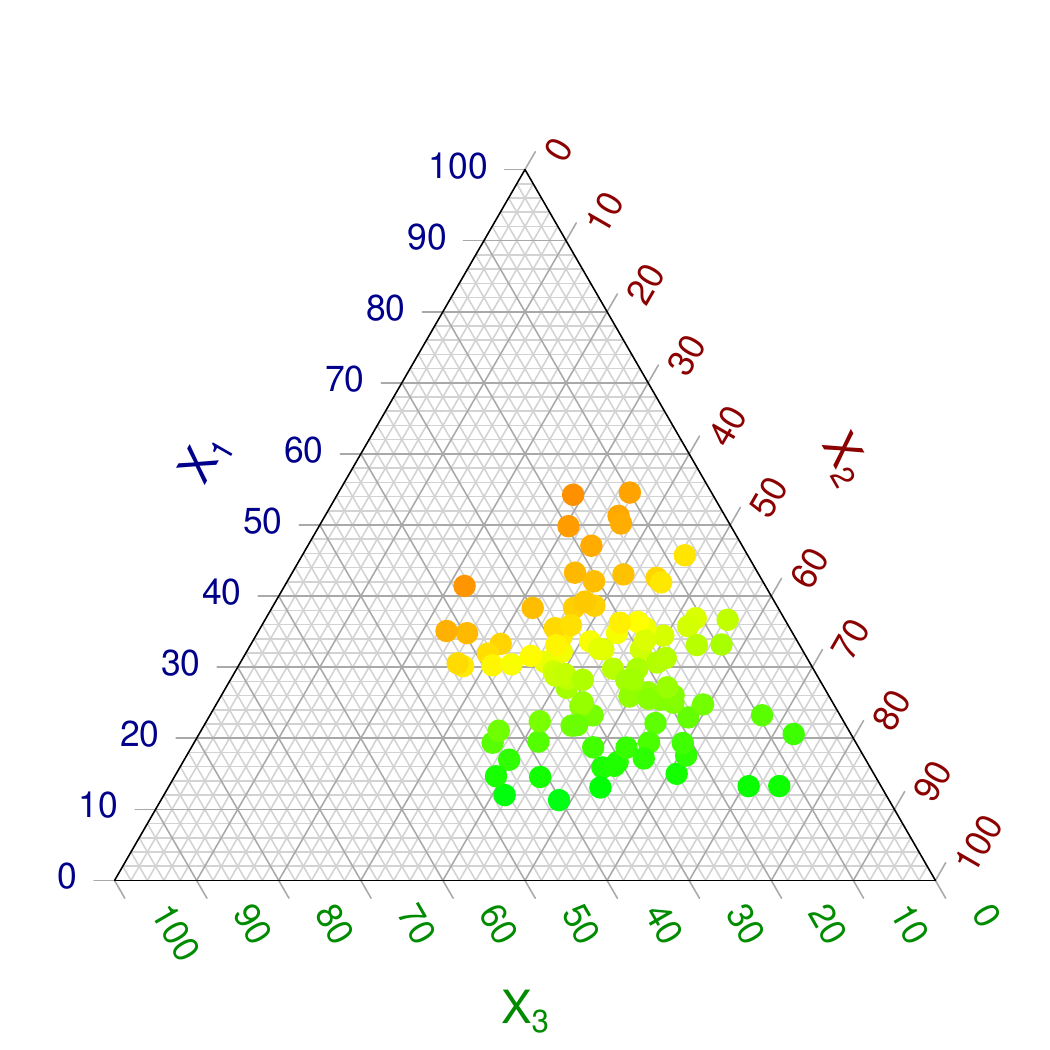} &
	 		\includegraphics[scale=0.31]{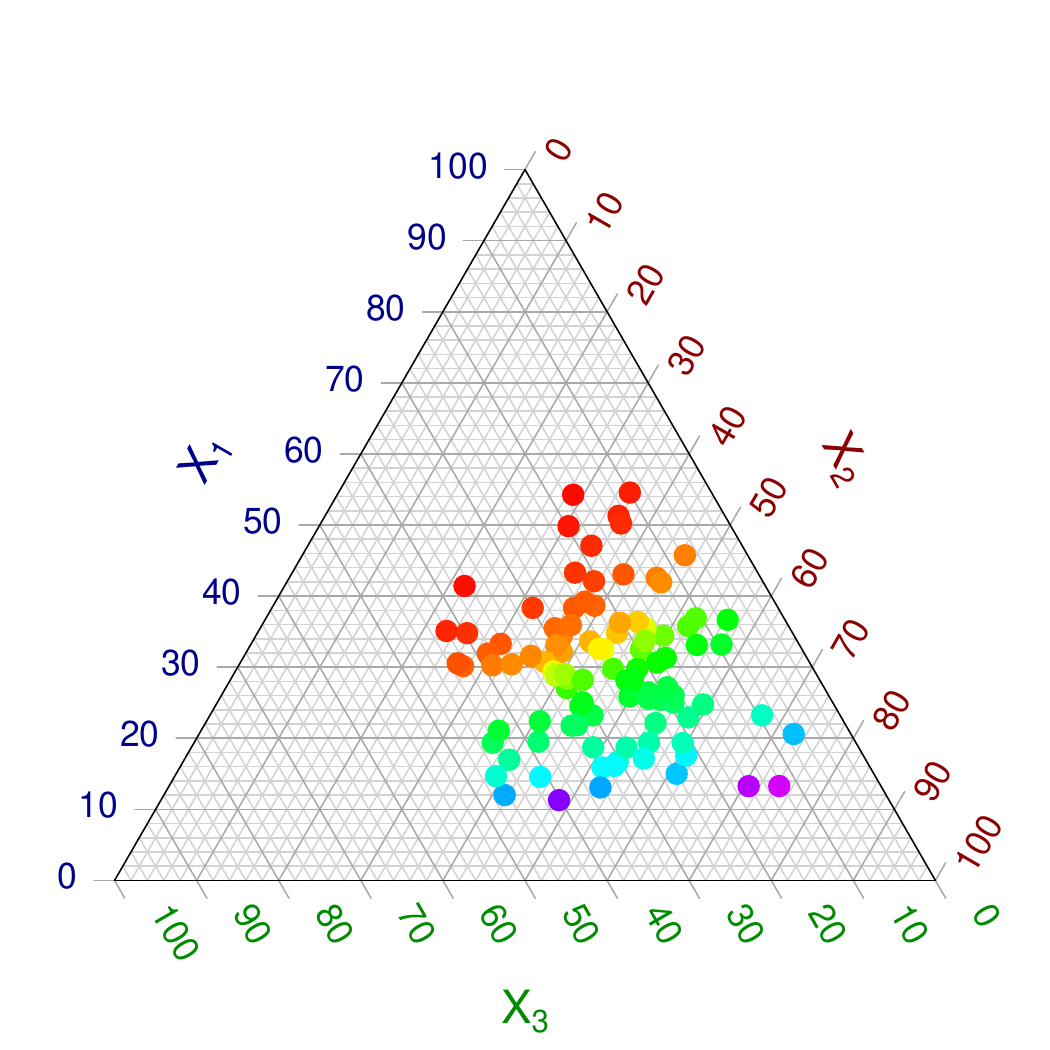}   
		\end{tabular}
		\vskip-0.1in \caption{ \small \label{fig:mhat.C1.10-5-7-4}  Ternary diagrams with the prediction points $\{\bx_{s}^{(0)}\}_{s=1}^M$ in different colours, when $\balfa=(5,7,4)\trasp$. The colour palette indicates the value of $\wm(\bx_{s}^{(0)})$ under $C_{1,0.10,10}$ and $m(\bx_{s}^{(0)})$.}
	\end{center} 
\end{figure}

To illustrate the effect of the considered contaminations on the estimators, Figures \ref{fig:mhat.C0} to \ref{fig:mhat.C1.10} depict the ternary diagram with the prediction points $\{\bx_{s}^{(0)}\}_{s=1}^M$ in different colours to show an intensity plot when $\balfa=(5,7,1)\trasp$, while Figures \ref{fig:mhat.C0-5-7-4} to \ref{fig:mhat.C1.10-5-7-4} display similar plots when $\balfa=(5,7,4)\trasp$. The colour palette indicates the value of $\wm(\bx_{s}^{(0)})$ at one replication. To facilitate comparisons we present in the left panel the ternary plot where the colours identify the real function at the grid points $m(\bx_{s}^{(0)})$. The warm colours (red and orange) correspond to small negative values, while as the colour spectrum becomes cooler (green, blue, violet) the values of the estimators or of the true function increase to the maximum of the four procedures and 1, which is the maximum of $m$.  The colours vary across figures due to the range of the obtained estimators, for that reason the true function is depicted in each Figure with different colours.

As expected, for clean samples the classical and robust estimators show a similar performance, since the plots in the upper and lower rows have a similar aspect. The advantage of the local linear over the local constant is evident when comparing the middle and right panels. In effect, the right and left panels of Figure \ref{fig:mhat.C0}  are quite similar, while the middle ones  lead to more biased estimators. Similar conclusions may be obtained when $\balfa=(5,7,4)\trasp$, see Figure \ref{fig:mhat.C0-5-7-4}.

Figures \ref{fig:mhat.C1.5} and \ref{fig:mhat.C1.10} and also Figures \ref{fig:mhat.C1.5-5-7-4} and \ref{fig:mhat.C1.10-5-7-4}   exhibit  the damaging effect of vertical outliers on the classical estimators in particular, when $\mu=10$. They also reflect  the stability of the robust local lineal procedure. 

\clearpage
\section{Real Data Analysis}\label{sec:realdata}

Clinical studies have investigated the impact of macronutrient composition on average glucose levels and glucose variability. The research highlights the importance of considering not just the quantity but also the quality of macronutrients (carbohydrates, lipids, and proteins) in managing and understanding glycemic control and cardiovascular disease risk, see  \citet{Wheeler:etal:2012} for a review on this topic.
Thus, we aimed to investigate the effect of diet composition on glucose average levels and glucose variability in subjects with and without diabetes.

The participants were a subset of those enrolled in the A Estrada Glycation and Inflammation Study (AEGIS), trial NCT01796184 at \url{www.clinicaltrials.gov}.   Our analysis   was performed over the $n=509$ non--diabetic  patients.  For six consecutive days, the participants were asked to record their diet at the time that their food and beverages were consumed. Energy intake, macronutrients (protein, carbohydrates and lipids) and micronutrients were calculated for each meal (breakfast, lunch and dinner); more details can be found in  \citet{Gude:etal:2016}.

 For each participant, the response $Y$ corresponds to a glycaemic index that was computed as  the area under the curve   of glycaemic excursions during the same period, see \citet{Service:2013} for more details.  
We performed the analysis with the following compositional covariates: proteins ($P$), lipids ($L$) and carbohydrates ($C$) percentages ingested in a week.   The assumed nonparametric model is an homoscesdastic one, i.e., $Y=m(\bX)+\sigma \epsilon$, where $\bX=(P,L,C)\trasp$ and the errors $\epsilon$ are assumed to be independent of the covariates and  have distribution symmetric around $0$.

\begin{figure}[ht!]
	\begin{center}
		\renewcommand{\arraystretch}{0.1}
		\newcolumntype{G}{>{\centering\arraybackslash}m{\dimexpr.34\linewidth-1\tabcolsep}}
		\hskip-0.6in
		\begin{tabular}{cc}
			Ternary Diagram &   ilr Plot \\ 
			\includegraphics[scale=0.3]{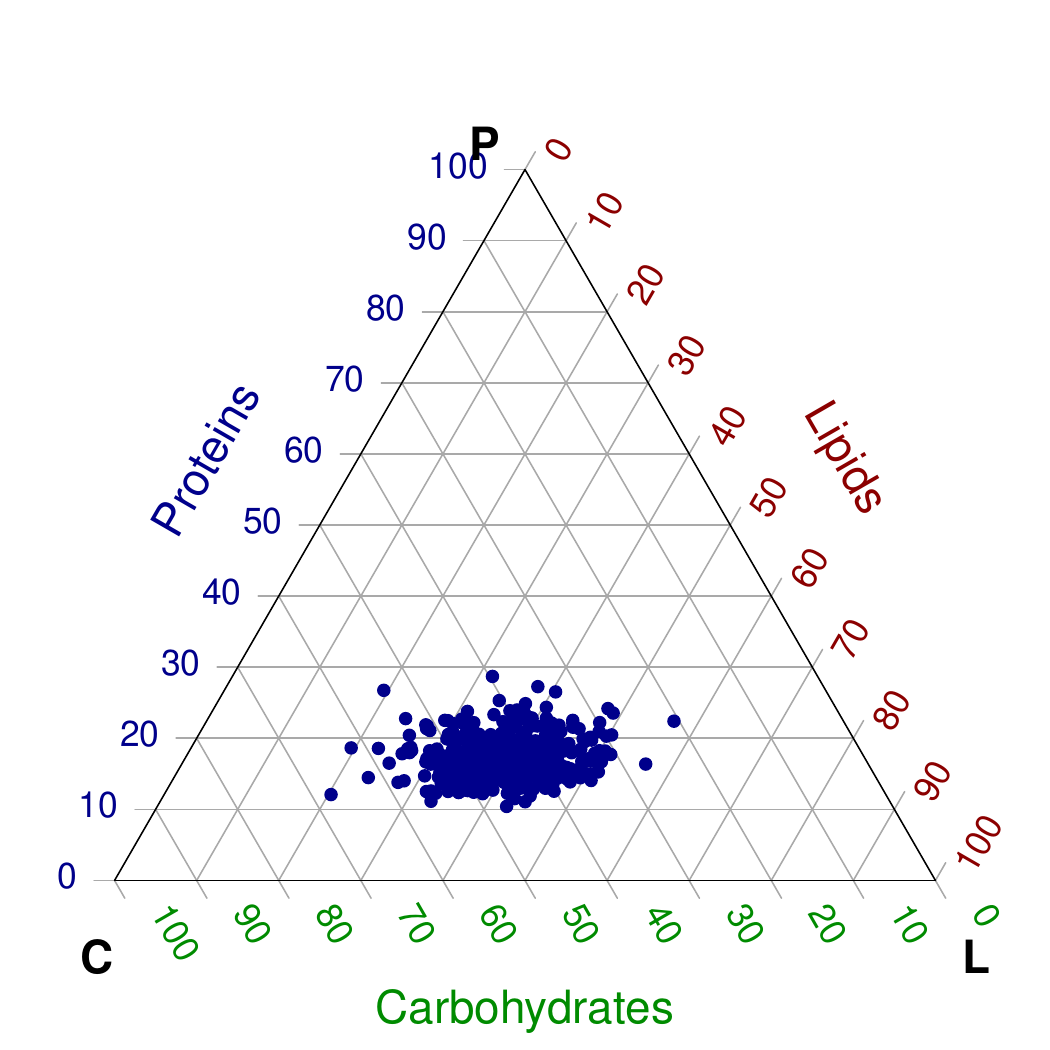} &	
			\includegraphics[scale=0.3]{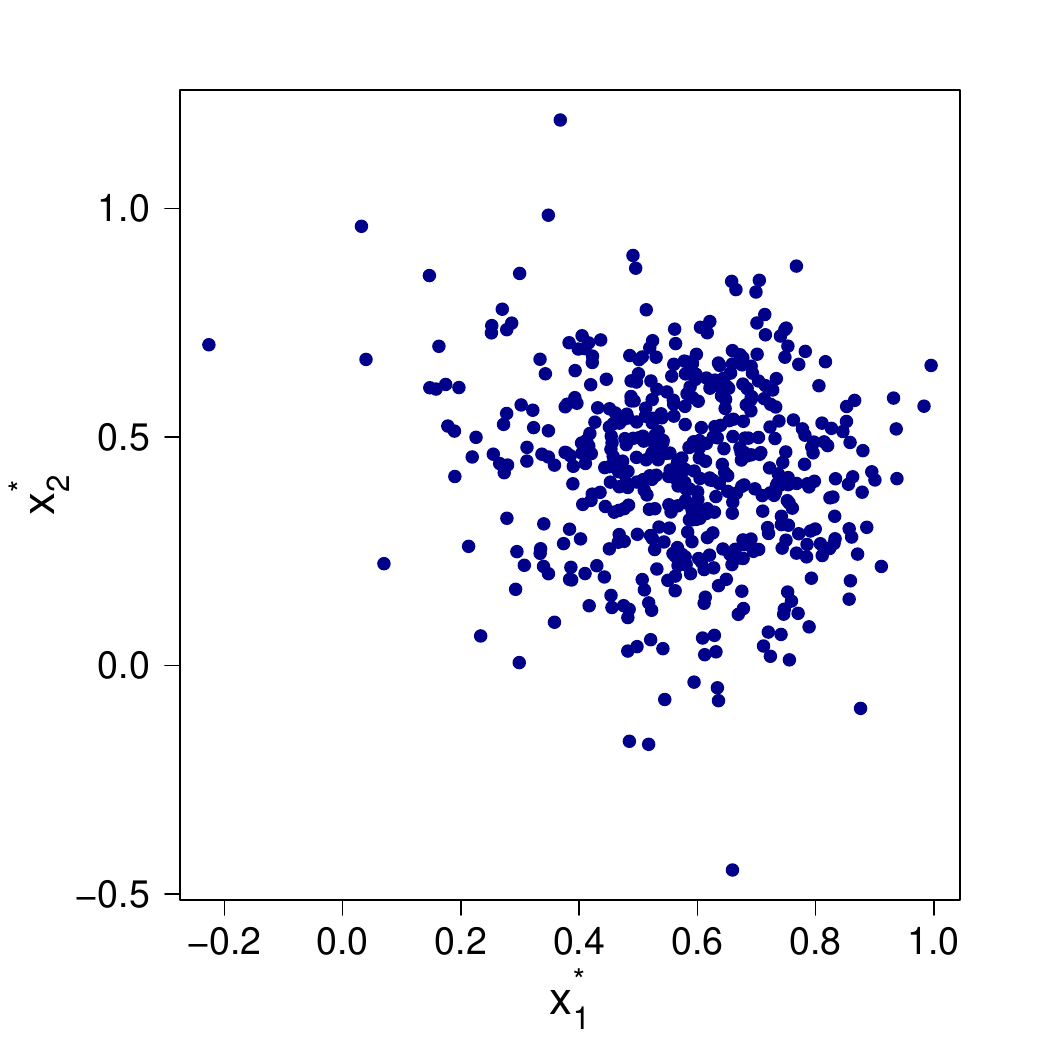}
		\end{tabular}
		\caption{ \small \label{fig:plot_ternary_nutricion}  Ternary and ilr representations for the nutrition data. }
	\end{center} 
\end{figure}

We choose as axis in the ilr representation given by $\bx^*=ilr(\bx)=(x_1^*,   x_2^* )\trasp$, those defined as
$$x_1^*=\sqrt{\frac 12}\log\left(\dst \frac{P}{L}\right)\qquad \qquad x_2^*=\sqrt{\frac 23}\log\left(\dst \frac{\sqrt{P\; L}}{C}\right)\;.$$
Figure \ref{fig:plot_ternary_nutricion} provides the ternary representation (left panel) together with the ilr one (right panel) for the considered data set.

The nonparametric regression estimators were computed using the same loss functions as in the simulation study, that is, for the classical estimator we use the squared loss function and for the robust ones, the scale parameter was obtained using the Tukey's bisquare function with tuning constants $c_0= 1.54764$ and $b = 1/2$, while the local $M-$smoother  was computed using the Tukey's bisquare function with tuning constant    $c_1=4.685$. Based on the simulation results we considered the local linear estimators.

\begin{figure}[ht!]
	\begin{center}
		\renewcommand{\arraystretch}{0.1}
		\newcolumntype{G}{>{\centering\arraybackslash}m{\dimexpr.34\linewidth-1\tabcolsep}}
		\hskip-0.6in
		\begin{tabular}{cc}
			Ternary Diagram &   ilr Plot \\
			\includegraphics[scale=0.3]{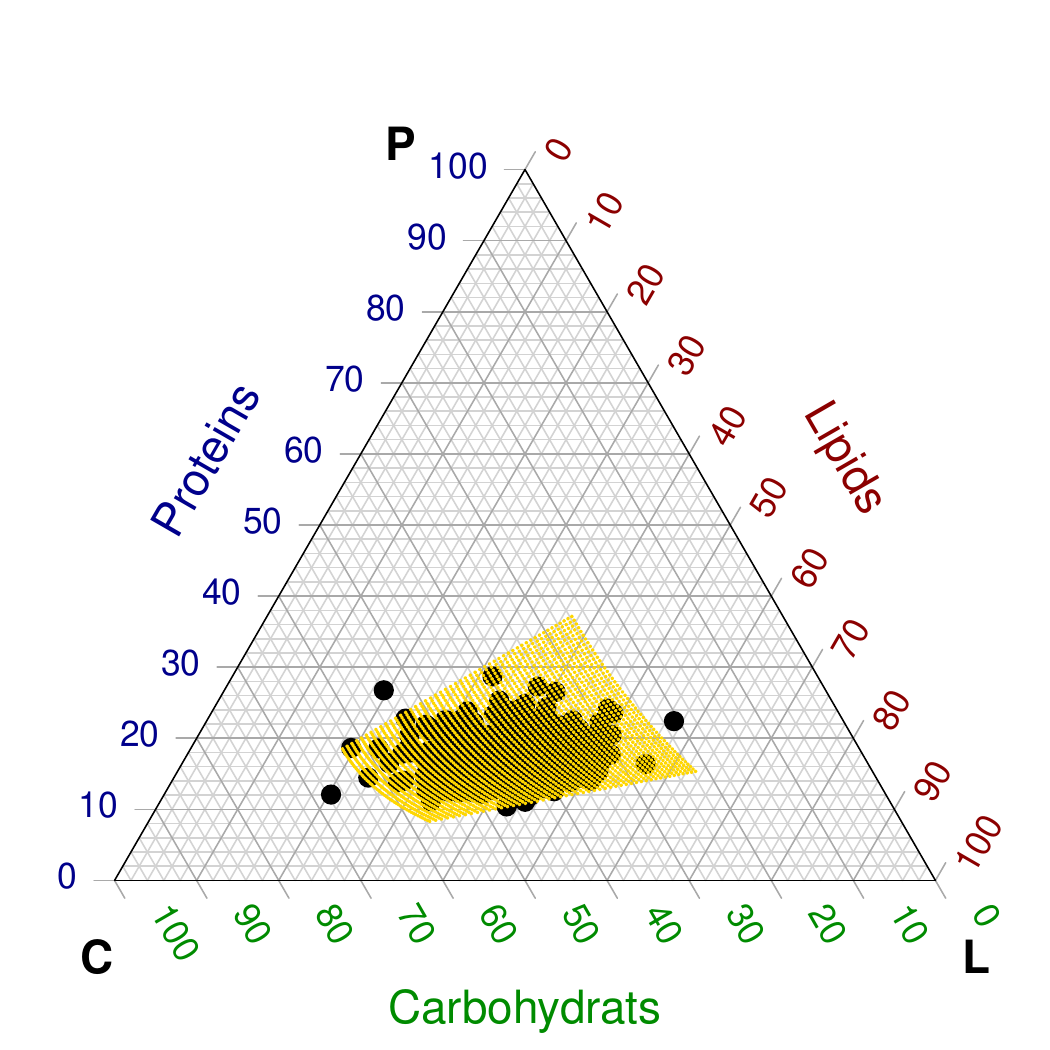} &
			\includegraphics[scale=0.3]{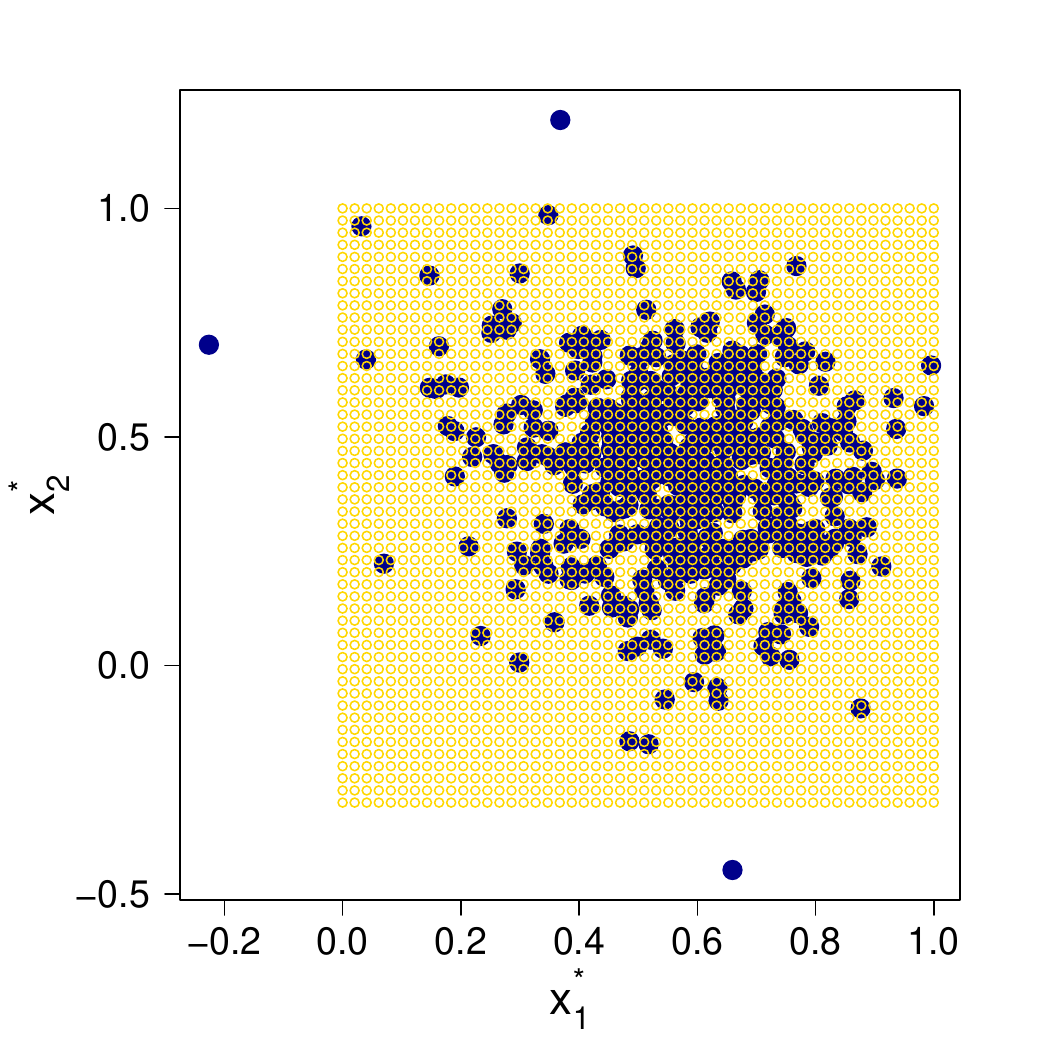}  
		\end{tabular}
		\caption{ \small \label{fig:grilla}  Grid points (in yellow) displayed on the ternary and ilr representations together with the data points. The left panel corresponds to the ternary diagram and the right one to the transformed data.}
	\end{center} 
\end{figure}

For the analysis, we evaluate the local linear estimators  at a grid of points, equally spaced in the ilr--space which was then transformed to obtain a grid of compositional points. The grid range in the ilr-space equals $[0,1]\times[-0.3,1]$ and has been chosen adapted to the  irl--space data range and the considered grid points are displayed in yellow in Figure \ref{fig:grilla} together with the data, in the ternary plot and the irl--space.  It is worth mentioning that the data points in the ilr--space   show a quite spherical shape, suggesting that the ilr--components may be independent and with the same dispersion.  

The estimators were  computed using as bandwidth matrix $\bH= h \identidad_2$, where the parameter $h$ was selected using the robust $K-$fold cross--validation procedure, as described in Section \ref{sec:band-selec}, with $K=5$ for the local linear $M-$smoother. For the   classical estimators based on the squared loss function defined in  \citet{Marzio:etal:2015},  $L^2-$ $K-$fold cross--validation was considered. The set of candidates $\itH$ for the bandwidth parameter corresponds to a  grid between 0.1 and 1 with step 0.1. The value 1 was selected since the maximum distance between points equal $1.66$ and more than 99\% of the distinct pair of points have distances lower or equal than 1. The grid was then refined around the obtained optimal bandwidth choosing a grid with step 0.02 around it.  The obtained bandwidths equal 0.44 for the classical procedure and 0.22 for the robust one. It is worth mentioning that the classical procedure leads to a larger bandwidth possible due to the effect of some vertical outliers.

\begin{figure}[ht!]
	\begin{center}
		\renewcommand{\arraystretch}{0.1}
		\newcolumntype{G}{>{\centering\arraybackslash}m{\dimexpr.34\linewidth-1\tabcolsep}}
		\hskip-0.6in
		\begin{tabular}{cc} 		
		(a) & (b) \\[-10ex]	
			\includegraphics[scale=0.5]{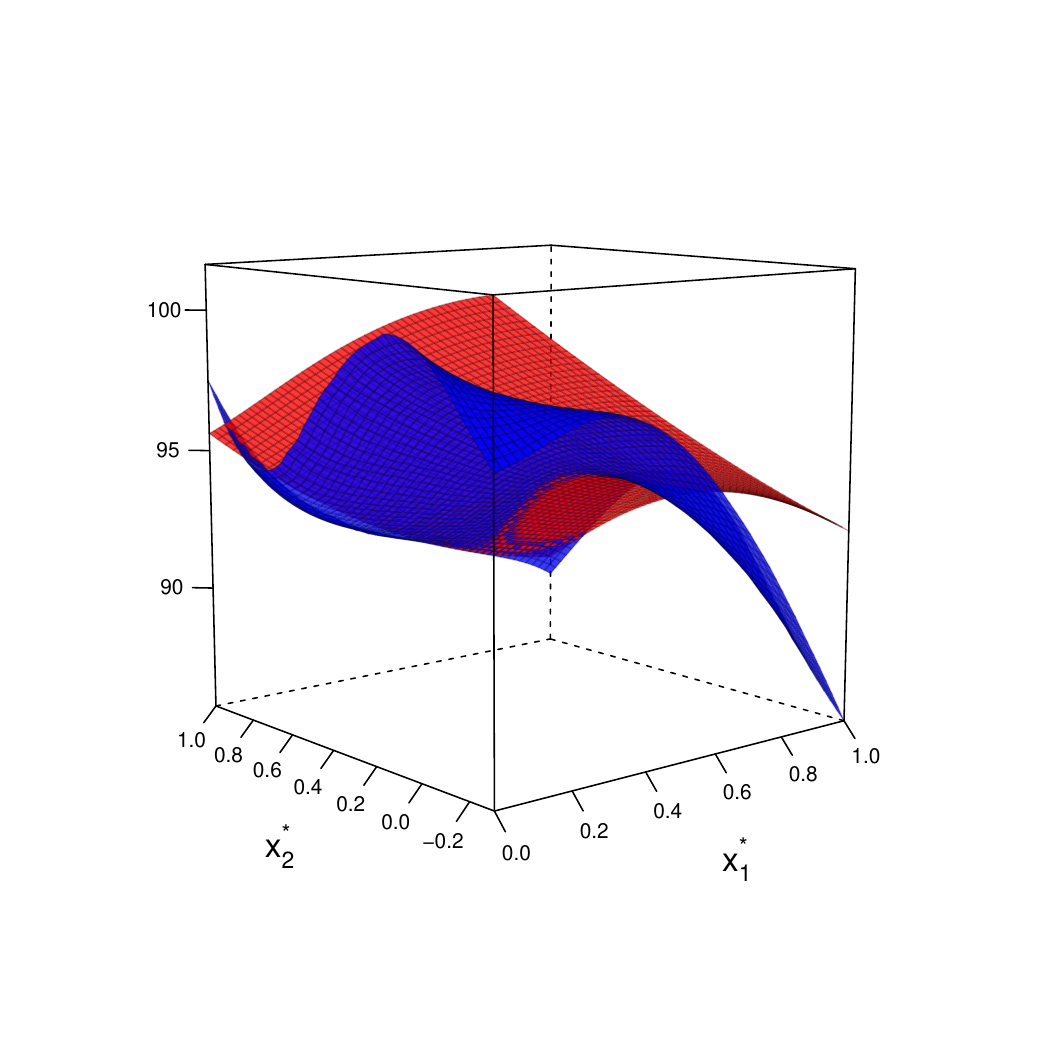} &
			\includegraphics[scale=0.5]{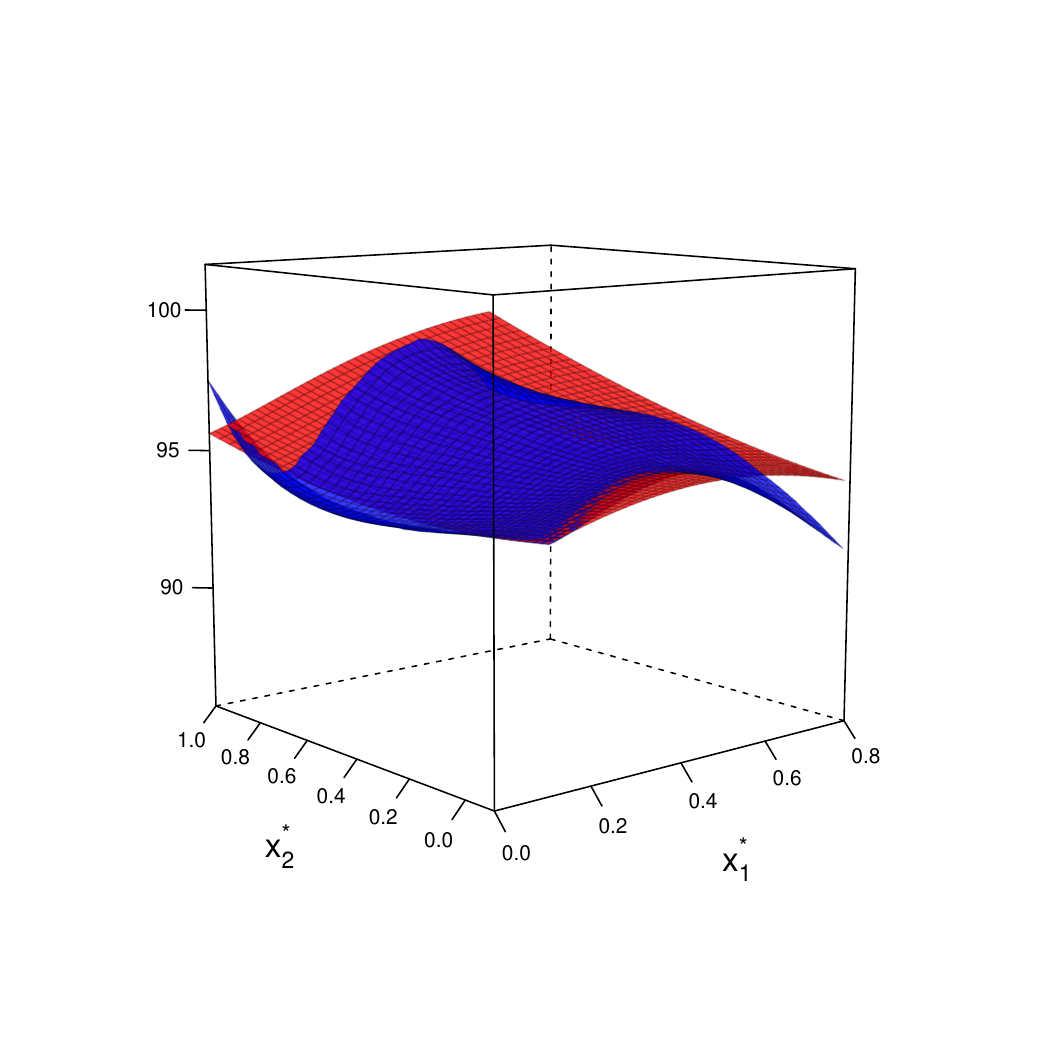} \\[-5ex]
			 \multicolumn{2}{c}{Rotated view}\\[-10ex]
			\includegraphics[scale=0.5]{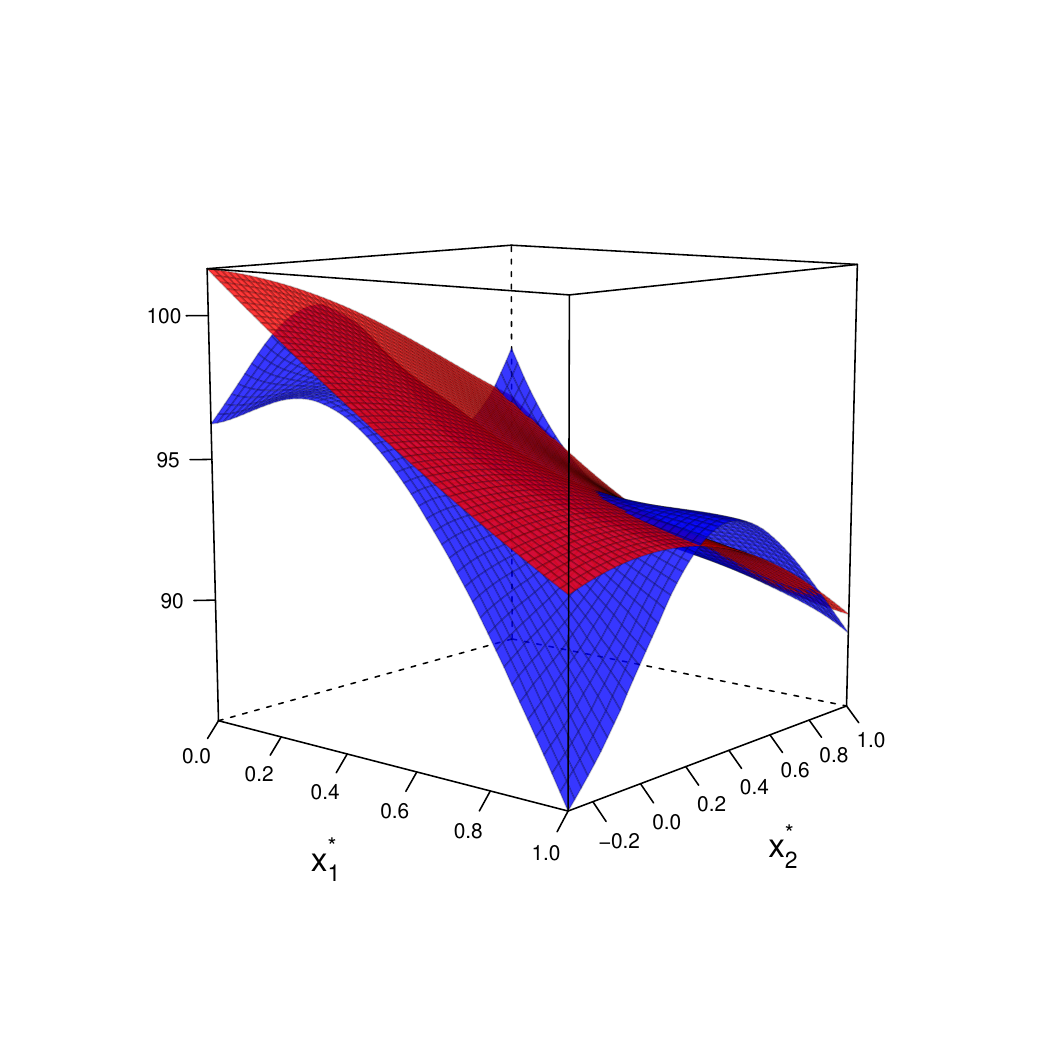}  &
			\includegraphics[scale=0.5]{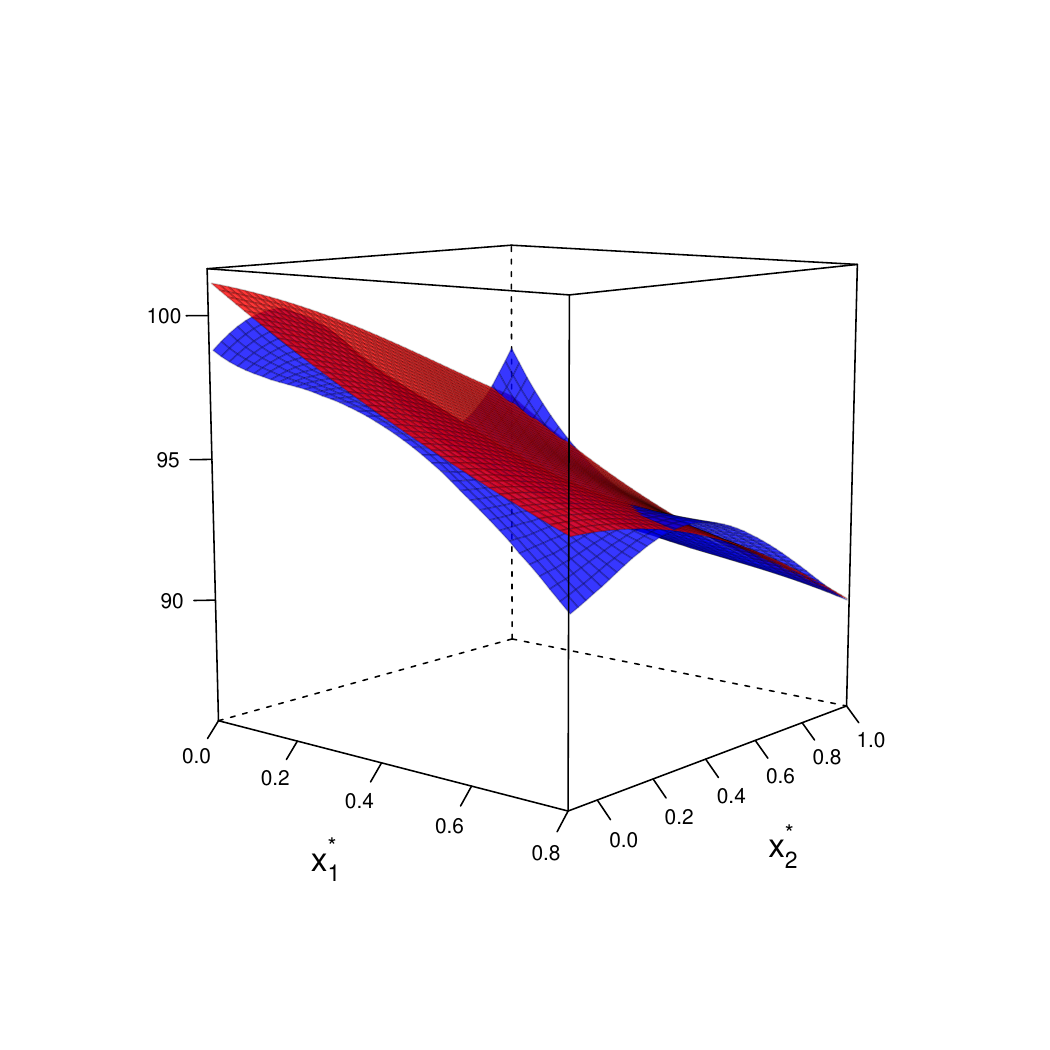}  
			
		\end{tabular}
				\vskip-0.3in
		\caption{ \small \label{fig:estimators-ilr}  Predicted surfaces in the ilr--space. The plots in the left column (a) correspond to the grid range $[0,1]\times[-0.3,1]$, while those in (b) are displayed on the reduced range $[0,0.8]\times[-0.1,1]$. The classical and robust estimators correspond to the red and blue surfaces, respectively.}
	\end{center} 
\end{figure}

The obtained estimators are displayed in Figure \ref{fig:estimators-ilr} in the ilr--space.  The classical and robust estimators correspond to the red and blue surfaces, respectively.

\begin{figure}[ht!]
	\begin{center}
		\includegraphics[scale=0.3]{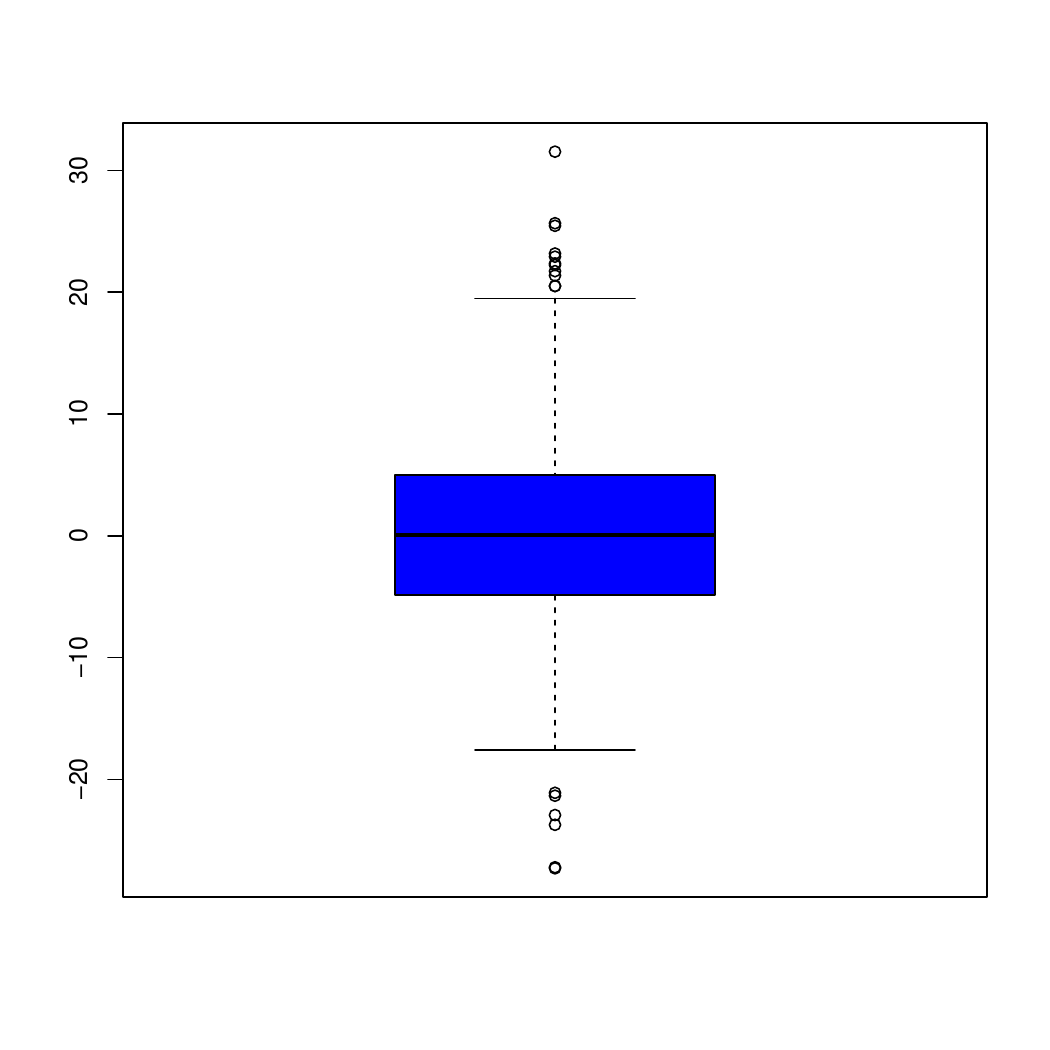}  
		\vskip-0.2in
				\caption{ \small \label{fig:boxplot} Boxplot of the residuals.}
	\end{center} 
\end{figure}

The robust estimators were also computed at the observed   covariates in order to  detect possible atypical observations by means of the boxplot of the residuals $\werre_i \, = \, y_i -  \wm_{\rob} (\bx_i)$,  $i=1, \dots, n$, where $\wm_{\rob}$ stands for the robust local linear smoother.  Figure \ref{fig:boxplot} displays the boxplot of the residuals which shows the presence of 17 atypical observations. Table \ref{tab:labe-atip}  reports the observation number ($N_{\outlier}$) and the glycaemic index  of the atypical observations detected. Going back to the records, it turned out that the patients whose observation number is presented in boldface in Table \ref{tab:labe-atip} may be   classified as prediabetic according to American Diabetes Association   using other characteristics such as the A1C criteria, see  \citet{ADA:2020}.
 
\begin{table}[ht!]
\caption{\small  Observation number  $N_{\outlier}$  of the atypical  detected observations and their corresponding glycaemic index value, $y_i$. Those in bold correspond to patients that were classified as pre--diabetic. \label{tab:labe-atip}}
	\begin{center}
 
\small	
	\setlength{\tabcolsep}{6pt}
 	 \begin{tabular}{c|c|c|c|c|c|c|c|c|c|}\hline
			$N_{\outlier}$ & \textbf{13} & 23 & \textbf{33} & \textbf{35} & \textbf{39} & 71 & \textbf{72}  &  98  & \\
		\hline
		$y_i$ &    \textbf{114.69} & 125.60 & \textbf{115.20} & \textbf{116.35} & \textbf{114.54} & 70.36 & \textbf{116.62} & 121.58    &
		\\
		\hline
		$N_{\outlier}$ & 107 & \textbf{112} & \textbf{135} & 164 & \textbf{175} & 245 & \textbf{357} & 413 & \textbf{478}  \\
		\hline
		$y_i$ &  68.10 & \textbf{117.96} & \textbf{115.16}  & 73.50 & 
		\textbf{66.48} & 72.94 & \textbf{116.69} & 72.64 & \textbf{117.05}    \\			\hline
		\end{tabular} 
\end{center} 
\end{table}	

\begin{figure}[ht!]
	\begin{center}
		\renewcommand{\arraystretch}{0.1}
		\newcolumntype{G}{>{\centering\arraybackslash}m{\dimexpr.34\linewidth-1\tabcolsep}}
		\hskip-0.6in
		\begin{tabular}{cc} 		
		(a) & (b) \\[-10ex]	
			\includegraphics[scale=0.5]{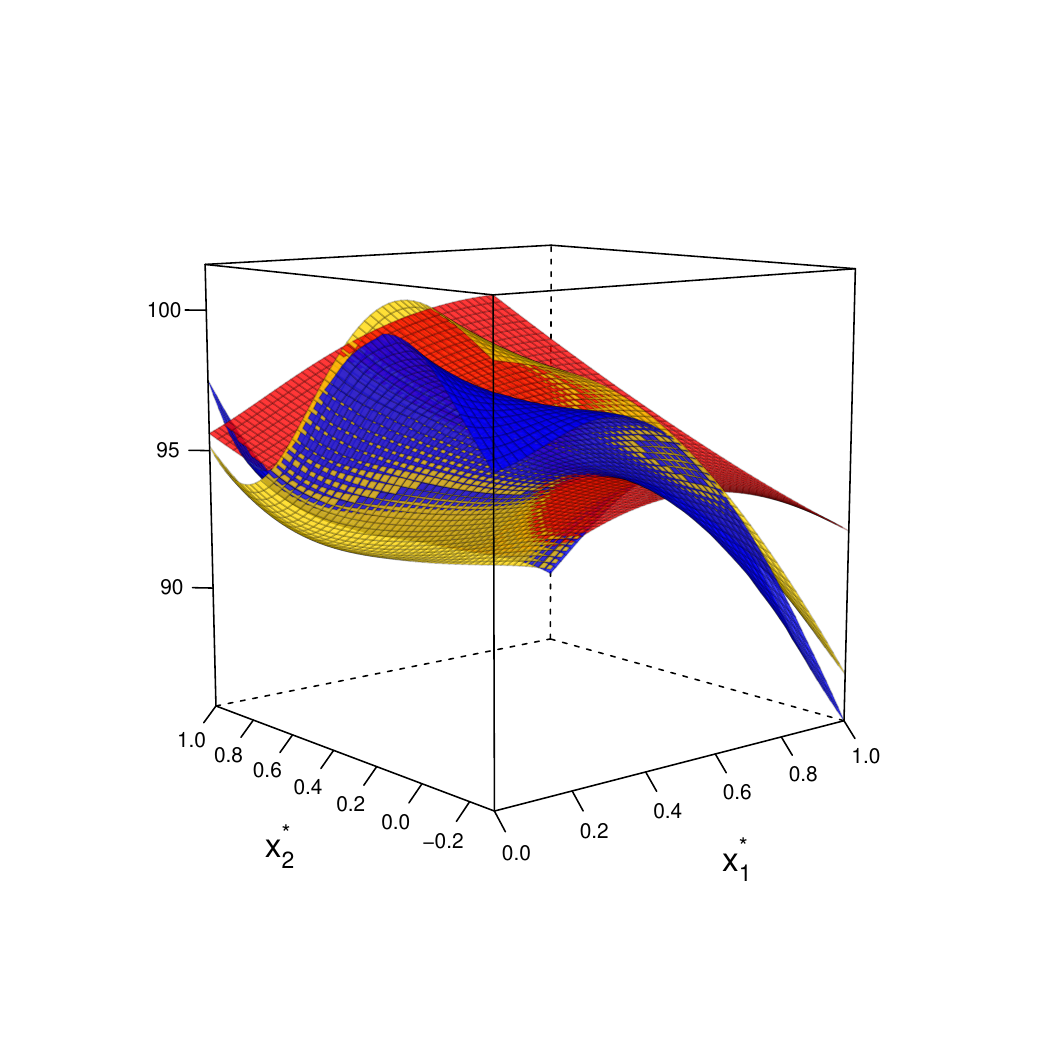} &
			\includegraphics[scale=0.5]{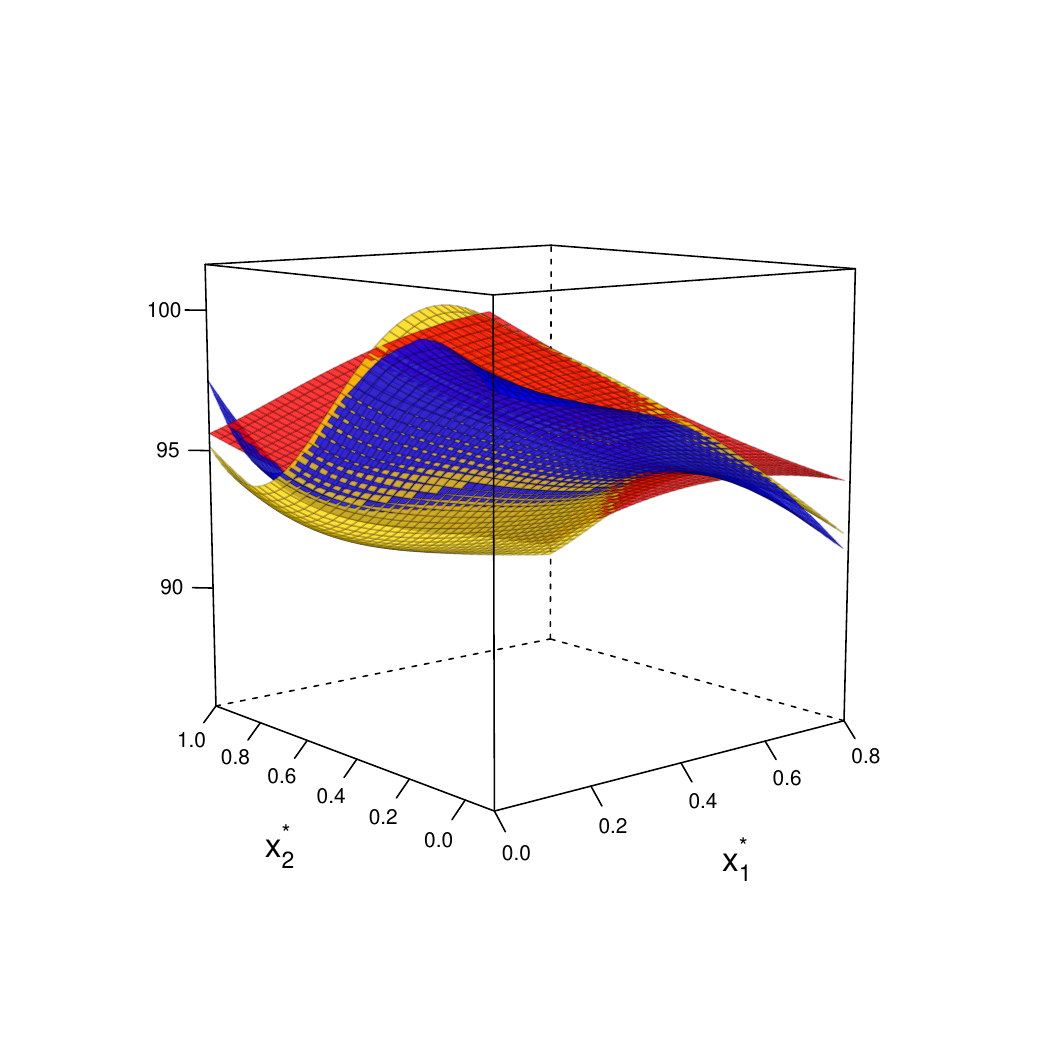} \\[-5ex]
			 \multicolumn{2}{c}{Rotated view}\\[-10ex]
			\includegraphics[scale=0.5]{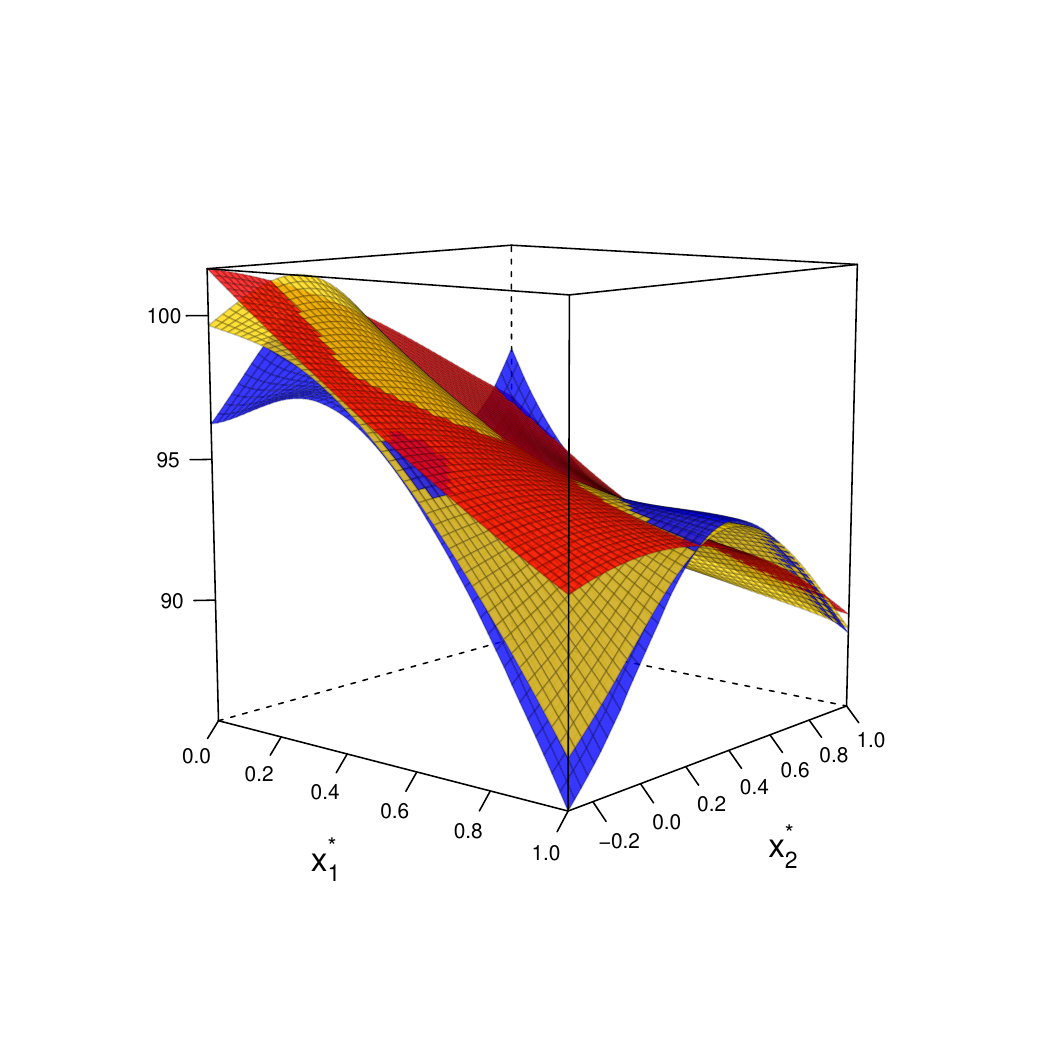}  &
			\includegraphics[scale=0.5]{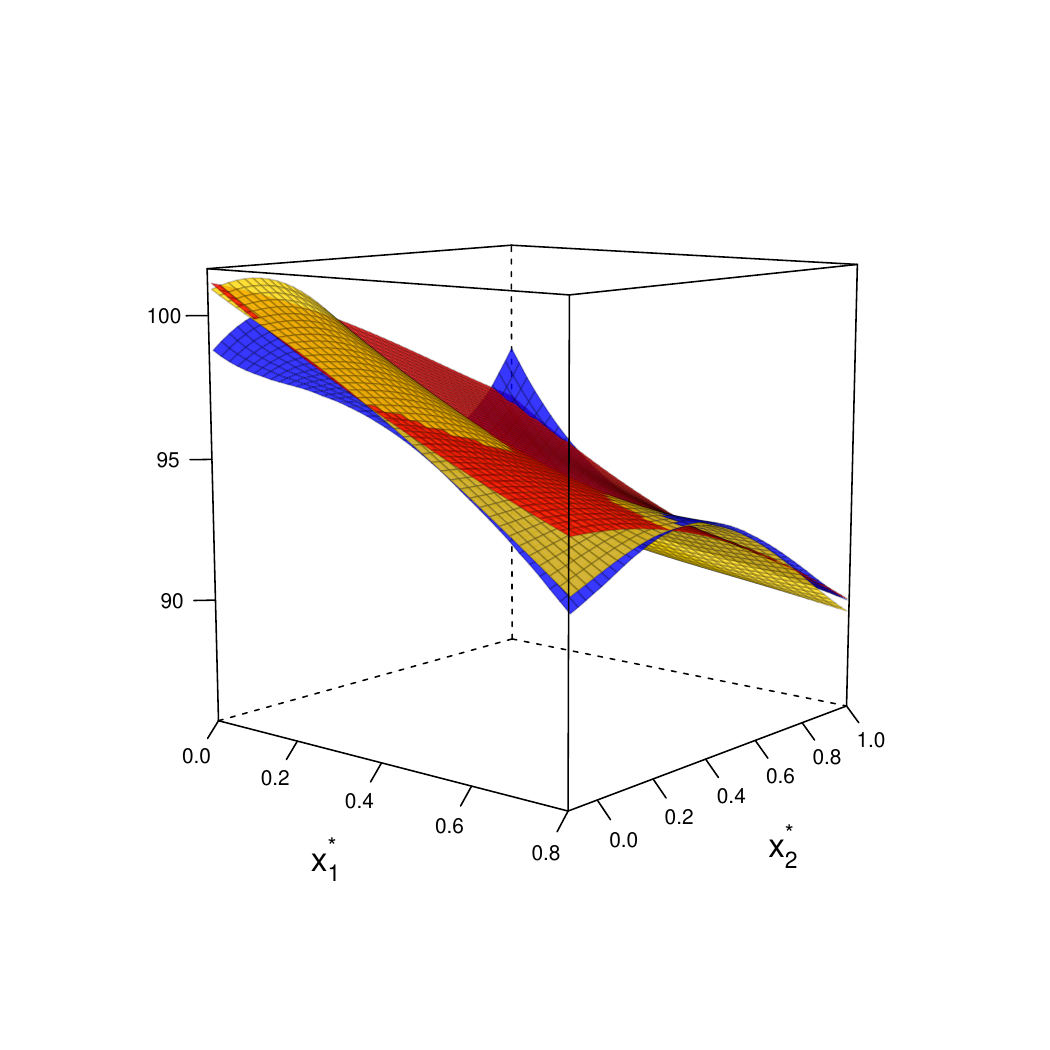}  
			
		\end{tabular}
				\vskip-0.3in
		\caption{ \small \label{fig:estimators-ilr-SO}  Predicted surfaces in the ilr--space. The plots in the left column (a) correspond to the grid range $[0,1]\times[-0.3,1]$, while those in (b) are displayed on the reduced range $[0,0.8]\times[-0.1,1]$. The classical and robust estimators computed with  the whole data set correspond to the red and blue surfaces, respectively, while the classical estimators  computed  without the atypical observations are displayed in yellow. }
	\end{center} 
\end{figure}

\begin{figure}[ht!]
	\begin{center}
		\renewcommand{\arraystretch}{0.1}
		\newcolumntype{G}{>{\centering\arraybackslash}m{\dimexpr.34\linewidth-1\tabcolsep}}
		\hskip-0.6in
		\begin{tabular}{cc} 		
		\includegraphics[scale=0.5]{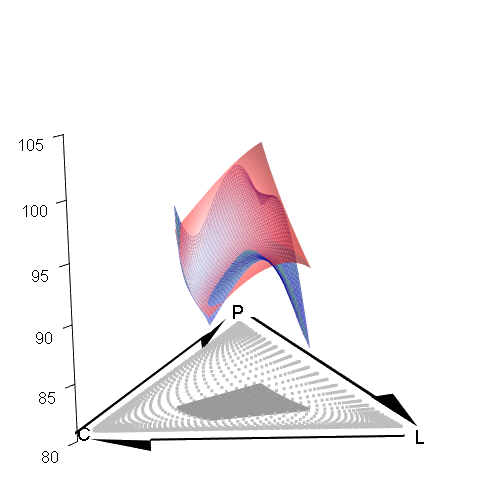} &
			\includegraphics[scale=0.5]{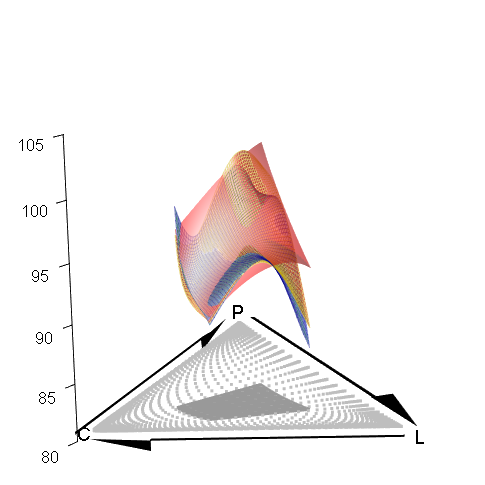} \\[-4ex]
			\multicolumn{2}{c}{\animategraphics[controls,width=0.74\linewidth,buttonsize=1.2em]{10}{AUC_cl_rob_clSO_KCV_3dplot}{0}{200}}
		\end{tabular}
				\vskip-0.1in
		\caption{ \small \label{fig:estimators-tern}  Predicted surfaces in the ternary diagram.   The classical and robust estimators computed with   the whole data set are displayed in the left panel and correspond to the red and blue surfaces, respectively. In the right panel these estimators are displayed with the classical estimators  computed  without the atypical observations that are displayed in yellow. The lower panel presents an animation to visualize the surfaces from different angles.}
	\end{center} 
\end{figure}

The classical estimators  computed  without these possible atypical observations are displayed in yellow in Figure \ref{fig:estimators-ilr-SO} in the ilr-space. The classical $K-$fold cross--validation bandwidth equals now 0.26 which is a value closer to the one obtained by the robust cross--validation with the whole data which was 0.22. Besides, Figure \ref{fig:estimators-tern} shows the estimators in   the ternary diagram. Again, the classical and robust estimators computed with   the whole data set  correspond to the red and blue surfaces, respectively, while the   classical estimators  computed  without the atypical observations   are displayed in yellow in the right panel.  The selected grid points is displayed in dark gray within the ternary diagram. Note that the classical estimators computed without these potential
outliers are very close to the robust ones. In other words, the robust estimator behaves similarly to the
classical one if one were able to manually remove suspected outliers.

\noi \textbf{Acknowledgements.}{This research was partially supported by Grants   \textsc{pict} 2021-I-A-00260 from \textsc{anpcyt} and    20020170100022\textsc{ba} from the Universidad de Buenos Aires, Argentina  and also by the Spanish Project  from the Ministry of Science and Innovation (MCIU/ AEI /10.13039/501100011033) PID2020-116587GB-I00, Spain.}
\bibliographystyle{apalike}
\bibliography{Biblio_comp}

\begin{thebibliography}{}

\bibitem[ADA, 2020]{ADA:2020}
ADA (2020).
\newblock {\em 2. Classification and diagnosis of diabetes: standards of
  medical care in diabetes-2020. Diabetes care, 43 (Supplement 1), S14-S31}.
\newblock American Diabetes Association.

\bibitem[Aitchison, 1982]{Aitchison:1982}
Aitchison, J. (1982).
\newblock The statistical analysis of compositional data.
\newblock {\em Journal of the Royal Statistical Society. Series B
  (Methodological)}, 44:139--177.

\bibitem[Aitchison, 2003]{Aitchison:2003}
Aitchison, J. (2003).
\newblock {\em The {S}tatistical {A}nalysis of {C}ompositional {D}ata}.
\newblock The Blackburn Press.

\bibitem[Aitchison and Bacon-Shone, 1984]{Aitchison:BaconShone:1984}
Aitchison, J. and Bacon-Shone, J. (1984).
\newblock Log contrast models for experiments with mixtures.
\newblock {\em Biometrika}, 71:323--330.

\bibitem[Aitchison and Lauder, 1985]{Aitchison:Lauder:1985}
Aitchison, J. and Lauder, I.~J. (1985).
\newblock Kernel density estimation for compositional data.
\newblock {\em Journal of te Royal Statistical Society. Series C}, 34:129--137.

\bibitem[Aitchison and Shen, 1980]{Aitchison:1980}
Aitchison, J. and Shen, S.~M. (1980).
\newblock Logistic-normal distributions: some properties and uses.
\newblock {\em Biometrika}, 67:261--272.

\bibitem[Bianco and Boente, 2007]{Bianco:Boente:2007}
Bianco, A. and Boente, G. (2007).
\newblock Robust estimators under a semiparametric partly linear autoregression
  model: {A}symptotic behaviour and bandwidth selection.
\newblock {\em Journal of Time Series Analysis}, 28:274--306.

\bibitem[Boente and Fraiman, 1989]{boente:fraiman:1989}
Boente, G. and Fraiman, R. (1989).
\newblock Robust nonparametric regression estimation.
\newblock {\em Journal of Multivariate Analysis}, 29:180--198.

\bibitem[Boente et~al., 1997]{boente:etal:1997}
Boente, G., Fraiman, R., and Meloche, J. (1997).
\newblock Robust nonparametric regression estimation.
\newblock {\em Journal of Statistiscal Planning and Inference}, 57:109--142.

\bibitem[Boente and Martinez, 2017]{boente:martinez:2017}
Boente, G. and Martinez, A. (2017).
\newblock Marginal integration {$M-$}estimators for additive models.
\newblock {\em Test}, 26:231--260.

\bibitem[Boente and Rodriguez, 2008]{boente:rodriguez:2008}
Boente, G. and Rodriguez, D. (2008).
\newblock Robust bandwidth selection in semiparametric partly linear regression
  models: {M}onte {C}arlo study and influential analysis.
\newblock {\em Computational Statistics and Data Analysis}, 52:383--397.

\bibitem[Cantoni and Ronchetti, 2001]{Cantoni:Ronchetti:2001}
Cantoni, E. and Ronchetti, E. (2001).
\newblock Resistant selection of the smoothing parameter for smoothing splines.
\newblock {\em Statistics and Computing}, 11:141--146.

\bibitem[Chac\'on et~al., 2011]{Chacon:etal:2011}
Chac\'on, J.~E., Mateu-Figueras, G., and Mart\'{\i}n-Fern\'andez, J.~A. (2011).
\newblock Gaussian kernels for density estimation with compositional data.
\newblock {\em Computers and Geosciences}, 37:702--711.

\bibitem[Cleveland, 1979]{cleveland:1979}
Cleveland, W. (1979).
\newblock Robust locally weighted regression and smoothing scatterplots.
\newblock {\em Journal of the American Statistical Association}, 74:829--836.

\bibitem[Coenders and Pawlowsky-Glahn, 2020]{Coenders:Pawlowsly:2020}
Coenders, G. and Pawlowsky-Glahn, V. (2020).
\newblock On interpretations of tests and effect sizes in regression models
  with a compositional predictor.
\newblock {\em SORT. Statistics and Operations Research Transactions},
  44:201--220.

\bibitem[Connor and Mosimann, 1969]{Connor:Mosimann:1969}
Connor, R.~J. and Mosimann, J.~E. (1969).
\newblock Concepts of independence for proportions with a generalization of the
  dirichlet distribution.
\newblock {\em Journal of the American Statistical Association}, 64:194--206.

\bibitem[{Di~Marzio} et~al., 2015]{Marzio:etal:2015}
{Di~Marzio}, M., Panzera, A., and Venieri, C. (2015).
\newblock Non-parametric regression for compositional data.
\newblock {\em Statistical Modelling}, 15:113--133.

\bibitem[Egozcue et~al., 2012]{Egozcue:etal:2012}
Egozcue, J.~J., Daunis-I-Estadella, J., Pawlowsky-Glahn, V., Hron, K., and
  Filzmoser, P. (2012).
\newblock Simplicial regression: the normal model.
\newblock {\em Journal of Applied Probability and Statistics}, 6:87--108.

\bibitem[Egozcue et~al., 2003]{Egozcue:etal:2003}
Egozcue, J.~J., Pawlowsky-Glahn, V., Mateu-Figueras, G., and Barcel{\'o}-Vidal,
  C. (2003).
\newblock Isometric logratio transformations for compositional data analysis.
\newblock {\em Mathematical Geology}, 35:279--300.

\bibitem[Greenacre, 2021]{Greenacre:2021}
Greenacre, M. (2021).
\newblock Compositional data analysis.
\newblock {\em Annual Review of Statistics and Its Application}, 8:271--299.

\bibitem[Gude et~al., 2016]{Gude:etal:2016}
Gude, F., D{\'i}az-Vidal, P., R{\'u}a-P{\'e}rez, C., Alonso-Sampedro, M.,
  Fern{\'a}ndez-Merino, C., Rey-Garc{\'i}a, J., Cadarso-Su{\'a}rez, C.,
  Pazos-Couselo, M., Garc{\'i}a-L{\'o}pez, J.~M., and Gonzalez-Quintela, A.
  (2016).
\newblock Glycemic variability and its association with demographics and
  lifestyles in a general adult population.
\newblock {\em Journal of Diabetes Science and Technology}, 11:780--790.

\bibitem[H\"ardle and Tsybakov, 1988]{hardle:tsybakov:1988}
H\"ardle, W. and Tsybakov, A.~B. (1988).
\newblock Robust nonparametric regression with simultaneous scale curve
  estimation.
\newblock {\em Annals of Statistics}, 16:120--135.

\bibitem[Hijazi and Jernigan, 2009]{hijazi:jernigan:2009}
Hijazi, R.~H. and Jernigan, R.~W. (2009).
\newblock Modelling compositional data using dirichlet regression models.
\newblock {\em Journal of Applied Probability and Statistics}, 4:77--91.

\bibitem[Hubert and Vandervieren, 2008]{Hubert:Vandervieren:2008}
Hubert, M. and Vandervieren, E. (2008).
\newblock An adjusted boxplot for skewed distributions.
\newblock {\em Computational Statistics and Data Analysis}, 52:5186--5201.

\bibitem[Jiang and Mack, 2001]{jiang:mack:2001}
Jiang, J. and Mack, Y. (2001).
\newblock Robust local polynomial regression for dependent data.
\newblock {\em Statistica Sinica}, 11:705--722.

\bibitem[Leung, 2005]{Leung:2005}
Leung, D. (2005).
\newblock Cross-validation in nonparametric regression with outliers.
\newblock {\em Annals of Statistics}, 33:2291--2310.

\bibitem[Leung et~al., 1993]{Leung:etal:1993}
Leung, D., Marriott, F., and Wu, E. (1993).
\newblock Bandwidth selection in robust smoothing.
\newblock {\em Journal of Nonparametric Statistics}, 4:333--339.

\bibitem[Maronna et~al., 2019]{Maronna:etal:2019}
Maronna, R., Martin, D., Yohai, V., and Salibi{\'a}n-Barrera, M. (2019).
\newblock {\em Robust {S}tatistics: {T}heory and {M}ethods (with
  {\texttt{R}})}.
\newblock John Wiley and Sons.

\bibitem[Mateu-Figueras and Pawlowsky-Glahn,
  2008]{MateuFigueras:Pawlowsky:2008}
Mateu-Figueras, G. and Pawlowsky-Glahn, V. (2008).
\newblock A critical approach to probability laws in~geochemistry.
\newblock {\em Mathematical Geosciences}, 40:489--502.

\bibitem[Pawlowsky-Glahn et~al., 2015]{Pawlowsky:etal:2015}
Pawlowsky-Glahn, V., Egozcue, J.~J., and Tolosana-Delgado, R. (2015).
\newblock {\em Modeling and {A}nalysis of {C}ompositional {D}ata}.
\newblock Wiley, Chichester.

\bibitem[Salibi\'an-Barrera, 2026]{Salibian:2023}
Salibi\'an-Barrera, M. (2026).
\newblock Robust nonparametric regression: Review and practical considerations.
\newblock {\em Econometrics and Statistics}, 39:294--309.

\bibitem[Scheff{\'{e}}, 1958]{Scheffe:1958}
Scheff{\'{e}}, H. (1958).
\newblock Experiments with mixtures.
\newblock {\em Journal of the Royal Statistical Society: Series B
  (Methodological)}, 20:344--360.

\bibitem[Scheff{\'{e}}, 1963]{Scheffe:1963}
Scheff{\'{e}}, H. (1963).
\newblock The simplex-centroid design for experiments with mixtures.
\newblock {\em Journal of the Royal Statistical Society: Series B
  (Methodological)}, 25:235--251.

\bibitem[Service, 2013]{Service:2013}
Service, F.~J. (2013).
\newblock Glucose variability.
\newblock {\em Diabetes}, 62(5):1398–1404.

\bibitem[Tsagris et~al., 2023]{Tsagris:etal:2023}
Tsagris, M., Alenazi, A., and Stewart, C. (2023).
\newblock Flexible non-parametric regression models for compositional response
  data with zeros.
\newblock {\em Statistics and Computing}, 33:1--17.

\bibitem[Tsagris and Stewart, 2018]{Tsagris:Stewart:2018}
Tsagris, M. and Stewart, C. (2018).
\newblock A dirichlet regression model for compositional data with zeros.
\newblock {\em Lobachevskii Journal of Mathematics}, 39:398--412.

\bibitem[Wang and Scott, 1994]{Wang:Scott:1994}
Wang, F. and Scott, D. (1994).
\newblock The $l_1$ method for robust nonparametric regression.
\newblock {\em Journal of the American Statistical Association}, 89:65--76.

\bibitem[Wheeler et~al., 2012]{Wheeler:etal:2012}
Wheeler, M.~L., Dunbar, S.~A., Jaacks, L.~M., Karmally, W., Mayer-Davis, E.~J.,
  Wylie-Rosett, J., and Yancy, W.~S. (2012).
\newblock Macronutrients, food groups and eating patterns in the management of
  diabetes: a systematic review of the literature, 2010.
\newblock {\em Diabetes Care}, 35 (2):434--445.

\end{thebibliography}

\end{document}